\documentclass[12pt,a4paper]{article}
\usepackage{authblk}
\usepackage{indentfirst}
\usepackage{graphicx}
\usepackage{epsfig}

\usepackage{latexsym}
\usepackage{amsmath}
\usepackage{amssymb}
\usepackage{amsfonts}
\usepackage{mathrsfs}
\usepackage{pifont}

\usepackage{amsthm}
\usepackage{float}
\usepackage{subfigure}
\usepackage{color}
\usepackage{multicol}
 \usepackage[flushleft]{threeparttable} 
\usepackage[colorlinks,linkcolor=blue,filecolor=black,citecolor=blue]{hyperref}

\newtheorem{theorem}{Theorem}[section]

 \newtheorem{remark}[theorem]{Remark}

\usepackage{ulem}
\title{A Tridomain Model for Potassium Clearance in Optic Nerve of Necturus 
\author[1]{Yi Zhu}
\author[2]{Shixin Xu\thanks{Corresponding author: shixin.xu@dukekunshan.edu.cn}}
\author[3]{Robert.S. Eisenberg}
\author[5,1,6]{Huaxiong Huang}

\affil[1]{\small Department of Mathematics and Statistics, York University, Toronto, Ontario, Canada.} 
\affil[2]{\small Duke Kunshan University, 8 Duke Ave, Kunshan, Jiangsu, China.}
\affil[3]{\small  Department of Applied Mathematics, Illinois Institute of Technology, Chicago IL 60616 USA.}
\affil[4]{\small Computer Science, University of Toronto, Toronto, Ontario, Canada.}
\affil[5]{\small Joint Mathematical Research Centre of Beijing Normal University and BNU-HKBU United International College, Zhuhai, China}
\affil[6]{\small Division of Science and Technology, BNU- HKBU United International College, Zhuhai, 519087,	China}}

\date{}
\begin{document}
\maketitle

\section*{Abstract}
The accumulation of potassium in the narrow space outside nerve cells is a classical subject of
biophysics that has received much attention recently. It may be involved in potassium
accumulation \textcolor{black}{including} spreading depression, perhaps migraine and some kinds of epilepsy, even
(speculatively) learning. Quantitative analysis is likely to help evaluate the role of potassium
clearance from the extracellular space after a train of action potentials. Clearance involves three structures that extend down the length of the nerve: glia, extracellular space, and axon and so need to be described as systems distributed in space in the tradition used for electrical potential in the `cable equations' of nerve since the work of Hodgkin in 1937. A three-compartment model is proposed here for the optic nerve and is used to study the accumulation of potassium and its clearance. The model allows the convection, diffusion, and electrical migration of water and ions. We depend on the data of Orkand et al to ensure the relevance of our model and align its parameters with the anatomy and properties of membranes, channels, and transporters: our model fits their
experimental data quite well. The aligned model shows that glia has an important role in
buffering potassium, as expected. The model shows that potassium is cleared mostly by
convective flow through the syncytia of glia driven by osmotic pressure differences. A simplified model might be possible, but it must involve flow down the length of the optic nerve. It is easy for compartment models to neglect this flow. Our model can be used for structures quite different from the optic nerve that might have different distributions of channels and
transporters in its three compartments. It can be generalized to include a fourth (distributed) compartment representing blood vessels
to deal with the glymphatic flow into the circulatory system.

\section{Introduction}
\label{introduction}
The now classical experiments of Hodgkin, Huxley, and Katz  \cite{hodgkin1952measurement,hodgkin1949ionic} were designed to
avoid the artifact of concentration polarization, the (significant) change of
concentration of ions as the currents maintaining their voltage clamp flowed across
electrodes inside the axon and its membranes. Change of concentration of potassium
was seen nonetheless after a few milliseconds of outward potassium currents (\cite{hodgkin1952currents}: p.
482, 485, 489, 494; \cite{frankenhaeuser1956after}), but the change was slow enough to be ignored in their
calculations of the action potential \cite{hodgkin1952quantitative}.

The Kuffler group at Harvard was interested in the role of glia \cite{kuffler1966physiology,kuffler1966physiological} in the central
nervous system and showed that the membrane potential of the optic nerve of
\textbf{\textit{Necturus}} reported the potassium concentration in the narrow extracellular space just
outside axons, between glia and axon. Orkand et al \cite{orkand1966effect} used this membrane potential
of glia to report the change in potassium concentration—the polarization of
concentration in the narrow extracellular space—as it accumulated during a train of
nerve action potentials. Earlier work \cite{hodgkin1952currents,frankenhaeuser1956after} had inferred this concentration change.
Orkand et al \cite{kuffler1966physiological} measured it quite directly. The artifact of concentration polarization
that so worried Hodgkin and Huxley became the experimental reality of potassium
accumulation \cite{hodgkin1952currents} in the central nervous system \cite{orkand1966effect}, that interested the Kuffler group.
Interest has only grown in the following fifty some odd years.

Potassium accumulation and flow in the extracellular space have been shown to have
important roles in many papers focused on aging, Alzheimers disease, anesthesia,
dementia, diabetes, epilepsy, migraine, sleep, stroke and traumatic brain injury.
Potassium accumulation and flow play an important role in the biology of the central
nervous system \cite{nedergaard2020glymphatic}  normal, and abnormal \cite{o2016effects,tuttle2019computational,ayata2015spreading,wei2014unification,ullah2015role,mori2015multidomain,miura2007cortical}. The glymphatic model has received
much interest in the last months. It links potassium accumulation, flows in the
extracellular space—particularly in sleep and diseases of aging—with general disposal
of waste through glial pathways to the circulatory system  \cite{nedergaard2020glymphatic,gakuba2018general,jiang2017impairment,abbott2018role,jessen2015glymphatic,mestre2020brain}.

The accumulation of ions in a narrow extracellular space, like that between nerve and
glia, or Schwann cell \cite{frankenhaeuser1956after}, depends on the diffusion, convection, and migration of ions in an electric field. Convection is likely to be important. Evolution uses the circulatory
system to provide convective transport close to nearly every cell in a mammal, the lens of the eye being a notable exception, so the
delays involved in electrodiffusion are overcome. Convection provides what diffusion
denies: speed.

The optic nerve, and the mammalian central nervous system in general, involves nerve,
glia and narrow extracellular space. It involves three types of flow, convection,
diffusion, and migration in the electric field in radial and longitudinal directions of a cylindrical structure. It involves (chiefly) three ions (K, Na, Cl), and a number of
different types of channels and pumps (voltage activated Na channels and at least two
types of K channels and the Na/K pump). The description and analysis, not to say the
numerical computation of the optic nerve, must deal with what is actually in the optic
nerve. It must deal with what evolution has actually built in the central nervous system
in general. Thus, the analysis must involve many forces, flows, structures, and
channels and transporters. Here we report the subset of our work that deals with the
accumulation of potassium in the narrow extracellular space of optic nerve.


The interactions between neuronal cells and glial cells have been included in
models of the important phenomenon of spreading depression \cite{tuckwell1978mathematical,postnov2007functional,chang2013mathematical,yao2011continuum} thought to be
related to epilepsy and migraine. Some \textcolor{black}{two-compartment} models for potassium
clearance (or spatial buffering) include interactions between neuron cells and
extracellular space \cite{bellinger2008submyelin} or interactions between glial cells and extracellular space \cite{chen2000spatial,reichenbach1993retinal,ostby2009astrocytic}. A tri-compartment model using ordinary differential equations (ODEs) was
introduced by Sibille \cite{sibille2015neuroglial} to study the role of $\mathrm{K_{ir4}}$ channels. It shows that the flows
play an important role in the central nerve system \cite{bellot2017astrocytic} via influx and efflux routes to help waste clearance, which has been called (with understandable enthusiasm if not
hyperbole) a final frontier of neuroscience \cite{nicholson2017brain}. Some models, including flow but not
electrodiffusion, were introduced to study the pressure effect on the flow \cite{hou2016intracranial,morgan2016cerebrospinal,killer2006cerebrospinal,band2009intracellular}.
Mori \cite{mori2015multidomain} proposed a multidomain model for cortical spreading depression, where
ionic electrodiffusion and osmosis between different \textcolor{black}{compartments} are considered. In this paper, 
we first extend those results and  the two-compartment structural analysis of the spherical lens \cite{zhu2019bidomain} to the
three-compartment cylindrical optic nerve fiber of \textbf{\textit{Necturus}}. 
Then we present some general conclusions based on the analysis
of a specific set of experiments Orkand et al \cite{kuffler1966physiological,orkand1966effect} using a model distributed in space in both radial and longitudinal directions and that involves action potentials generated by Hodgkin Huxley equations.

This paper is organized as follows. The full model for microcirculation of water and
ions are proposed based on conservation laws in Section \ref{sec:Model}. Then the
model is calibrated and aligned with the Orkand experiment results in Section \ref{sec:Calibration}. In Section \ref{sec:clearance}, the
calibrated model is used to study the flow and ion microcirculations during Potassium
clearance. A discussion on the parameters is presented in Section \ref{sec:discussion}. Then the conclusions and future work are given in Section \ref{sec:Conclusion}.

\section{Mathematical model} \label{sec:Model}
\label{Mathematical model}
In this section, we present a tridomain model for microcirculation of the optic nerve.
The model deals with two types of flow: the circulation of water (hydrodynamics) and
the circulation of ions (electrodynamics) in the
\begin{itemize}
	\item glial compartment $(\Omega_{gl})$;
	\item axon compartment $(\Omega_{ax})$;
	\item extracellular space $(\Omega_{ex})$.
\end{itemize}

\begin{figure}[hpt]
	\centering
	\includegraphics[width=3.25in,height=6.5cm ]{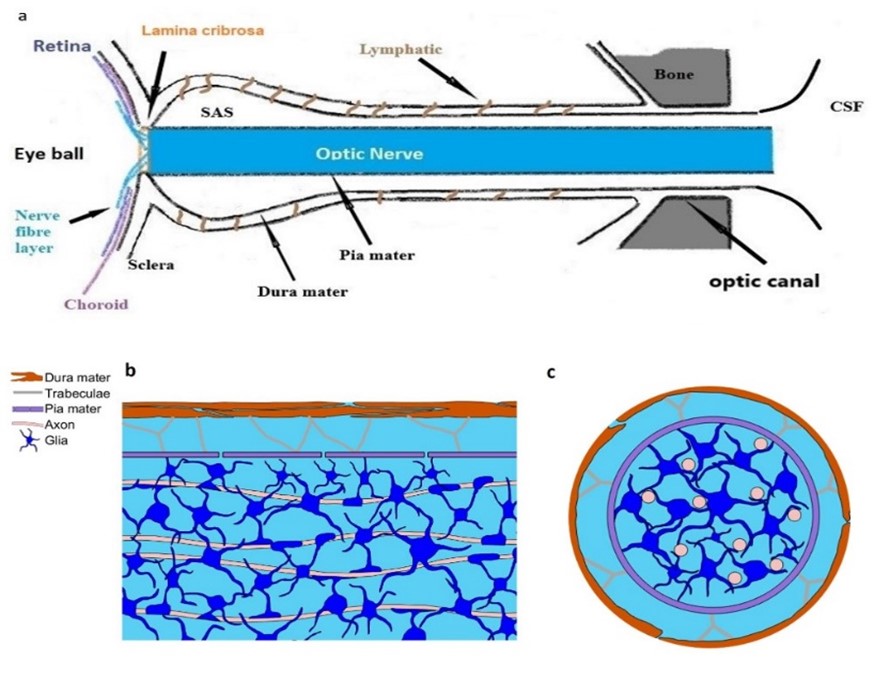}
	\caption{Optic nerve structure. a: Key features of the optic nerve region and subarachnoid space (SAS);
		b: Longitudinal section of the optic nerve; c: cross section of the optic nerve.\label{fig:schmatic}}
\end{figure}

The glial compartment and axon compartment \textcolor{black}{exist only} in the optic nerve, while extracellular space exists both in the optic nerve $\Omega^{OP}_{ex}$ and subarachnoid space $\Omega^{SAS}_{ex}$  (See Fig.\ref{fig:schmatic})

\begin{equation*}
\Omega_{OP}=\Omega_{ax} \cup \Omega_{gl} \cup \Omega_{ex}^{OP}, \quad \Omega_{SAS}=\Omega_{ex}^{SAS}.
\end{equation*}

The model is mainly derived from laws of conservation of ions and water for flow
through membranes between intracellular compartments and extracellular space \cite{nicholson2001diffusion}.

\begin{figure}[hpt]
	\centering
	\includegraphics[bb=5 55 580 380, clip=true, width=3.25in,height=4.5cm ]{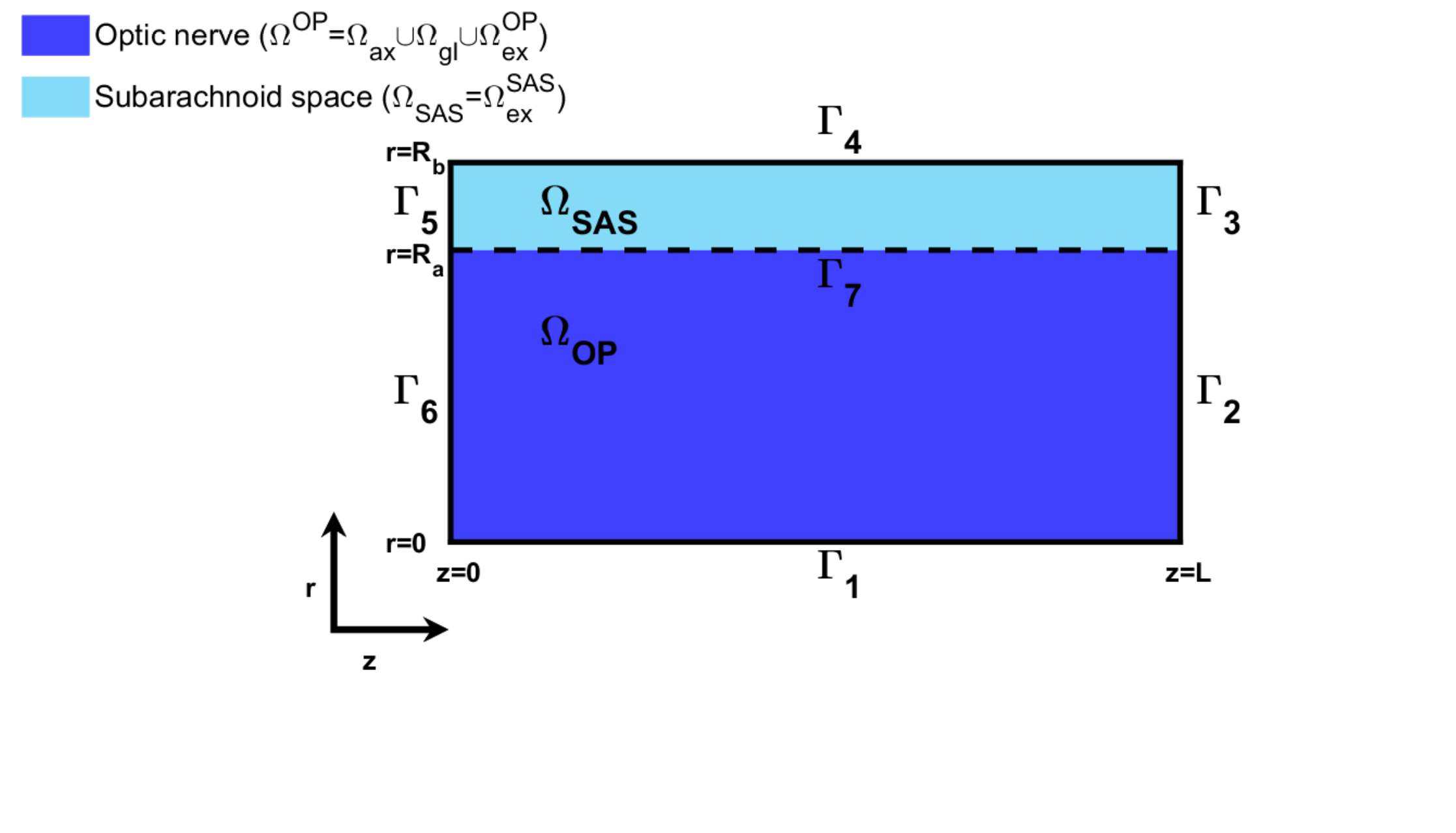}
	\caption{Domain of the axial symmetry model. The optic nerve  $\Omega_{OP}$ consist of axon compartment $\Omega_{ax}$, glial compartment $\Omega_{gl}$ and extracellular space $\Omega_{ex}^{OP}$.  The subarachnoid space only has extracellular space $\Omega_{ex}^{SAS}$.
		$R_{a}= \textcolor{black}{48} \  \mu m$  is the radius of optic nerve and $R_{b}= \textcolor{black}{60} \ \mu m $ is the radius from optical nerve center to the dura mater.  
		\label{fig:domain}}
\end{figure}

\subsection{Notations and Assumptions}
\label{Notations_and_domains}
We first introduce the following notations used in the paper, where $i=\mathrm{Na^{+},K^{+},Cl^{-}}$ for ion \textcolor{black}{species}, $l=ex,gl,ax$ for extracellular space, glial compartment and axon compartment, and $k=gl,ax$ for glial or axon membrane in the optic nerve.

Fig.\ref{fig:schmatic} b shows the model with the whole domain $\Omega$ that consists of the subarachnoid space (SAS) region $\Omega_{SAS}$  and optic nerve region $\Omega_{OP}$ (see Fig.\ref{fig:domain}),
\begin{equation*}
\Omega=\Omega_{SAS} \cup \Omega_{OP}.
\end{equation*}

In the optic nerve region $\Omega_{OP}$, the glial membrane and axon membrane separate domains $\Omega_{gl}$, $\Omega_{ax}$ from the extracellular region $\Omega_{ex}^{OP}$, respectively (also see Fig. \ref{fig:schmatic}). Based on the structure of the optic nerve, we have the following global assumptions for the model:

\noindent\textbf{\textbullet Charge neutrality:} In each domain, we assume that \textcolor{black}{there is} electroneutrality
\begin{subequations}
	\label{Charge_nuetrality}
	\begin{align}
	\eta_{gl} \sum_{i} z^{i} C_{gl}^{i}+z^{gl} \eta_{gl}^{re}  A_{gl} &=0, \label{Charge_gl} \\
	\eta_{ax} \sum_{i} z^{i} C_{ax}^{i}+z^{ax} \eta_{ax}^{re}  A_{ax} &=0, \label{Charge_ax}\\
	\sum_{i} z^{i} C_{ex}^{i} &=0, \label{Charge_ex}
	\end{align}
\end{subequations}
where $A_{l}>0$ with $l=ax,gl$ is the density of proteins in axons or glial cells. The proteins are negatively charged, but  the charge density is customarily  described by a positive number.  The $\eta_{ax}$ and $\eta_{gl}$ are the volume fraction of axon and glial compartments in the optic nerve and $\eta_{ax}^{re}$ and $\eta_{gl}^{re}$ are the resting state volume fractions. 

\noindent \textbf{\textbullet Axial symmetry:} For simplicity, axial symmetry is assumed. The model can be straightforwardly extended to three dimensions when data and needs justify the considerable extra computational resources needed to analyze such models. 

\noindent\textbf{\textbullet Isotropy of glial compartment and extracellular space:} 
\begin{itemize}
	\item[1)] The extracellular space forms a narrow structure of branching clefts surrounding the glial cells and nerve axons. 
	\item[2)] The glial cells are connected to each other by connexins and form a syncytium. 
	\item[3)] The extracellular space is continuous and forms a syncytium. 
\end{itemize}
Both syncytia are assumed isotropic here, until we know better. The axons are not connected to each other. For $l=gl,ex,$ and $ i=\mathrm{Na^+,K^+,Cl^-}$, the ion flux and water flow velocity are in the following forms  
\begin{align}
\boldsymbol{j}_{l}^{i}&=j_{l,r}^{i} \hat{\boldsymbol{r}} +j_{l,z}^{i} \hat{\boldsymbol{z}}, \\
\boldsymbol{u}_{l}&=u_{l}^{r} \hat{\boldsymbol{r}}+u_{l}^{z} \hat{\boldsymbol{z}}.
\end{align}

\noindent\textbf{\textbullet Anisotropy of axon compartment:} The axons are separated, more or less parallel cylindrical cells that do not form a syncytium. For $i=\mathrm{Na^+,K^+,Cl^-}$, the ion flux and water flow velocity are in the following forms  
\begin{align}
\boldsymbol{j}_{ax}^{i}&=j_{ax,z}^{i} \hat{\boldsymbol{z}},\\
\boldsymbol{u}_{ax}&=u_{ax}^{z} \hat{\boldsymbol{z}}.
\end{align}
\noindent\textbf{\textbullet  Communications between compartments:}  The communications between intracellular compartments and extracellular space are through membranes. There is no direct interaction between glial and axon compartments. Interactions occur only through changes in concentration, electrical potential, and flows in  the extracellular space \cite{sibille2015neuroglial}.

The interface of optic nerve and SAS is the pia mater denoted by $\Gamma_{7}$, 
\begin{equation*}
\Omega_{OP} \cap \Omega_{SAS}=\Gamma_{7}.
\end{equation*}
In our model, both pia mater $\Gamma_{7}$ and dura mater $\Gamma_{4}$ are modeled as macroscopic membranes and appropriate boundary conditions. The transmembrane water flow through pia mater \cite{filippidis2012permeability} depends on hydrostatic pressure, osmotic pressure and electric potential, while that through dura mater only depends on the hydrostatic pressure.


For the domain boundaries, $\Gamma_{1}$ is the radius center of the optic nerve; $\Gamma_{2}$ and $\Gamma_{3}$ are the far end (away from the eyeball) of the optic nerve which is connected to optic canal region \cite{hayreh1984sheath}. $\Gamma_{5}$ is used to model the dura mater connected to the sclera (the white matter of the eye) and assumed to be non-permeable \cite{hayreh2009ischemic}. $\Gamma_{6}$ is used to denote the lamina cribrosa where the optic nerve head exits the eye posteriorly through pores of the lamina cribrosa \cite{jonas2003anatomic}.

\subsection{Water circulation}
\label{water_cir}
We model water circulation with the following assumptions 
\begin{itemize}
	\item the loss or gain of water in axons and glial cells is only through membranes flowing into or out of the extracellular space. 
	\item the transmembrane water flux is proportional to the intra/extra-cellular hydrostatic pressure and osmotic pressure differences.
	\item the glial cell and axons can swell and shrink due to the water inflows and outflows.
\end{itemize}

\noindent We use $\eta_{l}(r,z,t)$ to describe the volume fraction of $l$ region $(l=gl,ax,ex)$, which varies over time and space due to transmembrane water flows. The conservation of water in each domain yields  

\begin{subequations}\label{incompressibility}
	\begin{align}
	&\frac{\partial \eta_{gl}}{\partial t}+\mathcal{M}_{gl} U^{m}_{gl}+\nabla \cdot\left(\eta_{g l} \boldsymbol{u}_{g l}\right) =0, \label{glincom}\\
	&\frac{\partial \eta_{ax}}{\partial t}+\mathcal{M}_{ax}U^{m}_{ax}+\frac{\partial}{\partial z}\left(\eta_{ax} u_{a x}^{z}\right) =0, \\
	&\nabla \cdot\left(\eta_{gl} \boldsymbol{u}_{gl}\right)+\nabla \cdot\left(\eta_{ex} \boldsymbol{u}_{e x}\right)+\frac{\partial}{\partial z}\left(\eta_{ax} u_{ax}^{z}\right)=0,
	\end{align}\end{subequations}
where we use the fact that density of water is constant. Here $\boldsymbol{u}_{l}$ with $l=gl,ax,ex$ is the velocity  in the glial cells, axons, and extracellular space, respectively. The transmembrane water flow $U^{m}_{k}$ with $k=gl,ax$ follows the Starling's law on the $k$th membrane, 
\begin{equation*}
\begin{aligned}
U^{m}_{gl}&= L_{gl}^{m}\left(P_{gl}-P_{ex}-\gamma_{gl} k_{B} T\left(O_{gl}-O_{ex}\right)\right),\\
U^{m}_{ax}&= L_{ax}^{m}\left(P_{ax}-P_{ex}-\gamma_{ax} k_{B} T\left(O_{ax}-O_{ex}\right)\right).
\end{aligned}
\end{equation*}

The $P_{l}$ with $l=gl,ax,ex$ are the hydrostatic pressure in the glial cells, axons, and extracellular space, respectively. And $k_BTO_{l}$ is the osmotic pressure \cite{xu2018osmosis,zhu2019bidomain} defined by 
\begin{equation*}
O_{ex}=\sum_{i} C_{ex}^{i}, \quad O_{l}=\sum_{i} C_{l}^{i}+A_{l} \frac{\eta_{j}^{re}}{\eta_{l}}, \quad l=gl, ax,
\end{equation*}
where $A_{l}\frac{\eta_{l}^{re}}{\eta_{l}}>0$ is the density of the permanent negatively charged protein in glial cells and axons that varies with the volume (fraction) of the region.  In this paper, we assume the permanent negatively charged protein is uniformly distributed within glial cells and axons and has valence $z^l$, $l=gl,ax$.  The $\mathcal{M}_{k}$ and $\gamma_{k}$, $k=gl,ax$ are the glial cells (or axons) membrane area per unit volume and membrane reflection coefficient \cite{feher2017quantitative} respectively. $L_{gl}^m$ and $L_{ax}^m$ are the surface membrane hydraulic
permeabilities of glial cells and axons.  The membrane reflection coefficient $\gamma_{k}\ (k=gl,ax)$ is the ratio between the observed osmotic pressure and theoretical osmotic pressure. $k_{B}$ is Boltzmann constant, and $T$ is temperature.

For the volume fraction $\eta_{l},\ l=gl,ax,ex$, we have 
\begin{equation}
\eta_{gl}+\eta_{ax}+\eta_{ex}=1, \quad \text { in } \Omega.
\end{equation}
\begin{remark}
	\normalfont
	Note the glial cells and axons are found only in the $\Omega_{OP}$ region. In other words, $\eta_{ax}=\eta_{gl}\equiv0$ and $\eta_{ex}\equiv1$ are fixed in $\Omega_{SAS}$. Therefore, the solution is incompressible in the $\Omega_{SAS}$, and we have
	
	\begin{equation}
	\nabla \cdot \boldsymbol{u}_{ex}=0, \quad \text{ in } ~\Omega_{SAS}.
	\end{equation} 
\end{remark}
\noindent The relation between the hydrostatic pressure $P_{l}$ and volume fraction $\eta_{l} \ (l=gl,ax,ex)$ is connected by the force balance on the membrane $\mathcal{M}_{k}\ (k=gl,ax)$.
\textcolor{black}{ The membrane force is balanced with the hydrostatic pressure difference on both sides of the semipermeable membrane \cite{xu2018osmosis,mori2015multidomain}.  Then the variation of volume fraction from the resting state is proportional to the variation of hydrostatic pressure difference from the resting state. }
\begin{subequations}
	\begin{align}
	\label{hydro_relation}
	K_{gl}\left(\eta_{gl}-\eta_{gl}^{re}\right) &=P_{gl}-P_{ex}-\left(P_{gl}^{re}-P_{ex}^{re}\right), \\
	K_{ax}\left(\eta_{ax}-\eta_{ax}^{re}\right) &=P_{ax}-P_{ex}-\left(P_{ax}^{re}-P_{ex}^{re}\right), \label{hydro_relation_ax}
	\end{align}
\end{subequations}
\noindent where $K_{gl,ax}$ is the stiffness constant and $\eta_{l}^{re}$ and $P_{l}^{re} \ (l=gl,ax,ex)$ are the resting state  volume fraction and hydrostatic pressure.  
\begin{remark}
	The osmotic pressure influences the volume fraction in an implicit way. Due to Eq. \eqref{incompressibility} and the definition of transmembrane water flow $U^{m}_{gl}$, the osmostic pressure  difference induces the transmembrane water flow going in and out of compartments, and the hydrostatic pressure changes correspondingly. 
\end{remark}

\begin{remark}
	In reality, both dura  mater $\Gamma_7$ and pia mater $\Gamma_4$ are deformable.   However,   the stiffness of them  are much larger than  glial cells' \cite{hua2018cerebrospinal,lu2006viscoelastic,feola2016finite}. For the sake of simplicity, we neglect  the deformation of dura mater and pia mater  so the  system could be solved in  \textcolor{black}{fixed} domains $\Omega$ and $\Omega^{OP}$.  We are aware that in some important clinical applications the deformation is of great importance and that our model will need to be extended to deal with them. The extensions do not produce analytical challenges but they are likely to add significant complexity to the numerical methods and make the computations much more involved and longer.
\end{remark}
Note that the concentrations of ions and effective concentration of water vary a great deal and so are described by equations in which the number density of ions and effective number density of water vary in both the radial and longitudinal directions according to conservation laws, without using compartments that may not have unique definitions or relations to anatomical structures. Indeed, the variation of concentration is one of the main determinants of the properties of ionic solutions. The solution is incompressible, the components are not \cite{eisenberg2010energy}.

\noindent Next, we define the velocity in each domain.

\noindent \textbf{Water Velocities in the Glial Compartment.} As we mentioned before, 
\textcolor{black}{the glial cells are connected through the narrow connections on the membranes and form a syncytium. The fluid velocity inside the compartment is limited by the transmembrane fluid velocity, which is determined by the difference of hydro pressure and osmotic pressure on both sides of the membrane \cite{xu2018osmosis,mori2015multidomain}.  On the macro-scale, by using averaging process \cite{zhu2019bidomain,malcolm2006computational,mathias1985steady}, the glial compartment is treated with membranes everywhere. So the velocity of fluid in glial syncytium $\boldsymbol{u}_{gl}$ depends on the gradients of hydrostatic pressure and osmotic pressure: 
}
\begin{subequations}
	\begin{align}
	\label{def_u_gl}
	u_{gl}^{r} &=-\frac{\kappa_{gl} \tau_{g l}}{\mu}\left(\frac{\partial P_{gl}}{\partial r}-\gamma_{gl} k_{B} T \frac{\partial O_{gl}}{\partial r}\right), \\
	u_{gl}^{z} &=-\frac{\kappa_{gl} \tau_{g l}}{\mu}\left(\frac{\partial P_{gl}}{\partial z}-\gamma_{gl} k_{B} T \frac{\partial O_{gl}}{\partial z}\right).
	\end{align}
\end{subequations}

Substituting Eq. \eqref{def_u_gl} into the incompressibility \eqref{glincom} yields a  Poisson equation for pressure in the  glial compartment.
The boundary conditions in the glial syncytium are as follows
\begin{equation}
\left\{
\begin{aligned}
&\frac{\partial P_{gl} }{\partial r}=0, & \text { on } \Gamma_{1}\cup\Gamma_{7}, \\
&\frac{\partial P_{gl} }{\partial z}=0, & \text { on } \Gamma_{2}\cup\Gamma_{6},
\end{aligned}
\right.
\end{equation}
\textcolor{black}{where homogeneous Neumann boundary condition is applied at all four of these boundaries.}


\textbf{Water Velocity in the Axon Compartment.}   Axons are arranged parallel in the longitudinal direction and are isolated from each other.  The fluid velocity in the axon compartment is defined along $z$ direction as 
\begin{subequations}
	\begin{align}
	\label{def_u_ax}
	u_{ax}^{r} &=0, \\
	u_{ax}^{z} &=-\frac{\kappa_{a x}}{\mu} \frac{\partial P_{ax}}{\partial z}.
	\end{align}
\end{subequations}

Similarly, substituting above velocity into the incompressibility equation gives the Poisson equation for pressure in the axon compartment. A homogeneous Neumann condition for pressure is used on the left and right boundaries of the axon compartment  
\begin{equation}
\frac{\partial P_{ax}}{\partial z}=0, ~~ \text { on } \Gamma_{2} \cup \Gamma_{6}.
\end{equation}

\noindent \textbf{Velocity in the Extracellular Space.} The extracellular space is a narrow connected domain, where the electro-osmotic effect \textcolor{black}{needs} to be considered \cite{zhu2019bidomain,wan2014self,vaghefi2012development}.  The extracellular velocity is determined by the gradients of hydrostatic pressure and electric field
\begin{subequations}
	\begin{align}
	\label{def_u_ex}
	u_{ex}^{r} &=-\frac{\kappa_{ex} \tau_{ex}}{\mu} \frac{\partial P_{ex}}{\partial r}-k_{e} \tau_{ex} \frac{\partial \phi_{ex}}{\partial r}, \\
	u_{ex}^{z} &=-\frac{\kappa_{ex} \tau_{ex}}{\mu} \frac{\partial P_{ex}}{\partial z}-k_{e} \tau_{ex} \frac{\partial \phi_{ex}}{\partial z},
	\end{align}\end{subequations}
where $\phi_{ex}$ is the electric potential in the extracellular space, $\tau_{ex}$ is the tortuosity of extracellular region \cite{nicholson2001diffusion,perez1995extracellular} and $\mu$ is the viscosity of water, $k_{e}$   describes the effect of electro-osmotic flow \cite{mclaughlin1985electro,vaghefi2012development,wan2014self}, $\kappa_{ex}$ is the permeability of the extracellular space. Here the hydro permeability $\kappa_{ex}$,  tortuosity $\tau_{ex}$ and electric-osmotic parameter $k_{e}$ have two distinguished values in the region $\Omega_{ex}^{OP}$ and $\Omega_{ex}^{SAS}$,
\begin{equation}
\begin{aligned}
&\kappa_{ex}=\left\{\begin{array}{l}
\kappa_{ex}^{OP}, \text { in } \Omega_{OP}, \\
\kappa_{ex}^{SAS},\text { in } \Omega_{SAS}, 
\end{array}\right. 
\tau_{ex}=\left\{\begin{array}{l}
\tau_{ex}^{OP}, \text { in } \Omega_{OP}, \\
\tau_{ex}^{SAS}, \text { in } \Omega_{SAS},
\end{array}\right. \\
&k_{e}=\left\{\begin{array}{l}
k_{e}^{OP}, \text { in } \Omega_{OP}, \\
k_{e}^{SAS}, \text { in } \Omega_{SAS},
\end{array}\right.
\end{aligned}
\end{equation}

\begin{remark}
	\textcolor{black}{By substituting the definitions of velocities Eqs. \eqref{def_u_ax}, \eqref{def_u_gl} and \eqref{def_u_ex}, and volume fractions Eq.\eqref{hydro_relation} into the mass conservation law Eq. \eqref{incompressibility}, it yields the evolution equations of hydrostatic pressure in different compartments, which are solved during the simulations in Session \ref{sec:clearance}.} 
\end{remark}

Since $\Gamma_{2} \cup \Gamma_{3}$ are the far end of optic nerve away from eyeball and   connect to the  optic canal, we assume the hydrostatic pressure of extracellular fluid  is equal to the cerebrospinal fluid pressure. On the other hand, the intraocular pressure (IOP) is imposed at $\Gamma_{6}$ where the extracellular space is connected to the retina. At the boundary $\Gamma_{5}$, we assume a non-permeable boundary. We are aware of the significance of the pressures and flows at these boundaries for clinical phenomena including glaucoma  \cite{band2009intracellular,gardiner2010computational,pache2006morphological} and will return to that subject in later publications.

The water flow across the semi-permeable membrane $\Gamma_4$  is  produced by the lymphatic drainage on the dura membrane, which depends on the difference between extracellular pressure and orbital pressure (OBP). We assume the velocity across the pia membrane $\Gamma_4$, is continuous and determined by the hydrostatic pressure and osmotic pressure. To summarize, the boundary conditions of the extracellular fluid are 
\small{
	\begin{equation} 
	\label{P_ex_boundary_cd}
	\left\{
	\begin{aligned}
	&\boldsymbol{u}_{ex}  \cdot \hat{\boldsymbol{r}}=0,  &  &\mbox{on $\Gamma_{1}$},\\
	&P_{ex}=P_{CSF}, &  &\mbox{on $\Gamma_{2}\cup \Gamma_{3}$},\\
	&\boldsymbol{u}^{SAS}_{ex} \cdot \ \hat{\boldsymbol{r}}=L^{m}_{dr}\left(P^{SAS}_{ex}-P_{OBP}\right),&&\mbox{on $\Gamma_{4}$},\\
	& \boldsymbol{u}_{ex} \cdot \hat{\boldsymbol{z}}=0, &  &\mbox{on $\Gamma_{5}$},\\
	&P_{ex}=P_{IOP},&&\mbox{on $\Gamma_{6}$},\\
	&\boldsymbol{u}^{OP}_{ex} \cdot  \hat{\boldsymbol{r}}=\boldsymbol{u}^{SAS}_{ex} \cdot  \hat{\boldsymbol{r}} &&\\
	&=L^{m}_{pia} \left(P^{OP}_{ex}-P^{SAS}_{ex}-\gamma_{pia}k_{B}T \left(O^{OP}_{ex}-O^{SAS}_{ex} \right) \right),&&\mbox{on $\Gamma_{7}$},
	\end{aligned}
	\right.
	\end{equation}}
where $P_{CSF}$ is the cerebrospinal fluid pressure \cite{band2009intracellular} and $P_{IOP}$ is the intraocular pressure  and $P_{OBP}$ is the orbital pressure on the dura mater.

\subsection{Ion Transport}
\label{ion_cir}
For ion circulation, we assume
\begin{itemize}
	\item only three types of ions are considered: $ \mathrm{Na^+, K^+} $ and  $\mathrm{Cl^-}$.
	\item the sodium-potassium ATP pump is present on both glial and axon membranes.
	\item ion channel conductance on glial cell membranes is a fixed constant, independent of the voltage and time. The sodium conductance is assumed small, and its channel origin unknown. The potassium conductance is large and comes from the $K_{ir4}$ channels \cite{bellot2017astrocytic,sigal2005factors}. When experimental evidence is available, other types of pumps and channels can be added to the model.
	\item  sodium channel and potassium channel conductance on axons are voltage-gated, while the conductance of chloride channel is fixed. 
\end{itemize}
The conservation of ions implies the following system of partial differential equations to describe the dynamics of ions in each region, for $i=\mathrm{Na^+,K^+,Cl^-}$
\begin{subequations}
	\label{ion_governing}
	\begin{align}
	&\frac{\partial\left(\eta_{gl} C_{gl}^{i}\right)}{\partial t}+\mathcal{M}_{gl}J^{m,i}_{gl}+\nabla \cdot\left(\eta_{g l} \boldsymbol{j}_{gl}^{i}\right)=0,  \label{C_gl} \\
	&\frac{\partial\left(\eta_{ax} C_{ax}^{i}\right)}{\partial t}+\mathcal{M}_{ax}J^{m,i}_{ax}+\frac{\partial}{\partial z}\left(\eta_{ax} j_{ax,z}^{i}\right)=0,   \label{C_ax}  \\
	&\frac{\partial\left(\eta_{ex} C_{ex}^{i}\right)}{\partial t}-\mathcal{M}_{ax}J^{m,i}_{ax}-\mathcal{M}_{gl}J^{m,i}_{gl}+\nabla \cdot\left(\eta_{ex} \boldsymbol{j}_{e x}^{i}\right)=0, \label{C_ex} 
	\end{align}
\end{subequations}
where the last equation reduces to the following in the $\Omega_{SAS}$ region, 
\begin{equation}
\frac{\partial C_{ex}^{i, SAS}}{\partial t}+\nabla \cdot \boldsymbol{j}_{ex}^{i, SAS}=0.
\end{equation}
The transmembrane ion flux  $J_{k}^{m,i} \ (k=gl,ax)$   consists of  active ion pump source  $a_{k}^{i}$  and  passive ion channel source $b_{k}^{i}$, for ion $i$ on the axons $(k=ax)$ or glial cells membranes $(k=gl)$. In the glial cell membranes,
\begin{equation*}
J_{k}^{m, i}=a_{k}^{i}+b_{k}^{i}, \quad k=gl,ax, \quad i=\mathrm{Na}^{+}, \mathrm{K}^{+}, \mathrm{Cl}^{-}.
\end{equation*}
In the glial cell membranes, $b_{gl}^{i}$ is defined as 
\begin{equation}
b_{gl}^{i}=\frac{g_{g l}^{i}}{z^{i} e}\left(\phi_{g l}-\phi_{e x}-E_{g l}^{i}\right), 
\end{equation}
where the Nernst potential is used to describe the gradient of chemical potential $E_{gl}^{i}= \frac{k_{B}T}{ez^{i}} \log\left(\frac{C_{ex}^i}{C_{gl}^i}\right)$ and the conductance $g_{gl}^{i}$ for each ion on the glial membrane is a fixed constant, independent of voltage and time. On the axon's membrane, $b_{ax}^{i}$ is defined as 
\begin{equation}
b_{ax}^{i}=\frac{g_{ax}^{i}}{z^{i} e}\left(\phi_{a x}-\phi_{e x}-E_{a x}^{i}\right), 
\end{equation}
where 

\begin{equation*}
\begin{array}{l}
g_{ax}^{Na}=\bar{g}^{Na} m^{3} h+g_{leak}^{N a} , \quad 
g_{ax}^{K}=\bar{g}^{K} n^{4}+g_{leak}^{K}, \\
E_{ax}^{i}= \frac{k_{B}T}{ez^{i}} \log\left(\frac{C_{ex}^i}{C_{ax}^i}\right). 
\end{array}
\end{equation*}
The time-dependent dynamic of open probability, often loosely called `gating' is governed by the Hodgkin-Huxley model \cite{song2018electroneutral,fitzhugh1960thresholds}
\begin{subequations}
	\label{HHMODEL}
	\begin{align}
	\frac{dn}{dt} &=\alpha_{n}(1-n)-\beta_{n} n, \\
	\frac{dm}{dt} &=\alpha_{m}(1-m)-\beta_{m} m, \\
	\frac{dh}{dt} &=\alpha_{h}(1-h)-\beta_{h} h.
	\end{align}
\end{subequations}
We assume that the only pump is the Na/K active transporter. We are more than aware that other active transport systems can and likely do move ions and water in this system. They will be included as experimental information becomes available.

In the case of the Na/K pump $a_{l}^{i}, l=ax,gl$, the strength of the pump depends on the concentration in the intracellular and extracellular space \cite{gao2000isoform,fitzhugh1960thresholds}, i.e. 
\begin{equation}
\label{pump_eq}
a_{k}^{Na}=\frac{3I_{k}}{e}, \  a_{k}^{K}=- \frac{2I_{k}}{e}, \ a_{k}^{Cl}=0, \ k=gl,ax, 
\end{equation}
where 
\begin{equation*}
\begin{aligned}
I_{k}&=I_{k,1}\left(\frac{c_{k}^{N a}}{c_{k}^{Na}+K_{Na1}}\right)^{3}\left(\frac{c_{ex}^{K}}{c_{ex}^{K}+K_{K1}}\right)^{2}\\ &+I_{k,2}\left(\frac{c_{k}^{Na}}{c_{k}^{Na}+K_{Na2}}\right)^{3}\left(\frac{c_{ex}^{K}}{c_{ex}^{K}+K_{K2}}\right)^{2},
\end{aligned}
\end{equation*}
$I_{k,1}$ and $I_{k,2}$ are related to   $\alpha_{1}-$ and $\alpha_{2}-$ isoform of $\mathrm{Na/K}$ pump.

The definitions of ion flux in each domain are as follows, for $i=\mathrm{Na^+,K^+,Cl^-}$,
\begin{subequations}\label{eqflux}
	\begin{align}
	&\boldsymbol{j}_{l}^{i} =C_{l}^{i} \boldsymbol{u}_{l}-D_{ l}^{i} \tau_{l}\left(\nabla C_{l}^{i}+\frac{z^{i} e}{k_{B} T} C_{l}^{i} \nabla \phi_{l}\right), \ l=gl, ex, \\
	&j_{ax,z}^{i} = C_{ax}^{i} u_{ax}^{z}-D_{a x}^{i}\left(\frac{\partial C_{ax}^{i}}{\partial z}+\frac{z^{i} e}{k_{B}T} C_{ax}^{i} \frac{\partial \phi_{ax}}{\partial z}\right).
	\end{align}
\end{subequations}
For the axon compartment boundary condition, we have 
\begin{equation*}
C_{ax}^{i}=C_{ax}^{i,re}, \quad \text { on } \ \Gamma_{2} \cup \Gamma_{6},
\end{equation*}
and 
\begin{equation*}
\left\{\begin{array}{ll}
\textcolor{black}{ \frac{ \partial C_{gl}^i }{ \partial r } } =0, & \text { on } \Gamma_{1}, \\
C_{gl}^{i}=C_{gl}^{i,re}, & \text { on } \Gamma_{2} \cup \Gamma_{6}, \\
\boldsymbol{j}_{gl}^{i} \cdot \hat{\boldsymbol{r}}=0, & \text { on } \Gamma_{7},
\end{array}\right.
\end{equation*}
where the Dirichlet boundary conditions are used at locations $\Gamma_{2}\cup \Gamma_{6} $ for axons and glial cell, and a non-flux boundary condition is used for glial cells ions flux on \textcolor{black}{the radius center $\Gamma_{1}$} and pia mater $\Gamma_{7}$ .

For the extracellular space boundary condition, similar boundary conditions are imposed except on the pia mater $\Gamma_7$. The flux across the pia mater is assumed continuous and Ohm’s law is used \cite{zhu2019bidomain}. Additionally, a non-permeable boundary condition is used at location $\Gamma_{5}$ and a homogeneous Neumann boundary condition is applied at the location of the dura mater $\Gamma_4$, 

\begin{equation}
\label{C_ex_bc}
\left\{\begin{array}{ll}
\textcolor{black}{ \frac{ \partial C_{ex}^i }{ \partial r } }=0, & \text { on } \Gamma_{1}, \\
C_{ex}^{i}=C_{csf}^{i}, & \text { on } \Gamma_{2} \cup \Gamma_{3}, \\
\textcolor{black}{ \frac{ \partial C_{ex}^i }{ \partial r } }=0, & \text { on } \Gamma_{4}, \\
\boldsymbol{j}_{ex}^{i}  \cdot \hat{\boldsymbol{z}}=0, & \text { on } \Gamma_{5}, \\
C_{ex}^{i}=C_{IOP}^{i},  & \text { on } \Gamma_{6}, \\
\boldsymbol{j}_{ex}^{i, OP} \cdot \hat{\boldsymbol{r}}=\boldsymbol{j}_{ex}^{i,SAS} \cdot \hat{\boldsymbol{r}}=\frac{G_{pia}^{i}}{z^{i} e}\left(\phi_{ex}^{OP}-\phi_{ex}^{SAS}-E_{pia}^{i}\right), & \text { on } \Gamma_{7}.
\end{array}\right.
\end{equation}

Multiplying equations in (\ref{C_gl}-\ref{C_ex}) with $z_{i}e$ respectively, summing up, and using equation (\ref{Charge_gl}-\ref{Charge_ex}) and equations \eqref{eqflux},   we have following system for the electric  potential in $ax,gl,ex$ 
\footnotesize{
	\begin{subequations}
		\begin{align}
		\label{phi_governing}
		&\sum_{i} z^{i} e \mathcal{M}_{gl}J^{m,i}_{gl}+\sum_{i} \nabla \cdot\left(z^{i} e \eta_{g} \boldsymbol{j}_{gl}^{i}\right) =0, \\
		&\sum_{i} z^{i} e \mathcal{M}_{a x}J^{m,i}_{ax}+\sum_{i}  \frac{\partial}{\partial z}\left(z^{i} e\eta_{ax} j_{ax, z}^{i}\right) =0, \\
		&\sum_{i}  \nabla \cdot\left(z^{i} e\eta_{gl} \boldsymbol{j}_{gl}^{i}\right)+\sum_{i} \frac{\partial}{\partial z}\left( z^{i} e\eta_{a x} j_{ax,z}^{i}\right)+\sum_{i}  \nabla \cdot\left(z^{i} e\eta_{ex} \boldsymbol{j}_{ex}^{i}\right) =0,
		\end{align}\end{subequations}}
which describe the spatial distributions of electric potentials in three compartments.

In the subarachnoid space $\Omega_{SAS}$, the governing equation for extracellular  electric potential reduces  to 
\begin{equation}
\nabla \cdot\left(\sum_{i} z^{i} e\boldsymbol{j}_{ex}^{i,SAS}\right)=0.
\end{equation}
The boundary conditions for electric fields $\phi_{ax}$, $\phi_{gl}$ and $\phi_{ex}$ are given below.\\
In the axon compartment: 
\begin{equation*}
\left\{
\begin{aligned}
\textcolor{black}{\frac{ \partial \phi_{ax} } {\partial z }} =0, & \text { on } \Gamma_{2}, \\
\textcolor{black}{\frac{ \partial \phi_{ax} } {\partial z }} =0, & \text { on } \Gamma_{6},
\end{aligned}\right.
\end{equation*}
In the glial compartment: 
\begin{equation*}
\left\{
\begin{aligned}
&\textcolor{black}{\frac{ \partial \phi_{gl} } {\partial r }}=0, & \text { on } \Gamma_{1}, \\
&\textcolor{black}{\frac{ \partial \phi_{gl} } {\partial z }}=0, & \text { on } \Gamma_{2}, \\
&\textcolor{black}{\frac{ \partial \phi_{gl} } {\partial z }}=0, & \text { on } \Gamma_{6}, \\
&\textcolor{black}{\frac{ \partial \phi_{gl} } {\partial r }}=0,  & \text { on } \Gamma_{7},
\end{aligned}
\right.
\end{equation*}
and in the extracellular space: 
\begin{equation}
\label{phi_bd}
\left\{
\begin{aligned}
&\textcolor{black}{\frac{ \partial \phi_{ex} } {\partial r }}=0,  &&\text {on } \Gamma_{1}, \\
&\textcolor{black}{\frac{ \partial \phi_{ex} } {\partial z }}=0, &&\text {on } \Gamma_{2} \cup \Gamma_{3}, \\
&\textcolor{black}{\frac{ \partial \phi_{ex} } {\partial r }}=0, &&\text {on } \Gamma_{4}, \\
&\textcolor{black}{\frac{ \partial \phi_{ex} } {\partial z }}=0, &&\text {on } \Gamma_{5}, \\
&\textcolor{black}{\frac{ \partial \phi_{ex} } {\partial z }}=0, &&\text {on } \Gamma_{6}, \\
&\sum_{i} z^{i} e \boldsymbol{j}_{ex}^{i, OP} \cdot \hat{\boldsymbol{r}}=\sum_{i} z^{i} e \boldsymbol{j}_{e x}^{i, SAS} \cdot \hat{\boldsymbol{r}} && \\
&=\sum_{i} G_{pia}^{i}\left(\phi_{ex}^{OP}-\phi_{e x}^{SAS}-E_{pia}^{i}\right), &&\text {on } \Gamma_{7}.
\end{aligned}
\right.
\end{equation}

\section{Model calibration} \label{sec:Calibration}

Our work is possible because of, and was motivated by the paper of Orkand et al \cite{orkand1966effect,kuffler1966physiological} that measured the accumulation of potassium in the narrow extracellular space ($\Omega^{OP}_{ex}$) of the optic nerve of the amphibian salamander Necturus, in the spirit of the original work of Frankenhaeuser and Hodgkin \cite{frankenhaeuser1956after},  that first identified and analyzed accumulation of potassium outside a nerve fiber. The existence and qualitative properties of that accumulation of potassium were known to, and a cause for concern for Hodgkin, from his first work on the voltage clamp \cite{hodgkin1952measurement,hodgkin1949ionic}, if not earlier. Hodgkin described the phenomena as part of what was called [concentration] `polarization' and discussed it extensively with students sometime later (Eisenberg, personal communication, ~1962)

The key experiment in the Orkand paper \cite{orkand1966effect} measures the change in potential across the glial membrane produced by a train of action potentials. The glial membrane potential is used to estimate and report the potassium concentration in the narrow extracellular space, because the glial membrane is populated with more or less voltage independent potassium channels and not much else.

In the experiment, optic nerve has been put in three different $\mathrm{K}^+$ concentrations (1.5 mM, 3 mM, 4.5 mM) in the bathing solution to change the resting potential across the glia membrane. Then the axon was stimulated to give a train of action potentials. The action potentials increased $\mathrm{K}^+$ in $\Omega_{ex}^{OP}$. The accumulated $\mathrm{K}^+$ then made the glia membrane potential more positive. Stimuli were applied at both ends of a region of the optic nerve thereby producing a more uniform (in space) potential within that region. 

\begin{figure}
	\centering
	\includegraphics[width=80mm]{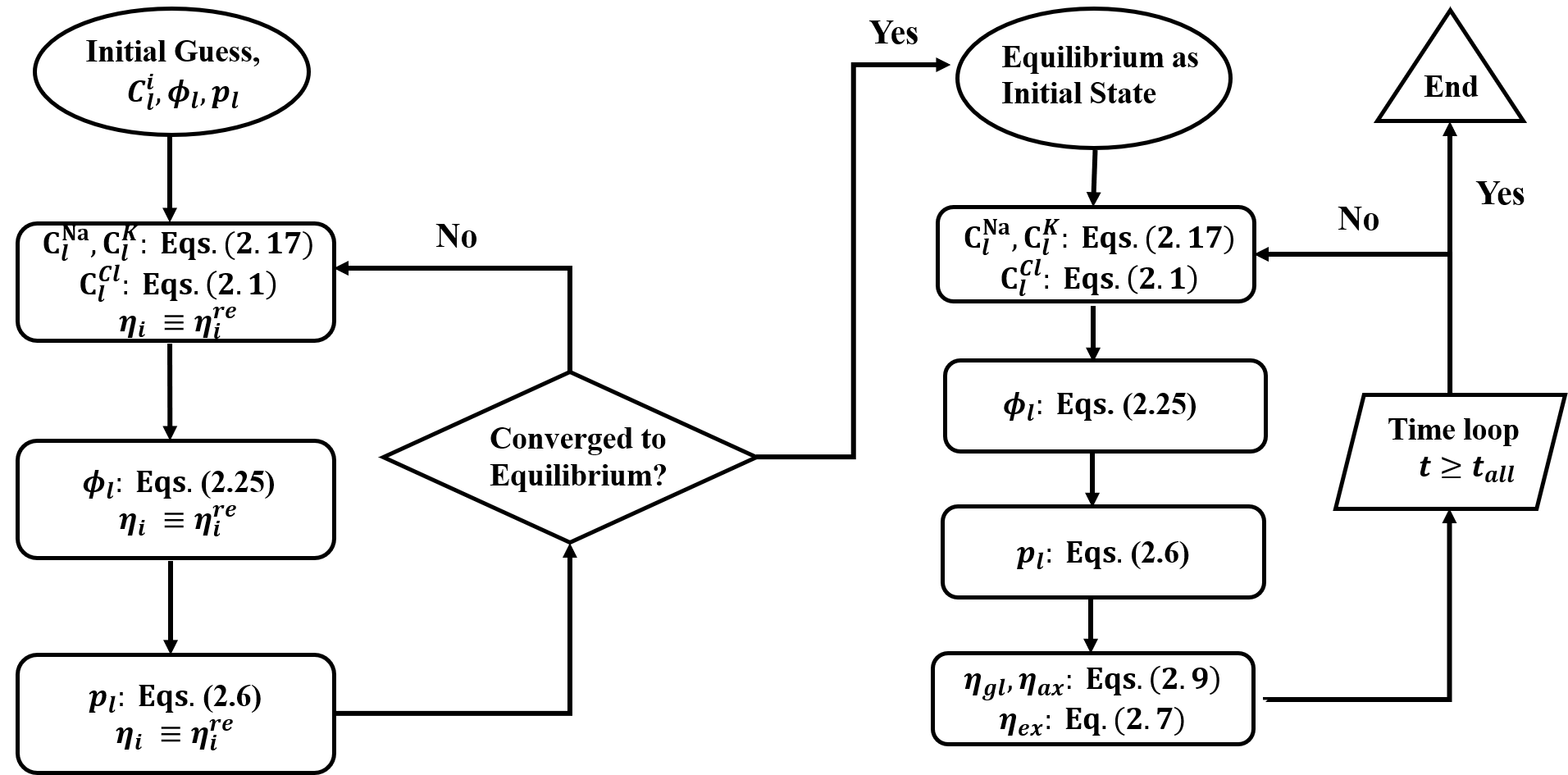}
	\caption{Flowchart for simulation process}
	\label{fig:flowchart}
\end{figure}

The model of this   system is solved by using the Finite Volume Method with mesh size $h=1/20$ and temporal size $t =1/10$ in dimensionless \textcolor{black}{units}. The code is written and executed in the Matlab environment. The  flowchart for the simulation  is shown in Fig. \ref{fig:flowchart}.  In the first step, we obtain the resting state of the system by iteration and fixed volume fraction  \cite{kuffler1966physiological}
\begin{equation*}
\eta^{re}_{ax} =0.5, \  \eta^{re}_{gl}=0.4, \ \eta^{re}_{ex}=0.1.
\end{equation*}
In the dynamic process, the  resting state values are taken as initial values.  Then, we first  solve the concentration governing equations, followed by electrical potentials equation and pressure equations. We update the volume fraction by using Eqs. (\ref{hydro_relation}) and (\ref{hydro_relation_ax}).

\begin{figure}[h]
	\centering
	\includegraphics[bb=40 5 920 445, clip=true,width=3.25in,height=5.6 cm ]{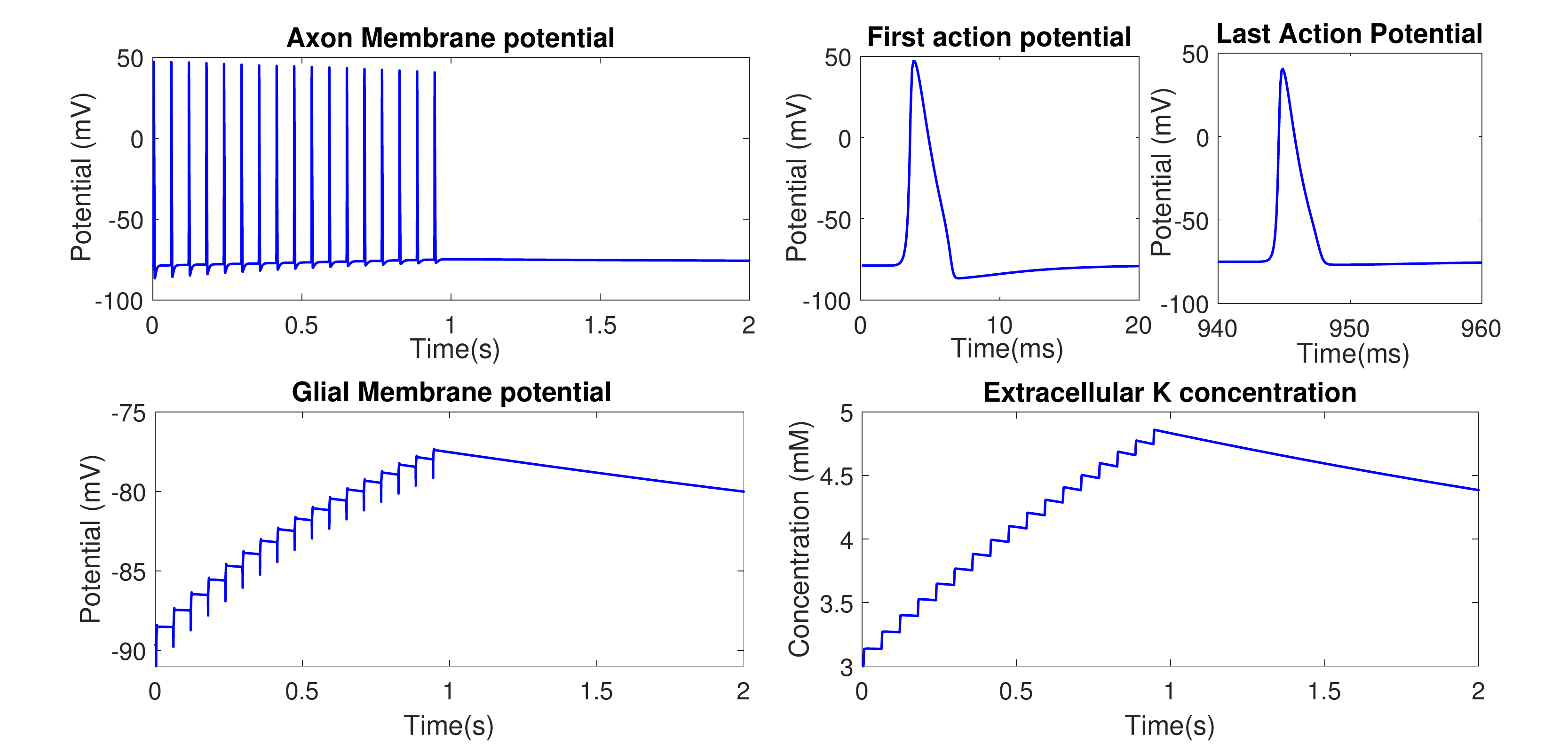}
	\caption{ Recording axon membrane potential, glial membrane potential and extracellular $\mathrm{K^+}$  at center axis point (where $r=0$ and $z=L/2$) when the extracellular solution with $3 \ \mathrm{mM}$ $\mathrm{K^+}$.}
	\label{fig::Orkand}
\end{figure}

\begin{figure}[h]
	\centering
	\includegraphics[bb=65 5 880 420,clip=true,width=3.25in,height=5.6cm ]{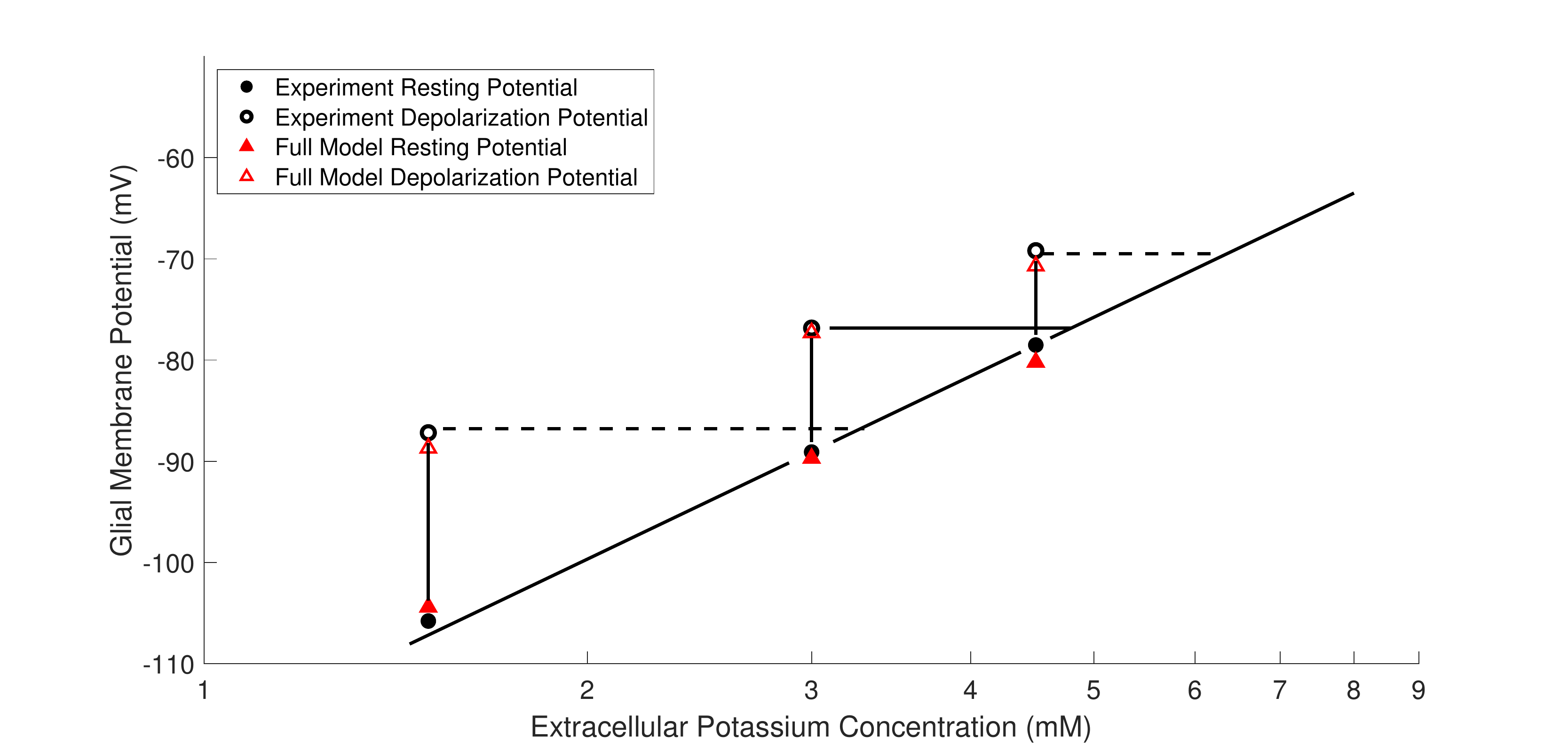}
	\caption{Comparison between the experiment in  \cite{orkand1966effect} and simulation on the effect of nerve impulses on the membrane potential of glial cells. The solid symbols are resting potentials and the open symbols are depolarization potentials with different $\mathrm{K^+}$ concentrations  \textcolor{black}{in $\Omega_{ex}^{OP}$} .}
	\label{fig::Orkand2}
\end{figure}

In our simulation study, we first set the ECS \textcolor{black}{($\Omega_{ex}^{OP}$)} concentration of  $\mathrm{K^+}$ to  be $3 \ \mathrm{mM}$ and obtained a resting potential across the glial membrane $(\sim-89\ \mathrm{mV})$. In Orkand's work \cite{orkand1966effect,kuffler1966physiological}, suction electrodes were used for stimulating; The two ends of the optic nerve were placed in suction electrodes as described in their Methods Section. We modeled the suction electrodes by applying a train of rectangular function (sometimes called a `box car' stimulus in the engineering literature) currents through the axon membrane at $z=2.25\  \mathrm{mm},13.5 \ \mathrm{mm}, \ \mathrm{and} \ 0< r < R_{a}=48 \ \mu m$. Each stimulus in the train lasted $3 \ \mathrm{ms}$ (as Orkand's paper indicated) with current strength  $3 \ \mathrm{mA/m^2} $. The stimulus was large enough to exceed the threshold and generate action potentials. After a train of stimuli with a frequency of $17/\mathrm{s}$ for 1 $\mathrm{s}$, the first panel of Figure \ref{fig::Orkand} shows the train of axon membrane action potentials and its return to the initial level $(\sim-89 \ \mathrm{mV})$ after $1 \ \mathrm{s}$. The profiles of the first action potential and the last action potential in the train are presented in the second panel. The third and fourth panels are used to illustrate the increase in glial cell membrane potential and extracellular potassium concentration during and after the train of stimuli. The fourth panel of Figure \ref{fig::Orkand} shows that during stimulus, the  $\mathrm{K^+}$ concentration  \textcolor{black}{in $\Omega_{ex}^{OP}$} keeps increasing due to the opening of the voltage-gated potassium channel of the axon membrane. As a result of the accumulated $\mathrm{K^+}$ in  \textcolor{black}{$\Omega_{ex}^{OP}$}, the membrane potential of glial cells also continues to increase  until the stimulus stops.
\\
Then, we vary the  $\mathrm{K^+}$ to be 1.5 mM, 3 mM, 4.5 mM  \textcolor{black}{in $\Omega_{ex}^{OP}$} and record the magnitude of the maximum glial membrane depolarized potential in each case as in the Fig. \ref{fig::Orkand2}. The black symbols are used for experimental data, \textcolor{black}{red} ones are the simulations results of our model, respectively. Fig. \ref{fig::Orkand2} shows that our model could match the experimental resting potentials (solid symbols) and depolarization potentials (open symbols) very well with different $\mathrm{K^+}$ concentrations \textcolor{black}{in $\Omega_{ex}^{OP}$}. 

Our work is limited by the lack of other types of data for calibration, although it is important to remember that the structural parameters, values of membrane capacitance, many conductance variables, and resistivities of bulk solutions are known quite well because of the work of generations of anatomists, physiologists, and biophysicists.

\section{Potassium clearance} \label{sec:clearance}
\label{K_Clearance}
In this section, we  compare  the potassium clearance under various conditions based on the full model. In Section \ref{alternative}, by applying the stimuli at alternative locations on the axon,  we show how the interaction between the extracellular pathway and glial transmembrane pathway helps potassium clearance. The microcirculation  patterns of water and ions between glial compartment and extracellular space is presented. 
In Section \ref{NKCC_PIA}, we have a glimpse of the effect of the glial membrane conductance changes as well as variations in the pia mater boundary conditions. We introduce the NKCC channel into the glial membrane and a non-selective pathway on the pia mater and compare the potassium clearance with the baseline model.  The NKCC channels on the glial membrane increase  potassium clearance  efficiency. An additional non-selective pathway in the pia mater does not have significant effect on potassium clearance due to the limited surface area of pia mater.

\subsection{Alternative Distribution of Stimulus Location}\label{alternative}

In this section, current is applied at several different radial locations. We wondered whether the choice of radial location would change our calculations of potassium clearance and fluid velocity. This approach is used for many variables to crudely estimate the sensitivity of our results to assumptions. 

To facilitate the discussion of the interaction between the extracellular pathway and the glial transmembrane pathway for $\mathrm{K}^{+}$ clearance, we define the following regions which potassium flux could pass through,
\begin{itemize}
	\item $M_{S}$: Glial membrane in stimulated region,
	\item $E_{T}$: Extracellular pathway   on transition interface.
	\item $M_{NS}$: Glial transmembrane in non-stimulated region.
	\item $G_{T}$: Glial pathway  on transition interface.
\end{itemize}
The normal directions on $M_{S}$ and $M_{NS}$ point to \textcolor{black}{extracellular space} and the normal direction of $E_{T}$ and $G_{T}$ is the radial direction.
We mainly focus on the two particular periods of time, (1) during a train of axon firing $([0,T_{sti}])$   (2) after axon firing $([T_{sti},T_{af}])$ \textcolor{black}{after the axon stimulation stops}.  In the simulations below, we take $T_{sti}=0.2 \ \mathrm{s}$ and $T_{af}=10 \ \mathrm{s}$. The frequency of the stimuli is $50 \ \mathrm{Hz} \ (T=0.02 \ \mathrm{s})$ and each single stimulus has current strength $I_{sti}=3\times 10^{-3} \ \mathrm{A/m^2} $ with duration  $3\ \mathrm{ms}$. 

\subsubsection{Inner and Outer radial regions stimulated}\label{inner_outer}
We first make a comparison between the inner radial region stimulated \textcolor{black}{case in} which the current \textcolor{black}{is} applied  at $z=z_0(=2.25\  \mathrm{mm})$ cross-section: $S_{sti}^{in}  \left( = \{(r,z) \vert r<\frac{R_a}{2},\ z=z_{0} \} \right)$  and outer radial region stimulated \textcolor{black}{case in} which $S_{sti}^{out} \left(=\{ (r,z)\vert  \frac{R_a}{2} < r < R_a,\ z=z_{0} \} \right)$. Since the axon signal \textcolor{black}{propagates} in \textcolor{black}{the} $z$ direction, for the inner radial region stimulated case, the stimulated region is  $V_{S}^{in}=\{ (r,z) \vert  r< \frac{R_a}{2}, z\in [0,L]\}$   and the non-stimulated region is $V_{NS}^{in}=\{ (r,z)\vert   \frac{R_a}{2}<r<R_a,\ z \in [0,L]\}$;  for the outer radial region stimulated case, $V_{S}^{out}=V_{NS}^{in}$ and $V_{ns}^{out}=V_{s}^{in}$.  The transition interface $S_{sti}=\{ (r,z)\vert  r=\frac{R_a}{2},\ z\in [0,L]\}$  is the same for both cases. 

\begin{figure}[hpt]
	\centering
	\includegraphics[bb=20 5 395 320, clip=true,width=3.25in,height=5.6 cm]{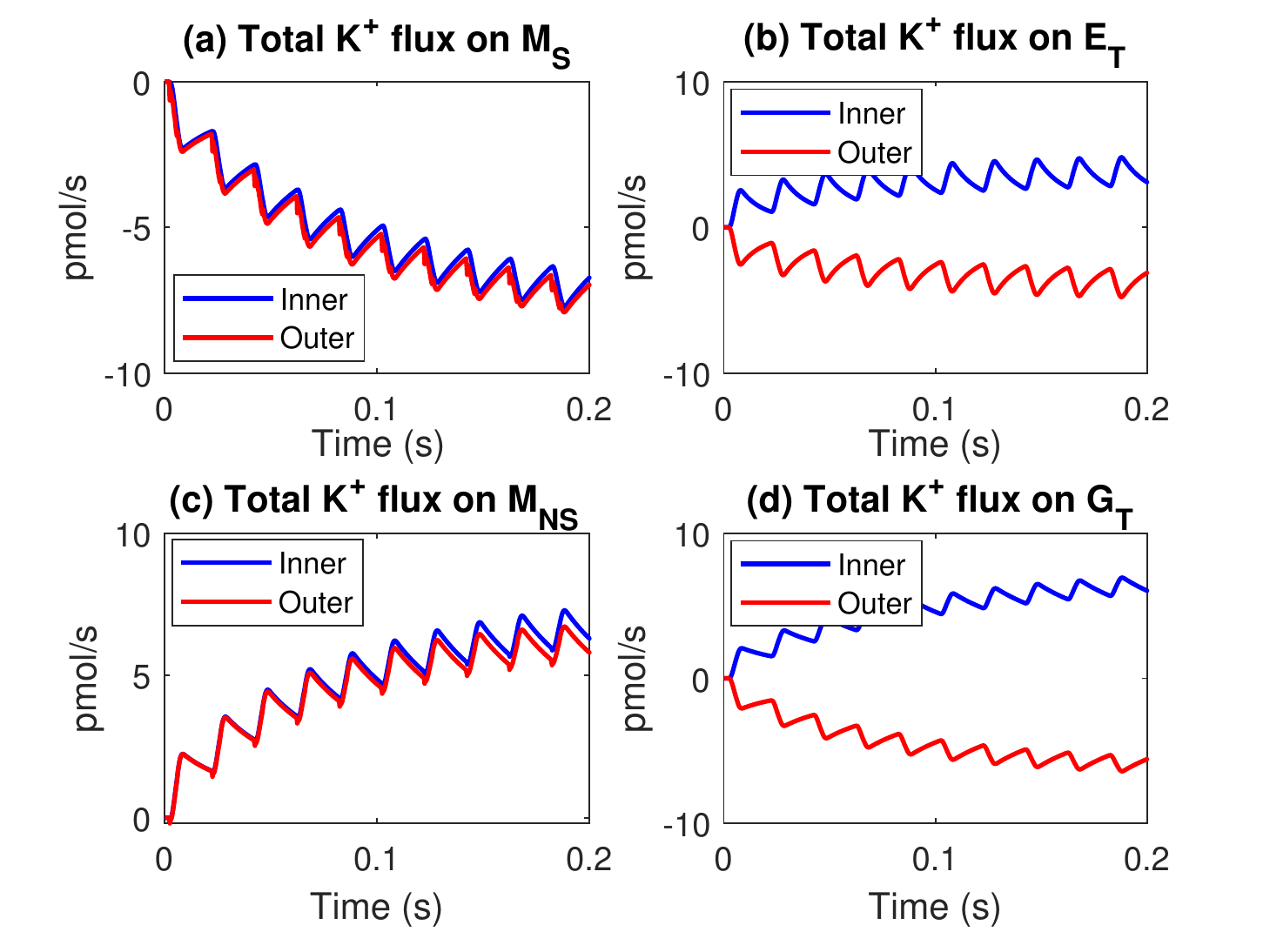}
	\includegraphics[bb=20 5 395 320, clip=true, width=3.25in,height=5.6 cm ] {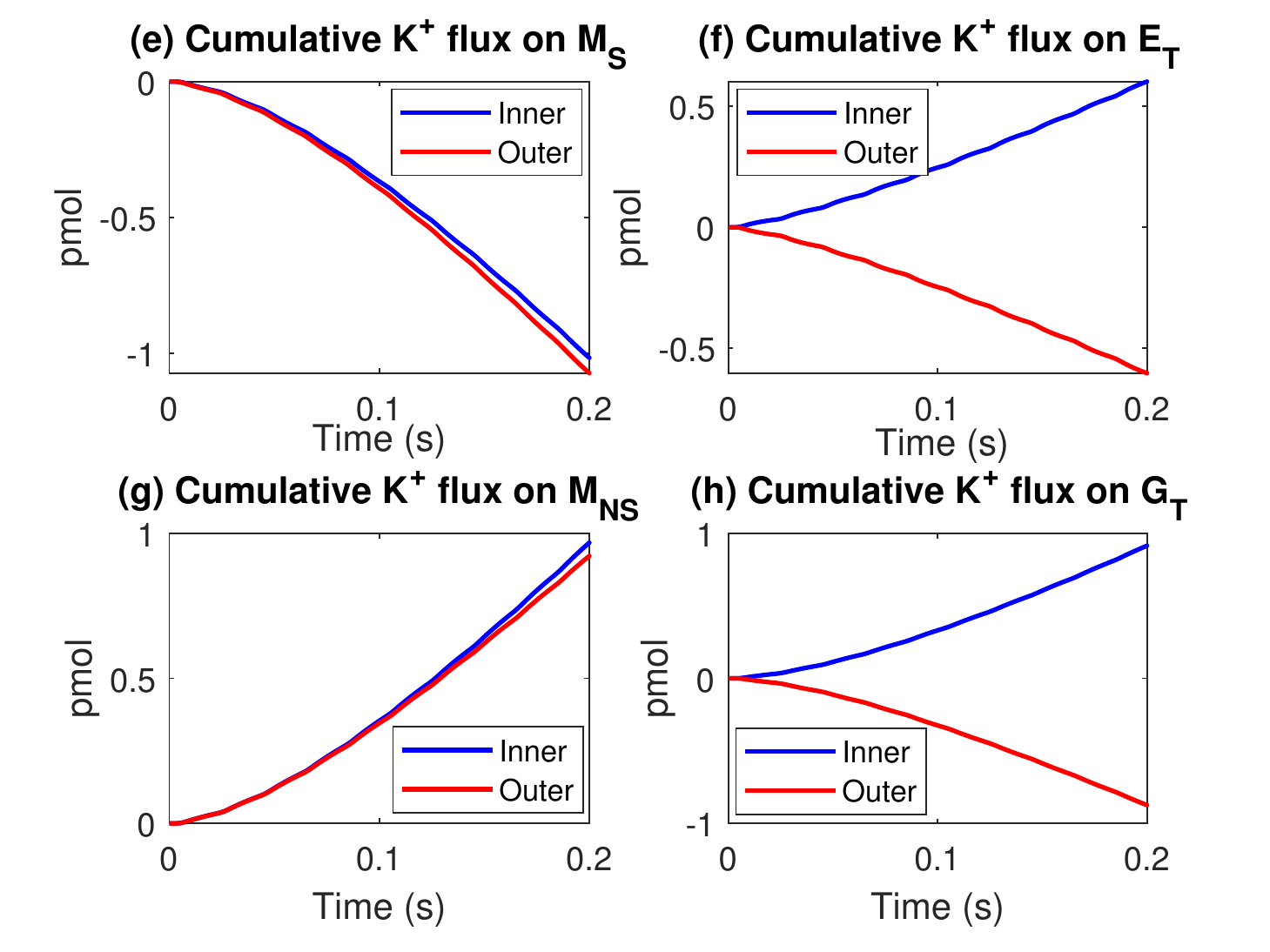}
	\caption{a-d: Potassium flux through $M_S,E_T,M_{NS}$ and $G_T$ during a train of axon firing. e-h: cumulative potassium flux during axon firing period $[0,T_{sti}]$.}
	\label{Fig::4d1}
\end{figure}

\noindent\textbf{(a) During a train of neuron firing}\\
In Fig. \ref{Fig::4d1}, we show the total potassium flux (potassium flux \textcolor{black}{density} integrated over area) and cumulative potassium flux (total potassium flux integrated over time) during the axon firing period $[0,T_{sti}]$.  In both cases, the figures show that the transmembrane flow from extracellular to glial and the communication inside the extracellular act together to help the potassium clearance. The strength of fluxes is gradually increased during axon firing period as in Fig. \ref{Fig::4d1}a$\&$b. The results  confirm that the potassium flux during stimulus  flows from the stimulus region to the non-stimulus region, in both the extracellular space and glial compartment. The cumulative potassium flux through the glial membrane $(M_S)$ is twice as large as that through the extracellular pathway in the transition interface $(E_T)$ as shown in Fig. \ref{Fig::4d1}e\&f. 

The glial compartment serves as an important and quick potassium transport device to remove potassium accumulated while the neuron fires action potentials. In the stimulated region, the accumulated potassium makes the potassium Nernst potential more positive. The change in the potassium Nernst potential induces potassium movement into the glial compartment from the extracellular space (Fig. \ref{Fig::4d1}a). This inflow makes the glial compartment electric potential more positive and moves potassium ions from the stimulated region to the unstimulated region (Fig. \ref{Fig::4d1}d). In the unstimulated region, the glial membrane potential also becomes more positive as it does in the stimulated region, because the glia is an electrical syncytium in the longitudinal and radial directions. However, the glial potassium Nernst potential in the unstimulated region is not very different from that in the resting state. These potentials produce the outward potassium flux from the glial compartment in the unstimulated region (Fig. \ref{Fig::4d1}c). Interacting regions of this sort depend on spatial variables and the properties of the glia as a syncytium in the longitudinal as well as radial directions. It is difficult to capture these effects in models that do not include radial and longitudinal directions as independent variables. Compartment models are possible but it is very difficult to uniquely define invariant parameters over the range of conditions of interest. If additional information becomes available experimentally, the ‘new’ conductances and pumps cannot be introduced in a unique way, without much thought. As the model is adapted to other structures in the brain, the parameters of a compartment model become quite difficult to specify. The distributed model uses structurally defined parameters with structural or specific biophysical meaning. These can be adjusted in a reasonably specific way in different versions of this model, appropriate for different systems in the brain. The schematic graph of potassium circulation in the optic nerve is summarized in the Fig. \ref{Fig::4d3}a.  

\begin{figure}[hpt]
	\centering
	\includegraphics[bb=20 0 910 430, clip=true,width=3.25in,height=5.6 cm]{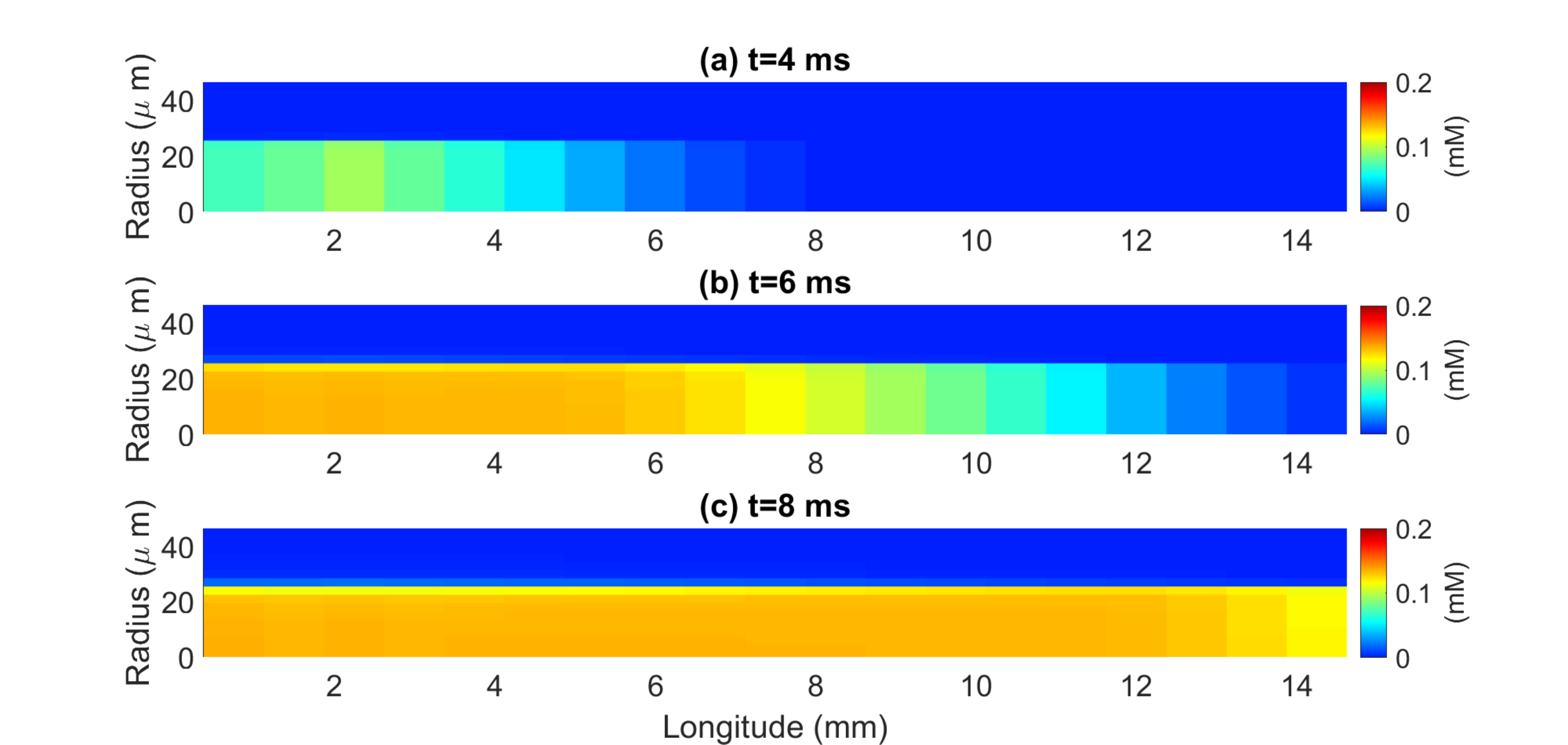}
	\caption{Spatial distribution of potassium changes from the resting state} during an action potential.
	\label{Fig::k_fast}
\end{figure}

\begin{figure}[hpt]
	\centering
	\includegraphics[bb=20 0 910 430, clip=true,width=3.25in,height=5.6 cm]{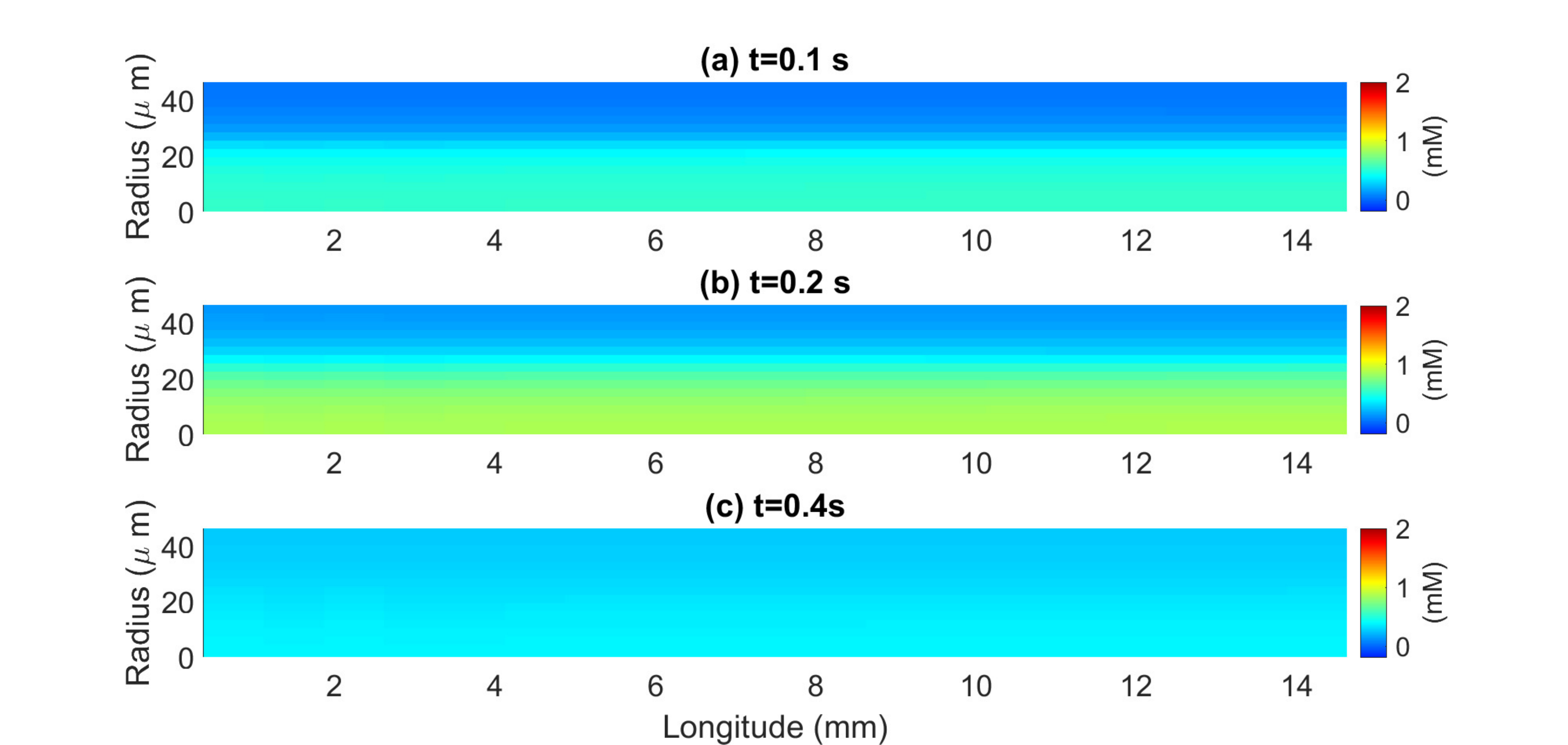}
	\caption{Spatial distribution of  potassium changes from the resting state during and after a train of stimuli.}
	\label{Fig::k_slow}
\end{figure}

The spatial distributions of  $\mathrm{K}^{+}$  concentration  changes from resting state over time are shown in the Fig. \ref{Fig::k_fast} and Fig. \ref{Fig::k_slow}.  \textcolor{black}{Fig. \ref{Fig::k_fast} shows that the $\mathrm{K}^{+}$ concentration varies in the stimulus region} along the longitudinal direction during one action potential, while there is \textcolor{black}{no} change of $\mathrm{K}^{+}$  in the non-stimulus region.   Fig. \ref{Fig::k_slow} a$\&$b show  there an obvious potassium concentration difference in the radial direction after a train of stimuli over time $T_{sti} = 0.2s$. In Fig. \ref{Fig::k_slow}c,  the potassium concentration difference vanishes due to the communication of the extracellular space and glial compartment shown in Fig \ref{Fig::4d4}fg.

\begin{figure}[hpt]
	\centering
	\includegraphics[bb=0 5 410 450, clip=true,width=3.25in,height=8 cm]{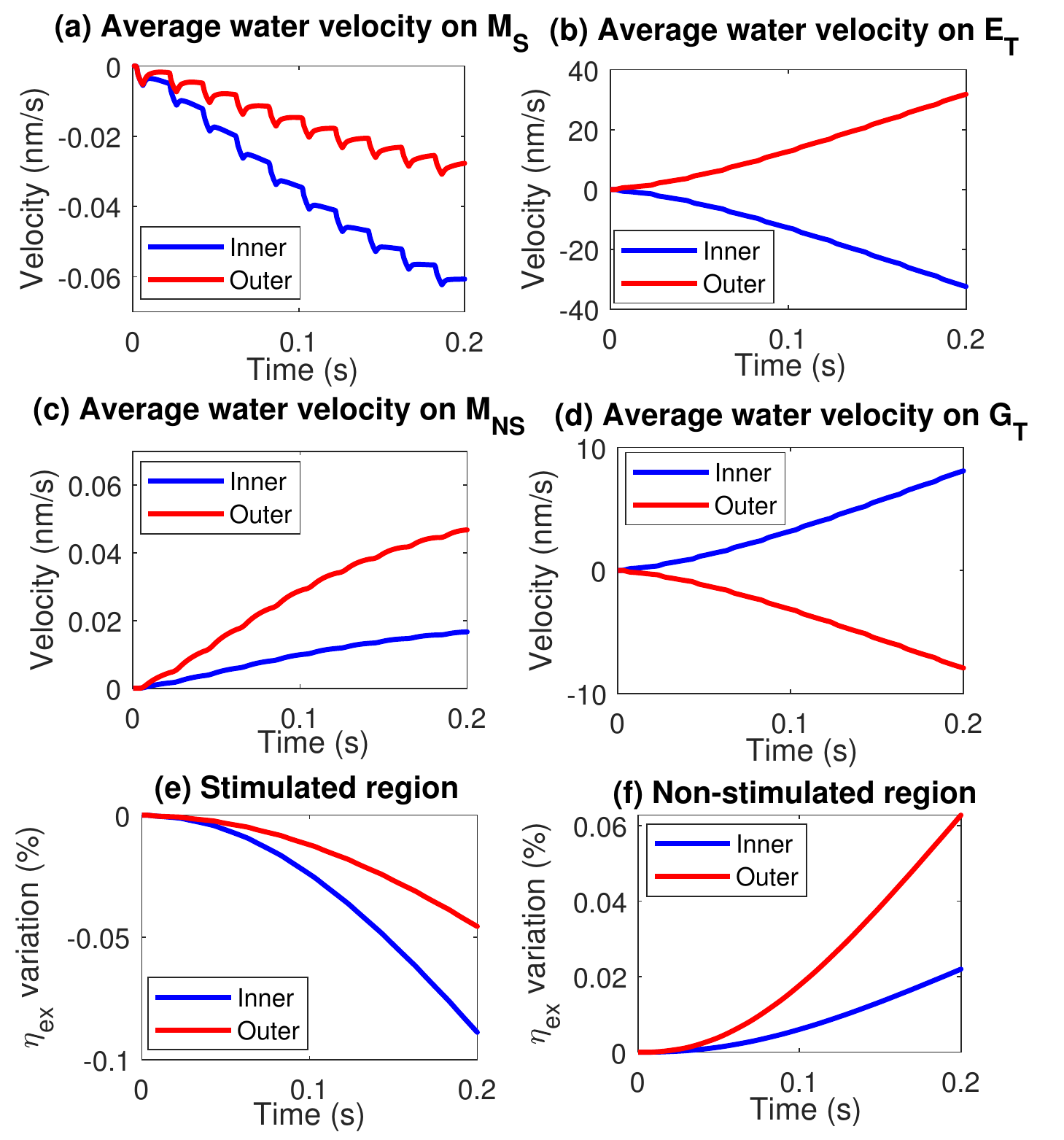}
	\caption{a-d: Average water velocity through $M_S,E_T,M_{NS}$ and $G_{T}$   during a train of axon firing   period $[0,T_{sti}]$. e-f: the extracellular volume fraction variation in the stimulated region and non-stimulated region.}
	\label{Fig::4d2}
\end{figure}

\begin{figure}[hpt]
	\centering
	\includegraphics[width=3.25in,height=5.2 cm]{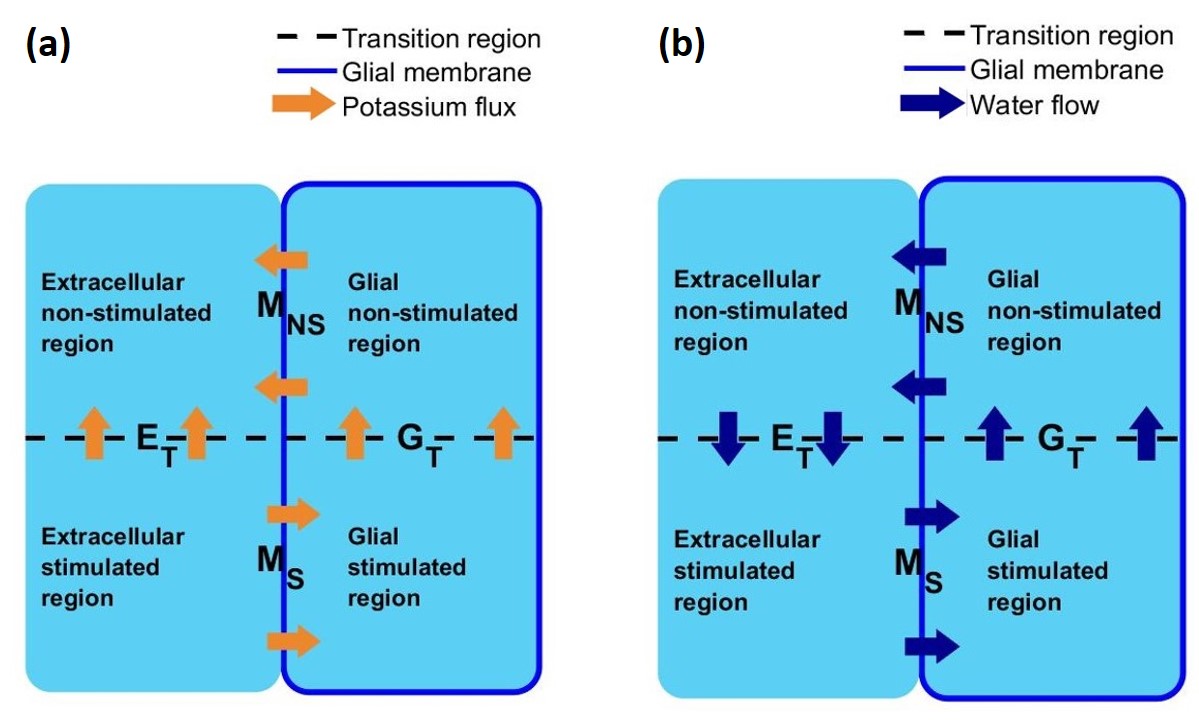}
	\caption{a: Schematic graph of the potassium flux when axon is stimulated. In the stimulated region, the potassium moves through the extracellular pathway and through the glial compartment by way of the glial membrane. In the non-stimulated region, the potassium leaks out to the extracellular space through the glial membrane. b: Schematic graph of the water circulation when the inner part of the  axon is stimulated. In the stimulated region, the glial transmembrane water flow goes from extracellular space into glial compartment as the effect of osmosis difference. In the extracellular space, water goes from non-stimulated region to stimulated region in the radial direction. In the glia compartment goes in the opposite direction. Note these graphs   summarize outputs of large numbers of calculations solving partial differential equations in longitudinal and radial spatial directions and time. They do not represent a compartmental model. They are the output of a model distributed in space.}
	\label{Fig::4d3}
\end{figure}

The water circulation in the optic nerve is driven by the gradient of osmotic pressure in the stimulated region. In the stimulated region, the extracellular osmotic pressure $k_{B} T O_{ex}$ decreased, and glial compartment osmotic flow of water is increased. This is because there is more potassium flux moving into the glial compartment through the glial membrane, but a smaller amount of sodium flux out of the glial compartment because of the ion channel conductance difference between the potassium and sodium channels in the glial membrane. In the glial compartment in the stimulated region of the optic nerve fiber, there is an increase in the water flux into the glial compartment from the extracellular space (Fig. \ref{Fig::4d2}a). The volume fraction of the glial compartment and the hydrostatic pressure have also increased. The increased hydrostatic pressure in the stimulated region also raises the hydrostatic pressure in locations far away from the stimulated region. The increased pressure also drives the flow from stimulated region to the unstimulated region because the glial compartment is a connected space (Fig. \ref{Fig::4d2}d), a longitudinal syncytium. In the unstimulated region, the water flows out of the glial compartment into the unstimulated region because of the increased hydrostatic pressure in the glial compartment (Fig \ref{Fig::4d2}c). Then, because the fluid is incompressible, the fluid in the unstimulated region flows back to the stimulated region (Fig. \ref{Fig::4d2}b). The schematic graph of water circulation in the optic nerve is summarized in Fig. \ref{Fig::4d3}b. This is a summary of our results. It is not a compartment model. Our models are distributed. 

In Fig. \ref{Fig::4d2}e\&f, we show the volume change of extracellular space. In the stimulated region, the extracellular space decreases because the water flows into the glial cell; while in the unstimulated region, the extracellular space swells because of   spatial buffering water flow \cite{kofuji2004potassium}. 

In sum, during a train of action potentials in the axon, the potassium flux transport pattern and potassium flux strength across the glia membrane and through the extracellular pathway are the same for both radial regions, inner and outer. The glial compartment pathway is the dominant clearance mechanism of the potassium accumulated in the extracellular stimulated region, in both cases.

\noindent \textbf{(b) After axon firing period}\\
After the stimulus period, the main potassium clearance mechanism is the passive flow from extracellular space to glial compartment through the glial membrane. The potassium flux in extracellular region and glial compartment is negligible. In these calculations, the extracellular region and the glial compartment could be approximated as a single compartment. We show a schematic figure of potassium flux pattern in Fig. \ref{Fig::4d6}a. This is a summary and sketch of our results. We did not use a compartment model. Our model is distributed.

\begin{figure}[hpt]
	\centering
	\includegraphics[bb=20 5 390 320, clip=true,width=3.25in,height=5.6 cm]{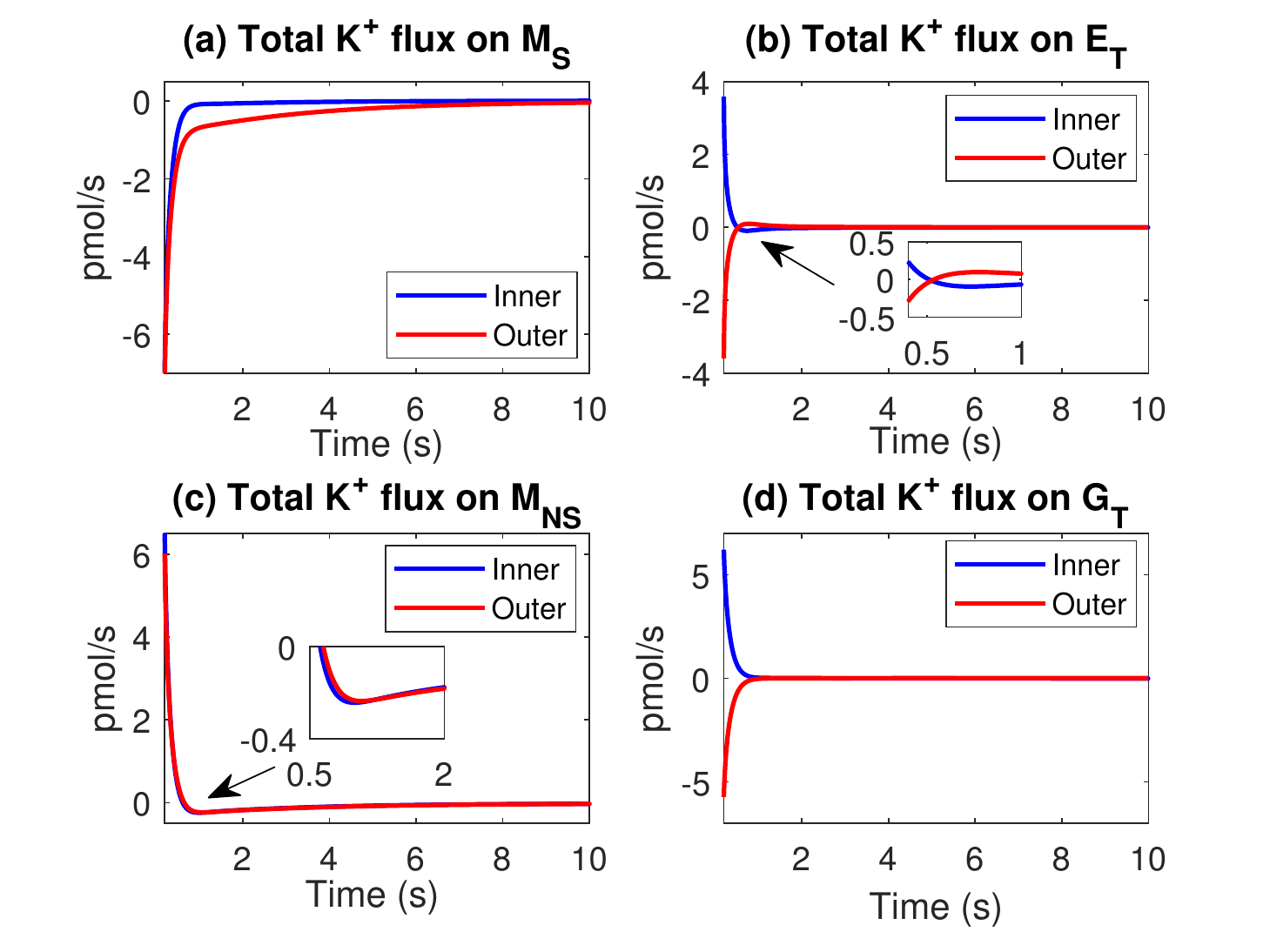}
	\includegraphics[bb=20 5 390 320, clip=true,width=3.25in,height=5.6 cm ] {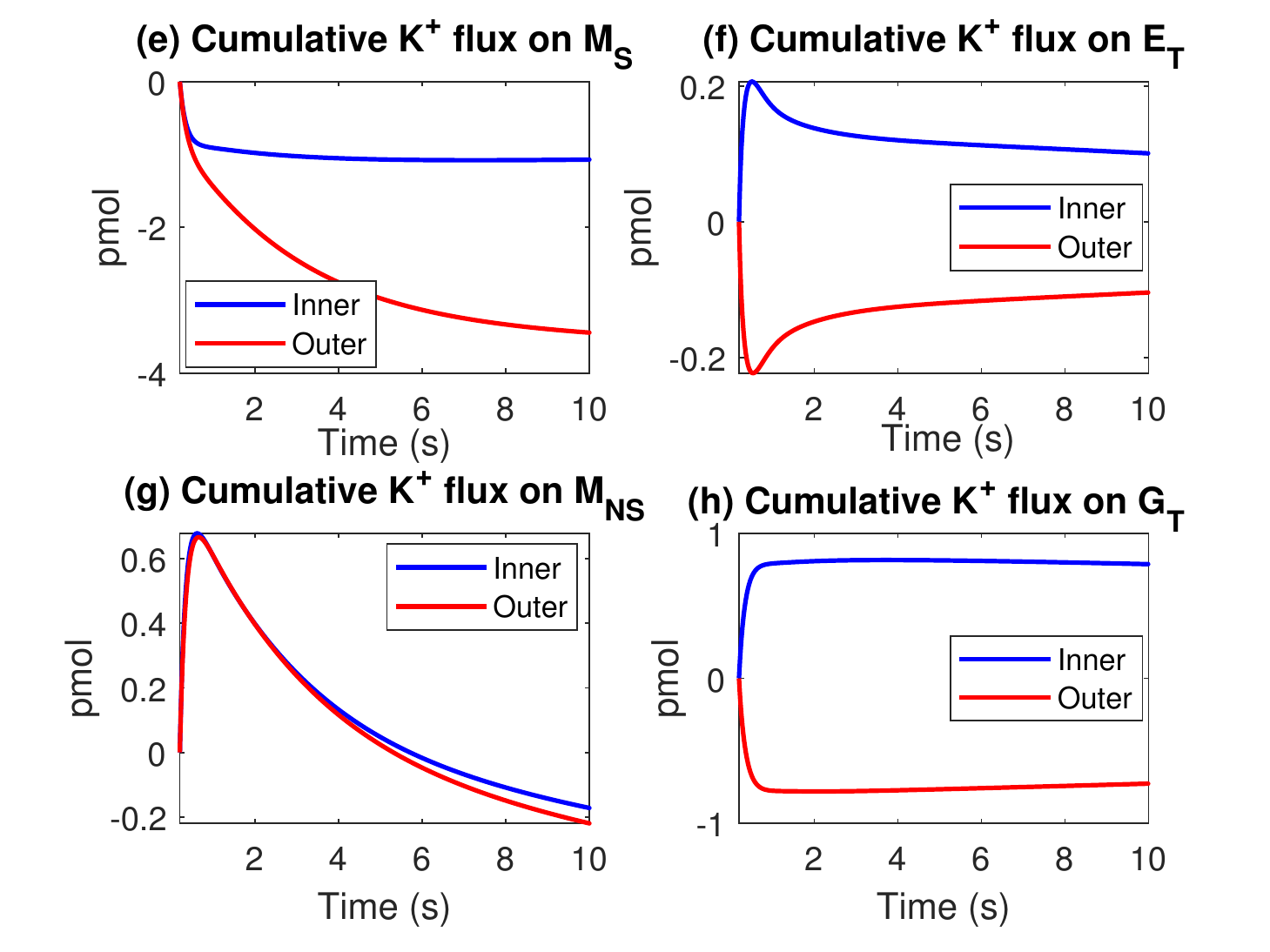}
	\caption{a-d: potassium flux through $M_S,E_T,M_{NS}$, and $G_{T}$ after a train action potentials. e-h: cumulative potassium flux after axon firing.}
	\label{Fig::4d4}
\end{figure}

In the Fig. \ref{Fig::4d4}, we show the total potassium flux and cumulative potassium flux through $M_S,E_T,M_{NS}$, and $G_T$  after axon firing period $[T_{sti},T_{af}]$.  For both cases, the strength of potassium fluxes through the glial transmembrane pathway $(M_S)$ and extracellular pathway have dramatically decreased after the axon stopped firing. 

Fig.  \ref{Fig::4d4}e shows that in both cases (inner stimulus and outer stimulus), the potassium flows into the stimulated glial compartment after the axon firing period. Fig.  \ref{Fig::4d4}g shows that the potassium flux through the glial membrane in the non-stimulated region reverses its direction for a short time after axon stop firing. This occurs because the extracellular potassium concentration becomes evenly distributed in both the stimulated and non-stimulated extracellular space.

\begin{figure}[hpt]
	\centering
	\includegraphics[bb=5 5 410 320, clip=true,width=3.25in,height=6.2 cm]{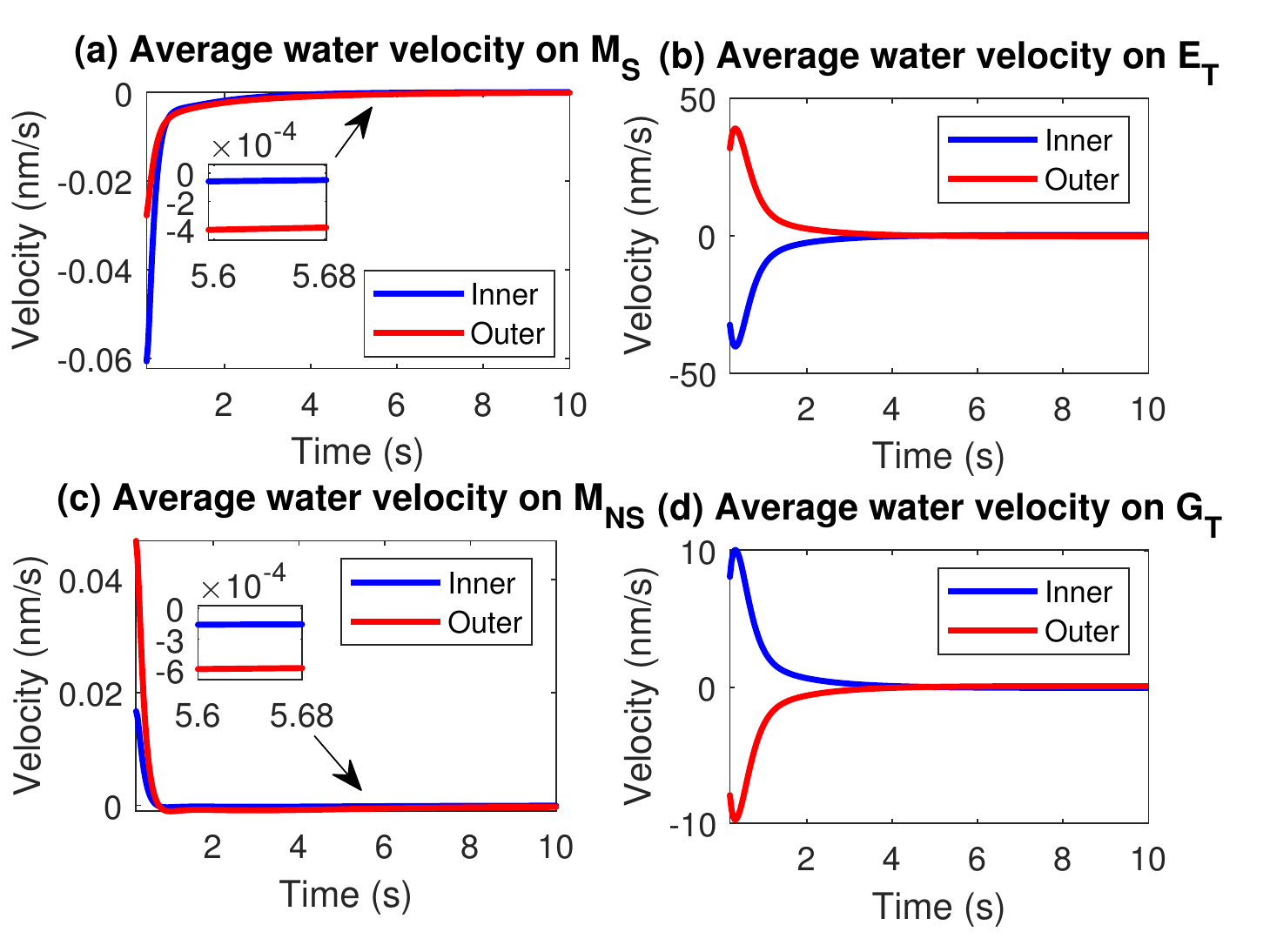}
	\caption{a-d: Average water velocity through $M_S,E_T,M_{NS}$ and $G_T$   after a train of axon firing.}
	\label{Fig::4d5}
\end{figure}

\begin{figure}[hpt]
	\centering
	\includegraphics[width=3.25in,height=5.2 cm]{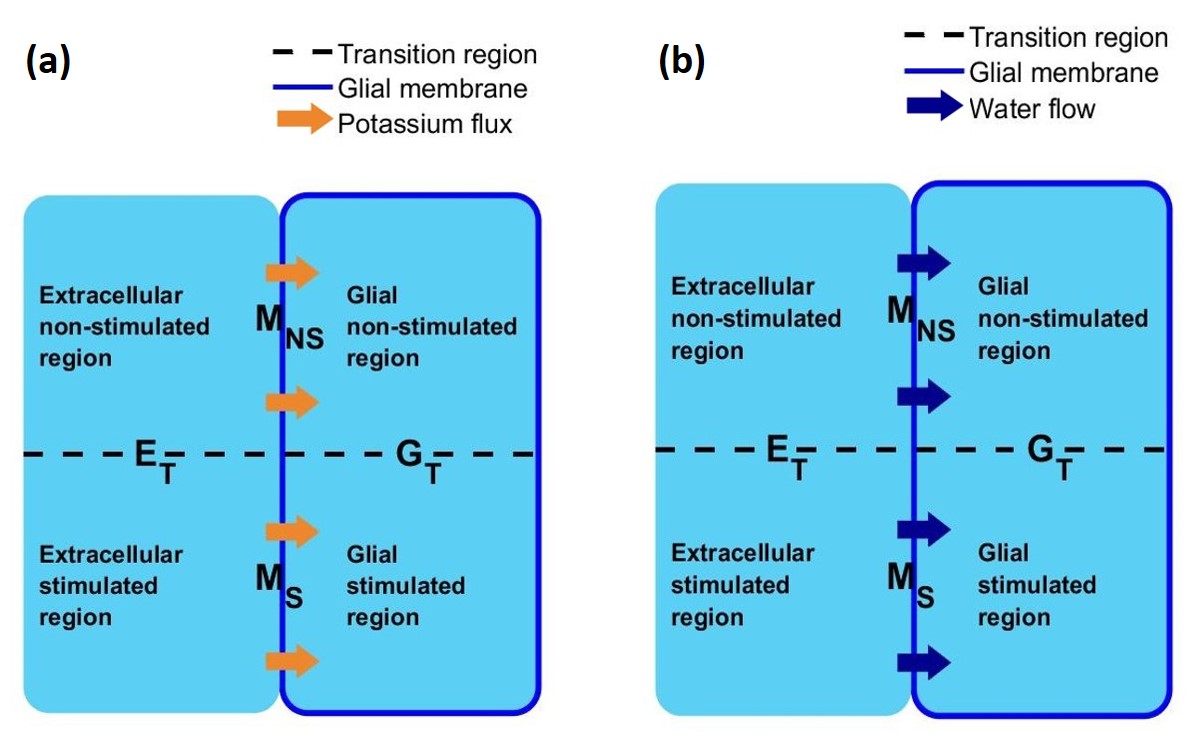}
	\caption{a: Schematic graph of the potassium flux after the axon was stimulated. The potassium flux leaks into the glial compartment from the extracellular space through the glial membrane in both stimulated and unstimulated regions. The potassium flux in the extracellular space and glial compartment is negligible. b: Schematic graph of the water flux after the axon was stimulated. Note these graphs are graphs that summarize outputs of large numbers of calculations solving partial differential equations in longitudinal and radial spatial directions and time. They do not represent a compartmental model. They are the output of a model distributed in space.}
	\label{Fig::4d6}
\end{figure}

In the Fig. \ref{Fig::4d4}f, in the extracellular space, potassium flows back to the stimulus region from the non-stimulus region via the extracellular pathway (in both cases). Accordingly, Fig. \ref{Fig::4d5} shows that after the axon stops firing, the water flow inside the compartments becomes almost zero. The water flows through both stimulated and unstimulated glial membrane into the glial compartment, which is the same as the schematic graph Fig. \ref{Fig::4d6}b.

In sum, after the axon stops firing action potentials, the extracellular potassium concentration is quickly transported in the glial compartment and extracellular pathway. The main clearance of potassium is through the glia membrane in both stimulated and unstimulated regions. The potassium flux through the extracellular pathway becomes weaker after the axon stops firing action potentials.

\begin{figure}[hpt]
	\centering
	\includegraphics[bb=30 105 510 260, clip=true, width=3.25in,height=2.8 cm]{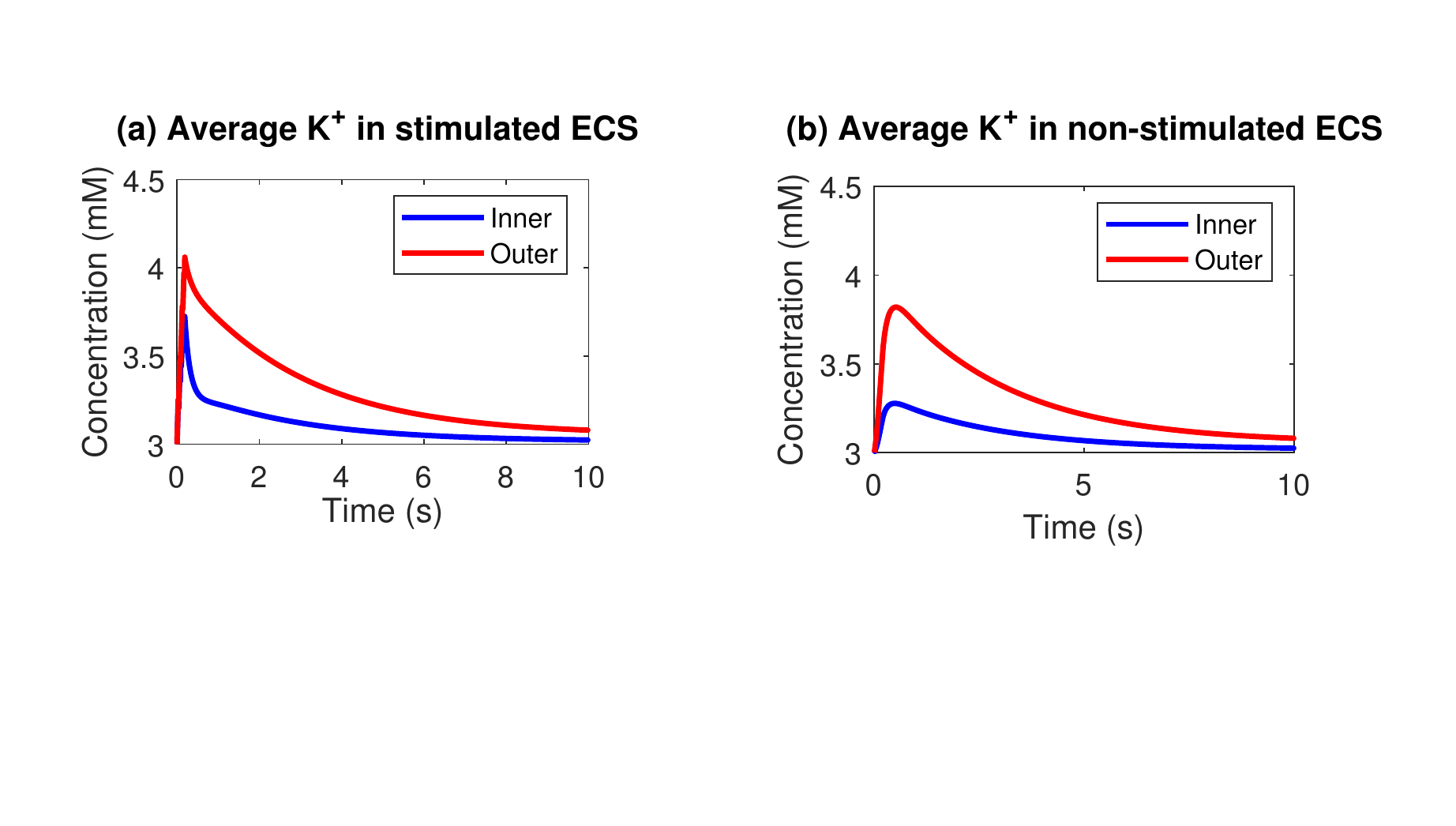}
	\caption{a-b: potassium concentration variation in the extracellular \textcolor{black}{stimulated} region and non-stimulated region.}
	\label{Fig::4d7}
\end{figure}

Fig. \ref{Fig::4d7}a\&b, we show the variation of the potassium concentration in the extracellular stimulated and unstimulated regions, respectively. The peak potassium concentration in the stimulated region is higher in the outer stimulated case compared to the inner stimulated case. As discussed previously, the strength of potassium clearance is the same for both cases, while in the outer radial stimulated case, there is three times as much potassium in the extracellular space during the axon firing as in the inner radial stimulated case. 

The equality of potassium clearance in both cases also explains why the potassium concentration drops faster in the inner radial stimulated case than the outer one after axon stop firing. We provide a decay timetable in the appendix.

\subsubsection{\textcolor{black}{Randomly} distributed \textcolor{black}{stimulation} }
\begin{figure}[hpt]
	\centering
	\includegraphics[bb=25 15 400 330, clip=true, width=3.25in,height=4.5 cm]{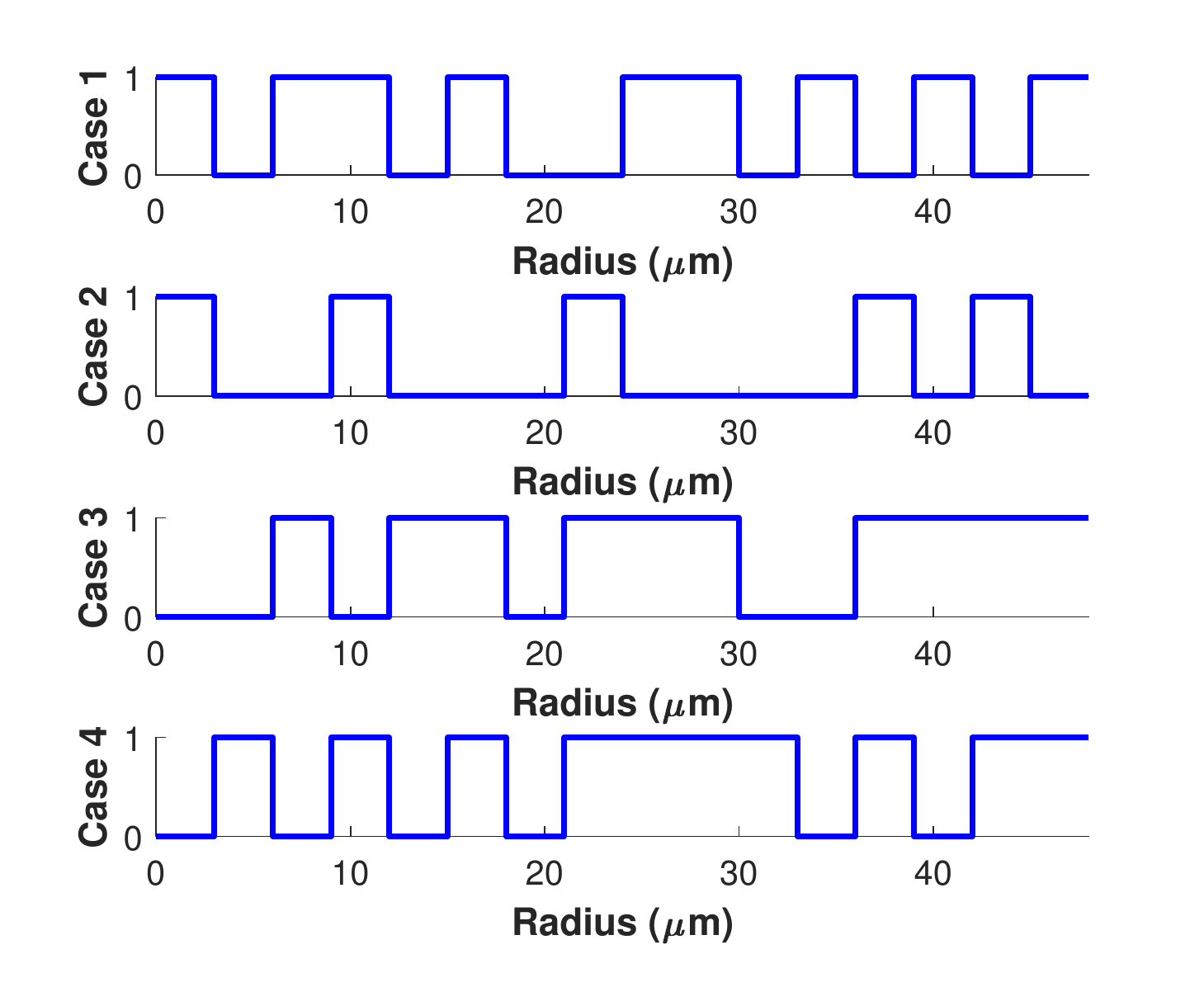}
	\caption{The stimulated radial segments in each case. The intervals with value 1 are stimulated segments and the intervals with value 0 are unstimulated segments.}
	\label{Fig::4d8}
\end{figure}

Spatially uniform patterns of stimulation might produce systematic artifacts as occur in Moire patterns and aliasing.
In this section, we study whether random stimulation patterns in space differ from spatially uniform patterns. We apply a train of \textcolor{black}{stimuli} to four randomly distributed stimulated regions in the radial direction. The strength and duration of the stimulus is same as in the \textcolor{black}{section \ref{inner_outer}} and the current is applied at the same longitudinal location $(z=z_{0})$. The details of the radial stimulated location in each case is shown in   Fig. \ref{Fig::4d8}. 

\begin{figure}[hpt]
	\centering
	\includegraphics[bb=20 5 390 320, clip=true,width=3.25in,height=5.6 cm]{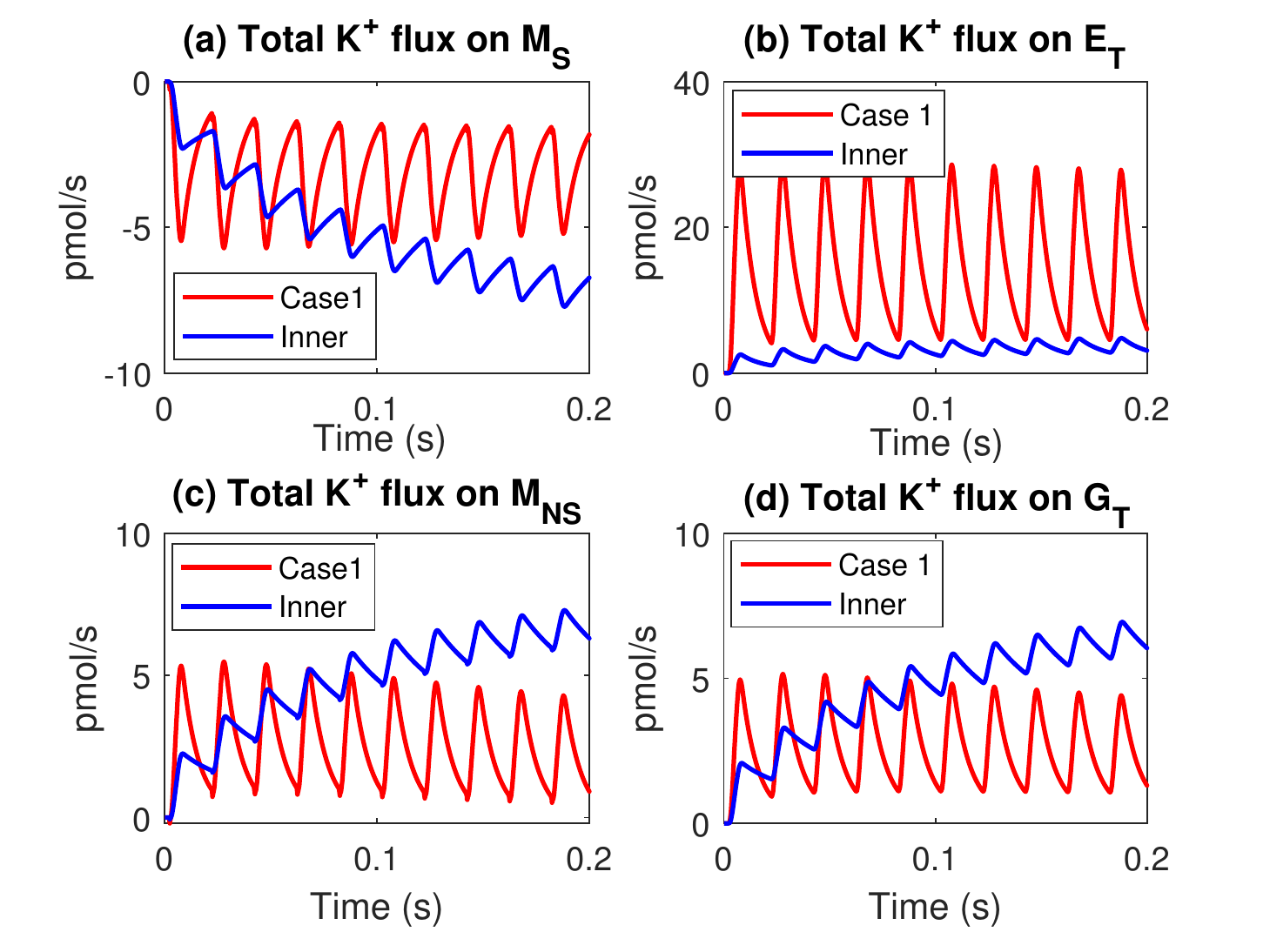}
	\includegraphics[bb=20 5 390 320, clip=true,width=3.25in,height=5.6 cm ] {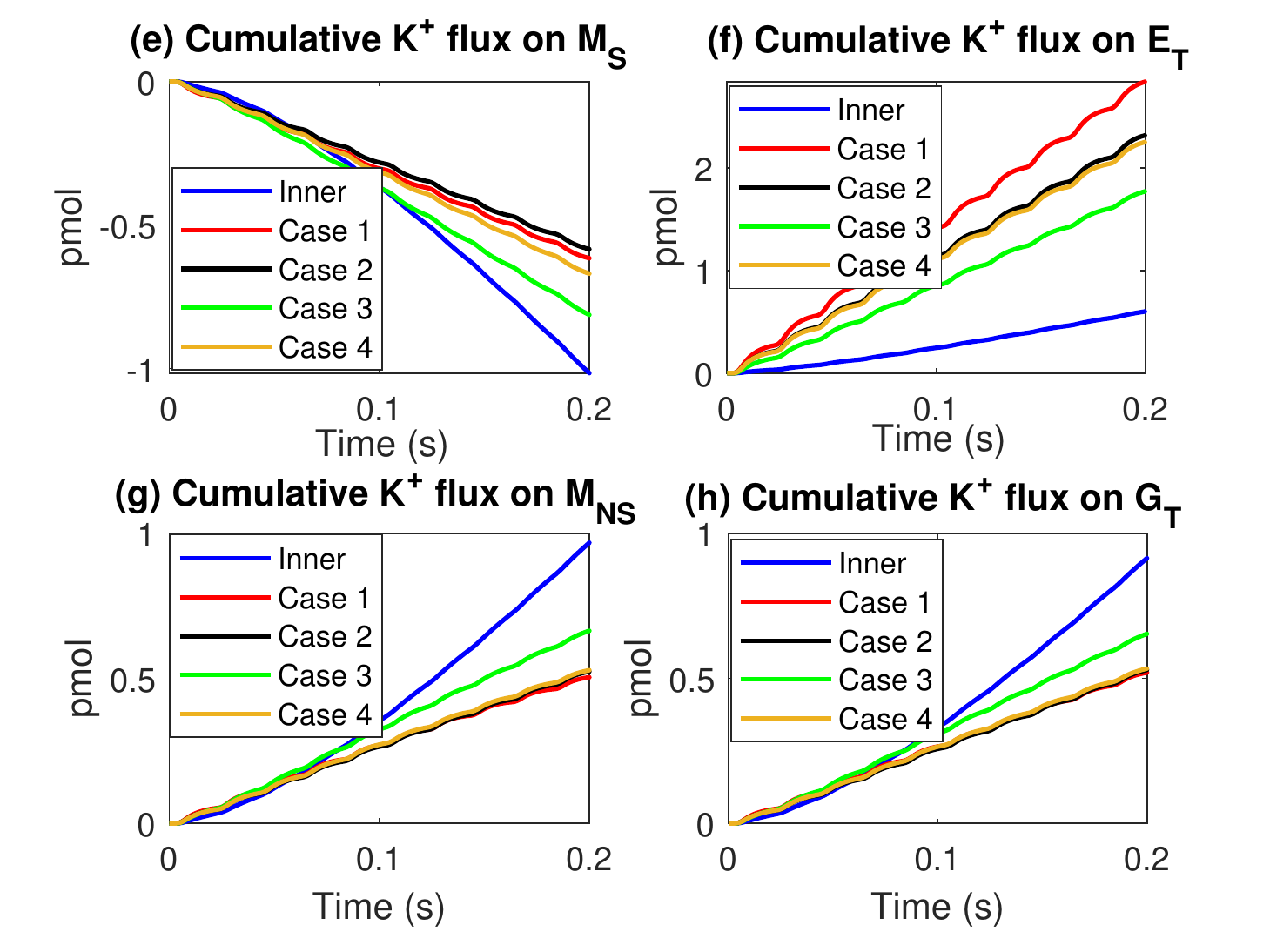}
	\caption{a-d: the comparison between the spatially random stimulated case with the spatially uniform radial (inner) case during a train of axon firing. e-f: the cumulative flux comparison in $M_S,E_T,M_{NS}$, and $G_T$  during a train of axon action potentials. }
	\label{Fig::4d9}
\end{figure}

\noindent \textbf{(a) During a train of axon firing}\\
We compare the spatially random stimulated case (case 1, the rest \textcolor{black}{of the} cases are similar and shown in an appendix) with the inner radial region stimulated case in the Fig. \ref{Fig::4d9}a-d. During a train of axon firing, Fig. \ref{Fig::4d9}a\&b, shows, in the randomly stimulated case, that the potassium clearance through the glial transmembrane $(M_S)$ has been reduced while the potassium flux through extracellular pathway in transition interface $(E_T)$ has dramatically increased. Fig. \ref{Fig::4d9}e\&f shows that the major clearance pathway in the randomly stimulated cases becomes the extracellular pathway in transition interface $(E_T)$. More potassium flux goes though the extracellular path in the transition interface \textcolor{black}{in comparison to how much} goes through glial membrane in the stimulated region. This differs from the outer and inner stimulated cases, where the glial transmembrane $(M_S)$ dominates as seen in Fig. \ref{Fig::4d1}e\&f.

Fig. \ref{Fig::4d9}a-d shows that in the inner stimulated case (blue line), the potassium flux strength gradually increased with small oscillation in each axon stimulus time period, while in the spatially random stimulated case (red line), the potassium flux strength shows a periodic pattern in time with larger oscillation in each stimulus period. The reason for this quite different potassium flux pattern is that the extra potassium in the extracellular stimulated region has been cleared quicker in the random selected case. So, in each stimulus period, the potassium flux decreases dramatically since the potassium goes back to its resting state. While in the inner stimulus case, for each stimulus period, the clearance of potassium is slower and there is accumulation of potassium in the stimulated extracellular space.

In sum, Fig. \ref{Fig::4d9}e\&f, during a train of axon firing, much more potassium flux goes through the extracellular pathway through transition interface $(E_T)$, which reduces the effect of glial compartment pathway in the stimulated region $(M_S)$.

\begin{figure}[hpt]
	\centering
	\includegraphics[bb=20 5 390 320, clip=true,width=3.25in,height=5.6 cm]{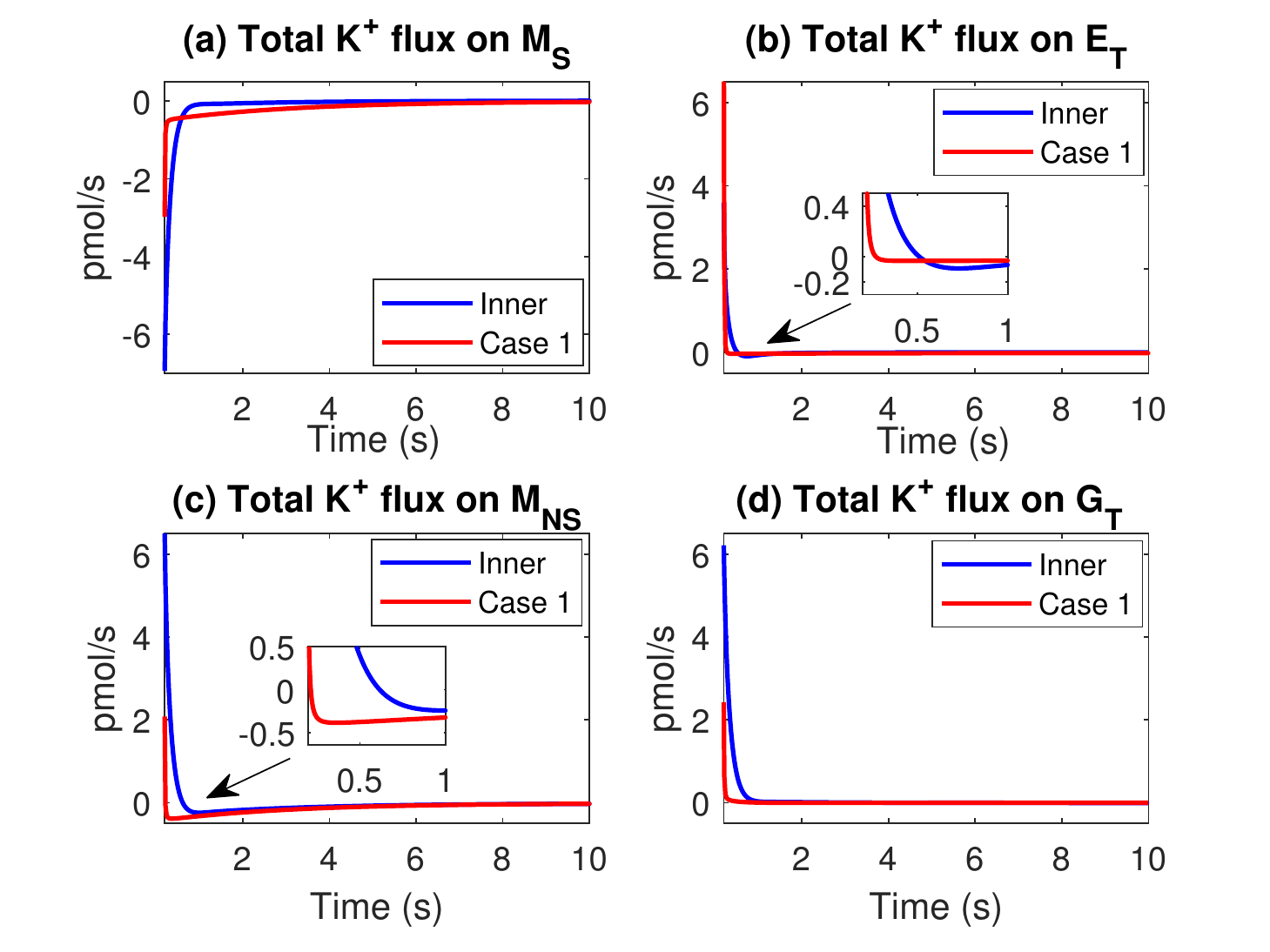}
	\includegraphics[bb=20 5 390 320, clip=true,width=3.25in,height=5.6 cm ] {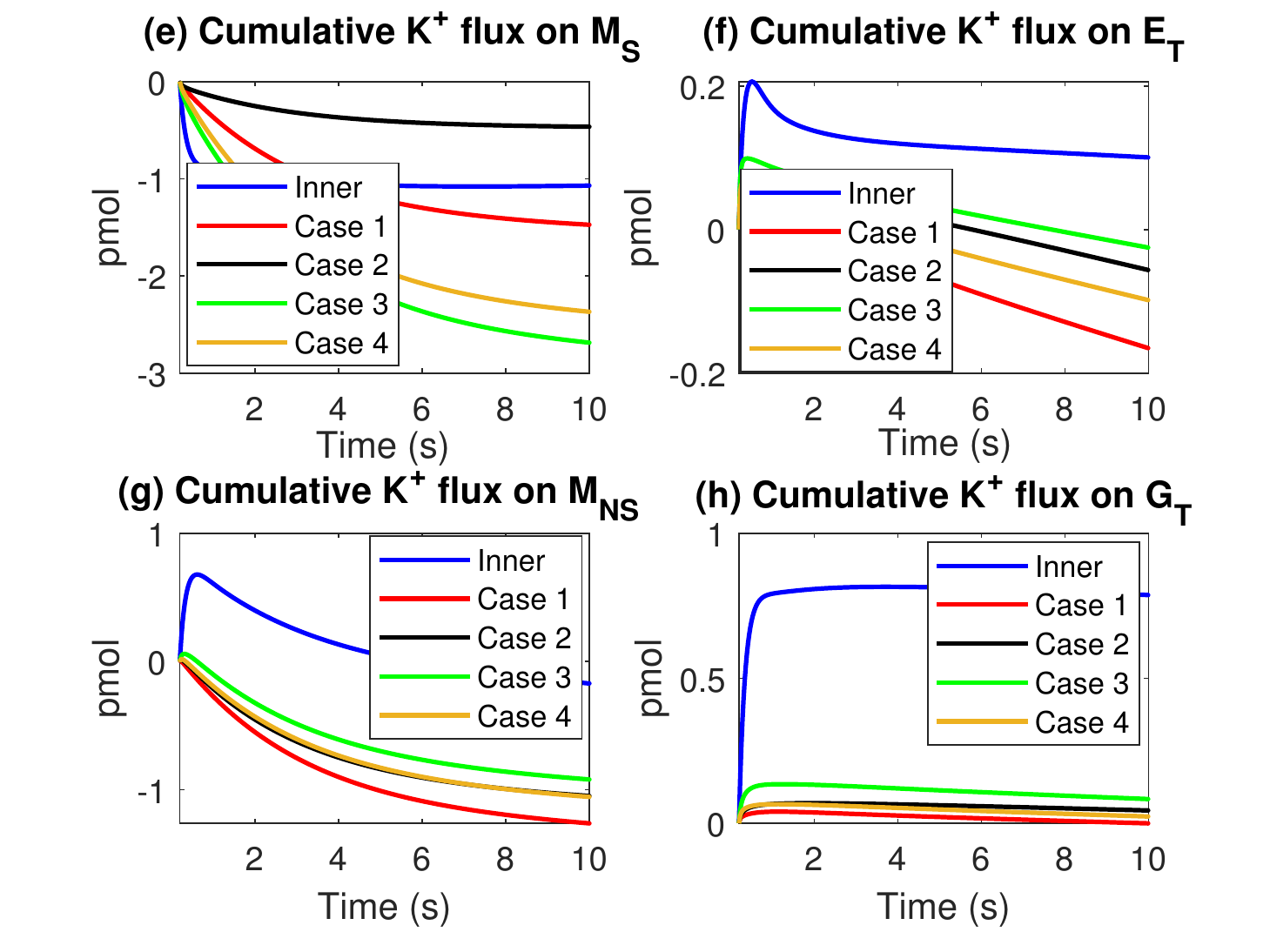}
	\caption{a-d: the comparison between the spatially random case with the uniform inner radial case after a train of action potentials Vector directions as defined previously. e-f: the cumulative flux comparison in $M_S,E_T,M_{NS}$, and $G_T$  after a train of action potentials.}
	\label{Fig::4d10}
\end{figure}

\noindent \textbf{(b) After axon firing period}\\
After the axon stops firing action potentials, both the spatially random case and the uniform inner case have similar reduced potassium fluxes, See $M_S,E_T,M_{NS}$, and $G_T$ In Fig. \ref{Fig::4d10}e\&g shows that the main pathway for potassium clearance (after the action potentials cease) is through the glial transmembrane pathway.

Fig. \ref{Fig::4d10}h shows a large difference between the spatially random and the spatially uniform case in some properties as we feared might occur. The cumulative transport of potassium flux through the glial compartment transition interface in the spatially random cases is much smaller than it is in the uniform inner case. This is because during axon firing period, the extracellular potassium in the stimulated region has been quickly removed in the random cases. The potassium concentration became more homogeneous and there was less difference between the stimulated region and non-stimulated regions. As a result, the glial compartment electric potential $\phi_{gl}$  becomes homogeneous in entire glial compartment and the electric drift flow in the glial transition interface was reduced for potassium. These results show the importance of checking for artifacts produced by artificial assumptions of spatial uniformity or periodicity.

\noindent\textbf{(c) Potassium clearance and fluid velocity in the extracellular space and glial compartment}\\

\begin{figure}[hpt]
	\centering
	\includegraphics[bb=20 0 390 320, clip=true,width=3.25in,height=5.6 cm]{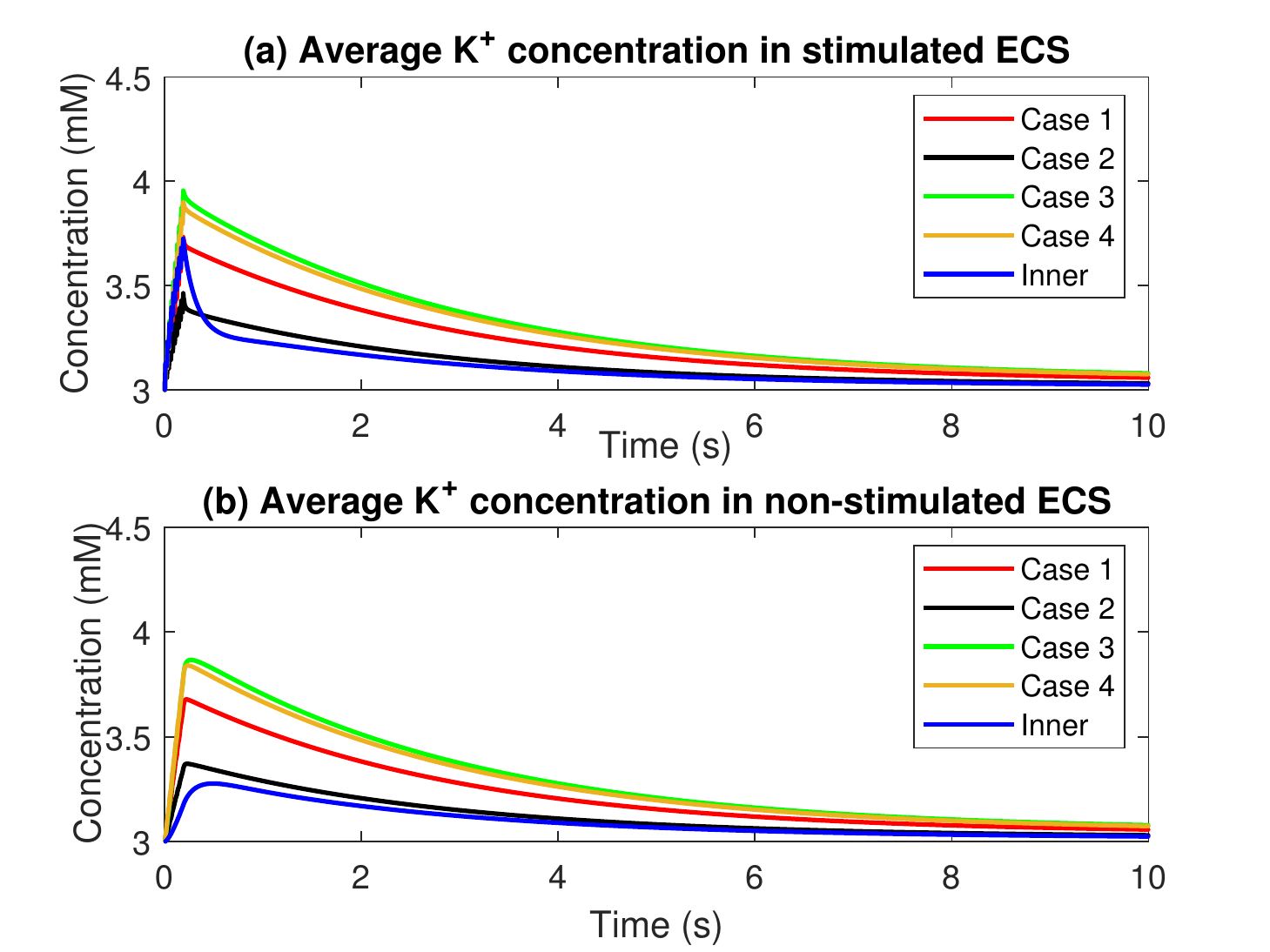}
	\includegraphics[bb=20 0 390 320, clip=true,width=3.25in,height=5.6 cm ] {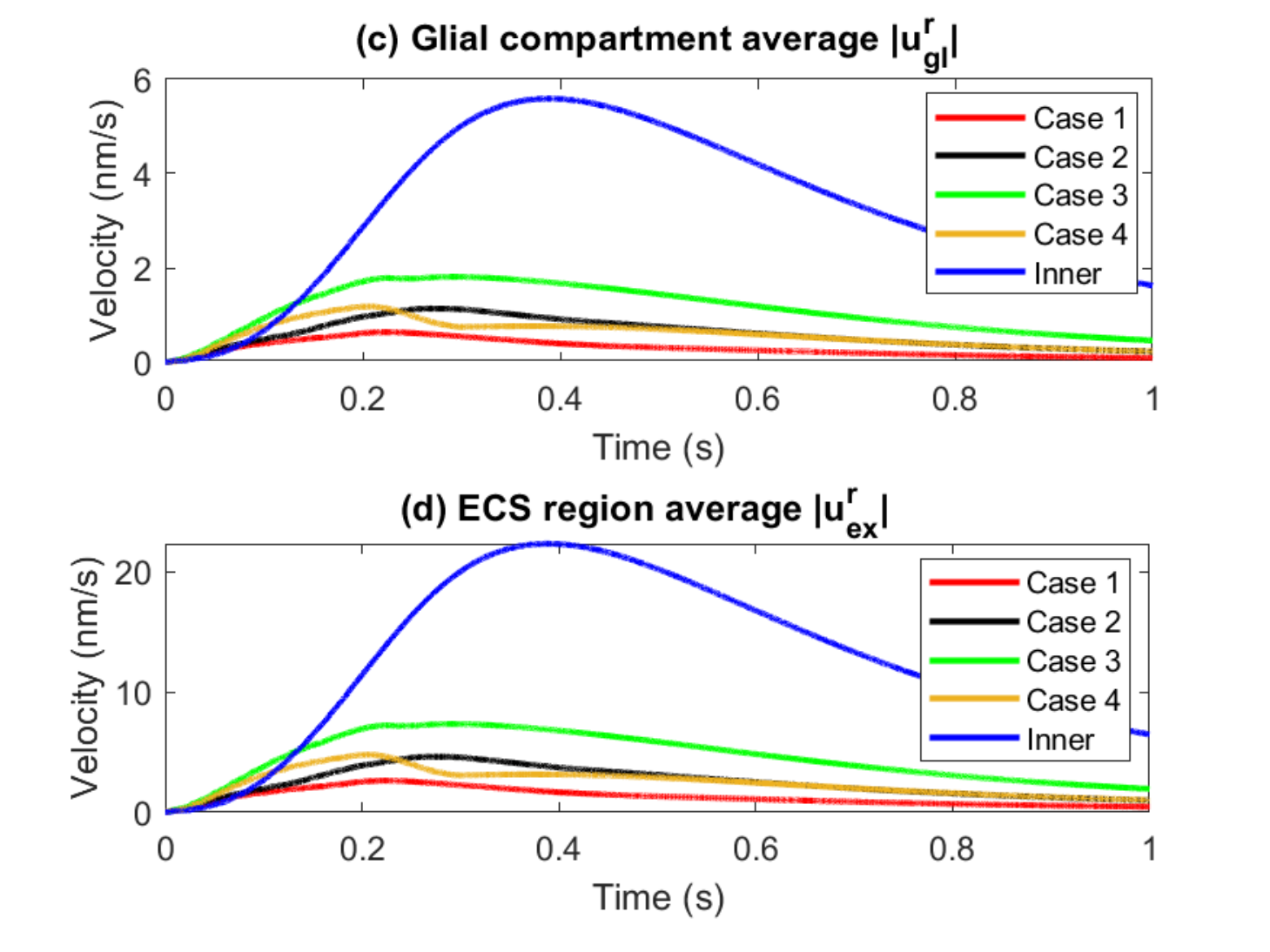}
	\caption{a-b: variation of potassium concentration in the extracellular simulated region and unstimulated regions. c-d: average glial compartment radial absolute velocity and extracellular space radial absolute velocity.}
	\label{Fig::4d11}
\end{figure}

Since the potassium goes less through the glial transmembrane pathway in the stimulated region, the osmosis in the extracellular $k_{B}T O_{ex}$ and in glial $k_B T O_{gl}$, and variation in the stimulated region has been reduced. Therefore, the strength of the velocities decreases in both glia and extracellular compartment since the decrease of the glial transmembrane water flow as \textcolor{black}{shown} in Fig. \ref{Fig::4d11}c\&d. The average potassium concentration changes in both stimulated region and non-stimulated region \textcolor{black}{as} shown in the Fig. \ref{Fig::4d11}a\&b and its decay time in each case is in the appendix.

\subsection{Effect of NKCC and non-selective pathway \label{NKCC_PIA}}
In this section, we consider the effect of glial membrane conductance and a non-selective pathway on the pia boundary on the potassium clearance process. We first introduce the NKCC channel into the glial membrane, as widely studied in the literature. We compare simulation results between the model with NKCC and without NKCC. In the second part, we consider the effect of a non-selective pathway in the pia mater boundary. That pathway allows the convection flux through an extracellular pathway out of the optic nerve. We investigate how these factors affect the potassium clearance. While the contribution of the NKCC channel is significant, it is important to realize that in the present state of knowledge, other relevant channels may not yet have revealed themselves. Some channels become activated only under special conditions that are hard to find. Some channels may become activated only to protect systems under severe stress, as in clinical situations like oxygen deprivation, swelling, and so on. It will be important to identify such channels as the model is applied to clinical situations, some of which are of considerable importance.

\subsubsection{Effect of NKCC on the glial membrane}
We introduce a model of NKCC channels in the glial membrane as in the \cite{ostby2009astrocytic,lauf2000k}, We describe the $\mathrm{K^+}$,$\mathrm{Na^+}$and $\mathrm{Cl^-}$ flux through the NKCC channel in the glial membrane as in  \cite{ostby2009astrocytic,lauf2000k}

\begin{equation}
\begin{array}{l}
J_{NKCC}^{K}=-\frac{I_{\max}^{NKCC}}{e z^{K}} \log \left(\frac{C_{ex}^{K}}{C_{gl}^{K}} \frac{C_{ex}^{Na}}{C_{gl}^{Na}}\left(\frac{C_{ex}^{Cl}}{C_{gl}^{Cl}}\right)^{2}\right), \\
J_{NKCC}^{Na}=-\frac{I_{\max}^{NKCC}}{ez^{Na}} \log \left(\frac{C_{ex}^{K}}{C_{gl}^{K}} \frac{C_{ex}^{Na}}{C_{gl}^{Na}}\left(\frac{C_{ex}^{Cl}}{C_{gl}^{Cl}}\right)^{2}\right), \\
J_{NKCC}^{Cl}=2 \frac{I_{\max}^{NKCC}}{ez^{Cl}} \log \left(\frac{C_{ex}^{K}}{C_{g l}^{K}} \frac{C_{ex}^{N a}}{C_{gl}^{Na}}\left(\frac{C_{ex}^{Cl}}{C_{gl}^{Cl}}\right)^{2}\right).
\end{array}
\end{equation}

To compare the effect of NKCC without NKCC, we keep the potassium and sodium concentration as well as the resting electric potential in the glial compartment and extracellular space the same. In the resting state, we set sodium and potassium current through NKCC channel to be comparable with respect to the Na/K pump current as in the paper \cite{ostby2009astrocytic,lauf2000k}. In the appendix, we provide the two sets of parameters (NKCCa and NKCCb) that balance the additional NKCC current through the glial membrane.  

\begin{figure}[hpt]
	\centering
	\includegraphics[bb=20 5 390 320, clip=true,width=3.25in,height=5.6 cm]{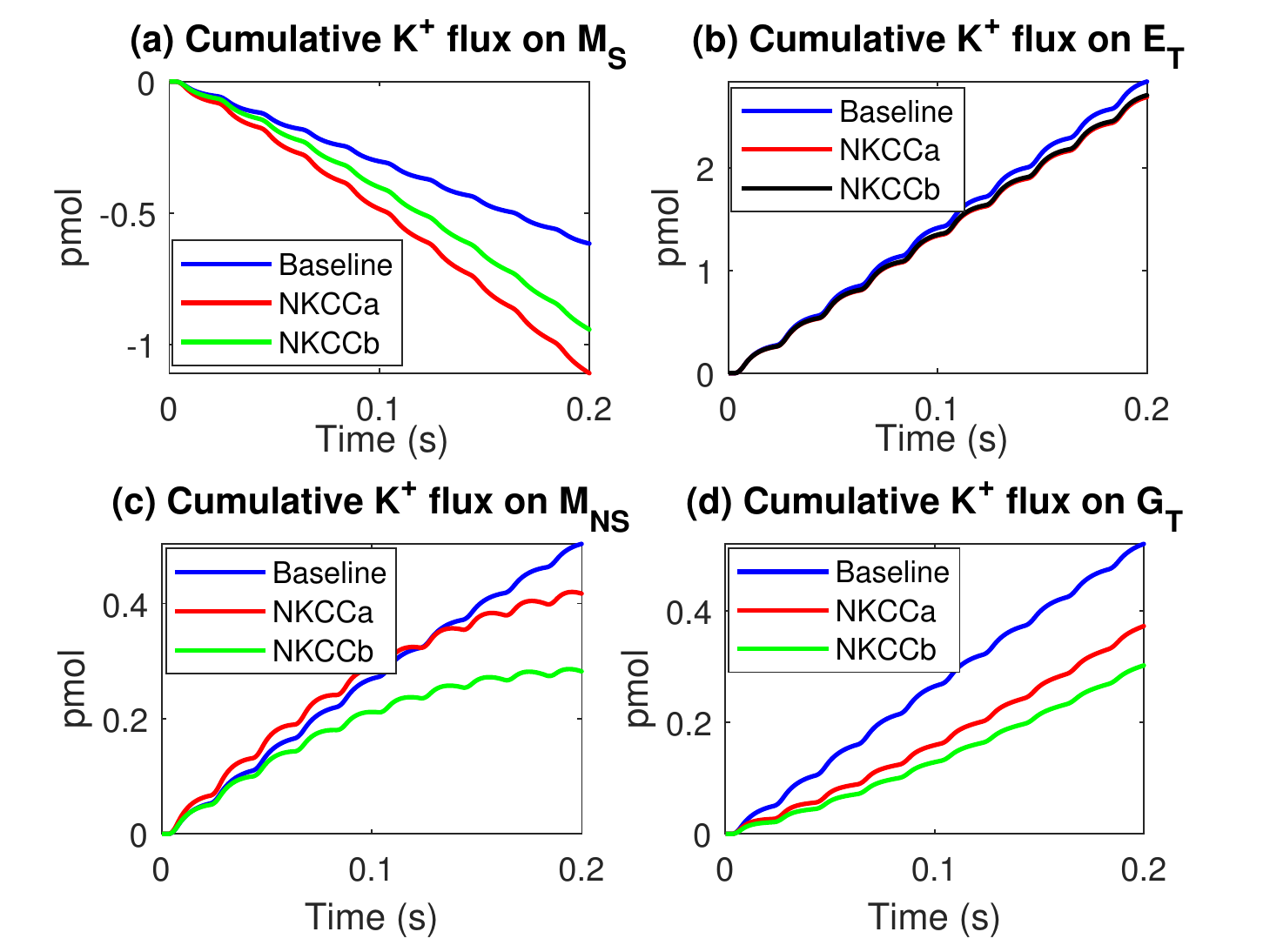}
	\includegraphics[bb=20 5 390 320, clip=true,width=3.25in,height=5.6 cm ] {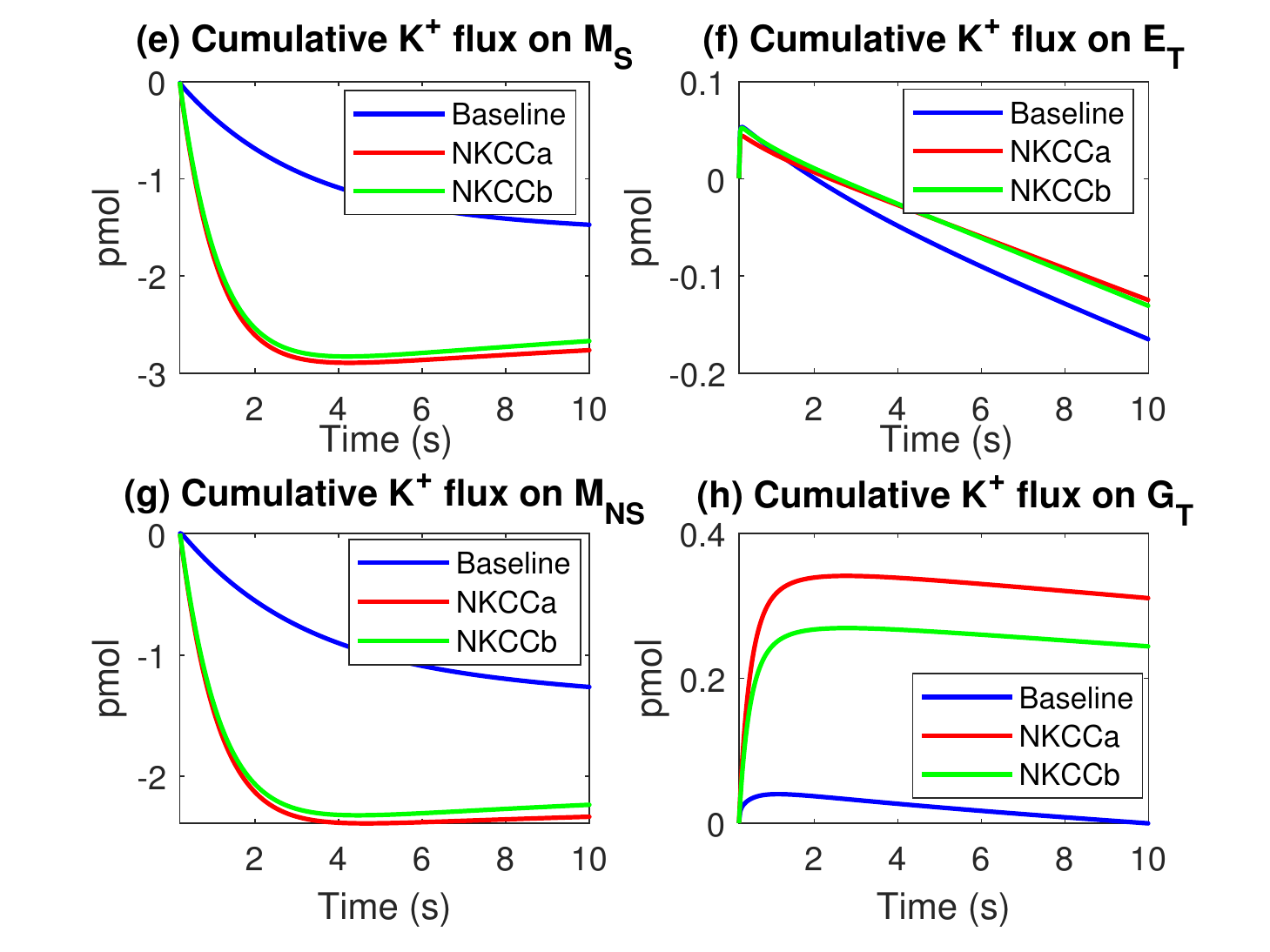}
	\caption{a-d: the cumulative flux comparison in $M_S,E_T,M_{NS}$, and $G_{T}$  during a train of  action potentials. e-h: the cumulative flux comparison in $M_S,E_T,M_{NS}$, and $G_{T}$  after a train of action potentials.}
	\label{Fig::4d12}
\end{figure}

In the simulation below, we compare the models with and without NKCC. The stimulated region Fig. \ref{Fig::4d12}, shows the cumulative potassium flux through the $M_S,E_T,M_{NS}$, and $G_T$ during and after stimuli. 

Fig. \ref{Fig::4d12}a shows that more potassium goes through the glial membrane in the simulated region during axon firing when NKCC is present, which is hardly surprising. The NKCC channel has enhanced the transport of potassium through the glial membrane after the axon stopped firing as in Fig. \ref{Fig::4d12}e\&g.

Fig. 4.13a\&b shows the variation of potassium concentration in the extracellular space in the stimulated and unstimulated cases. The potassium decay in much faster when NKCC is present presumably because NKCC allows larger movement of potassium into the glial compartment. We provide the decay timetable in the appendix. 

The quicker potassium is taken into glial compartment by the NKCC, the slower the return of potassium concentration back to resting state. Fig. 4.13c, shows the variation of the potassium concentration in the stimulated region. After action potentials cease, the potassium movement back to the axon compartment is reduced in the model with NKCC channel. Fig. 4.13c, shows that the average potassium concentration in the baseline model increases faster after the axon stops firing than in the model with NKCC.

\begin{figure}[!hbt]
	\centering
	\hspace*{-0.6cm} \includegraphics[bb=55 225 880 460, clip=true,width=3.25in,height=4cm]{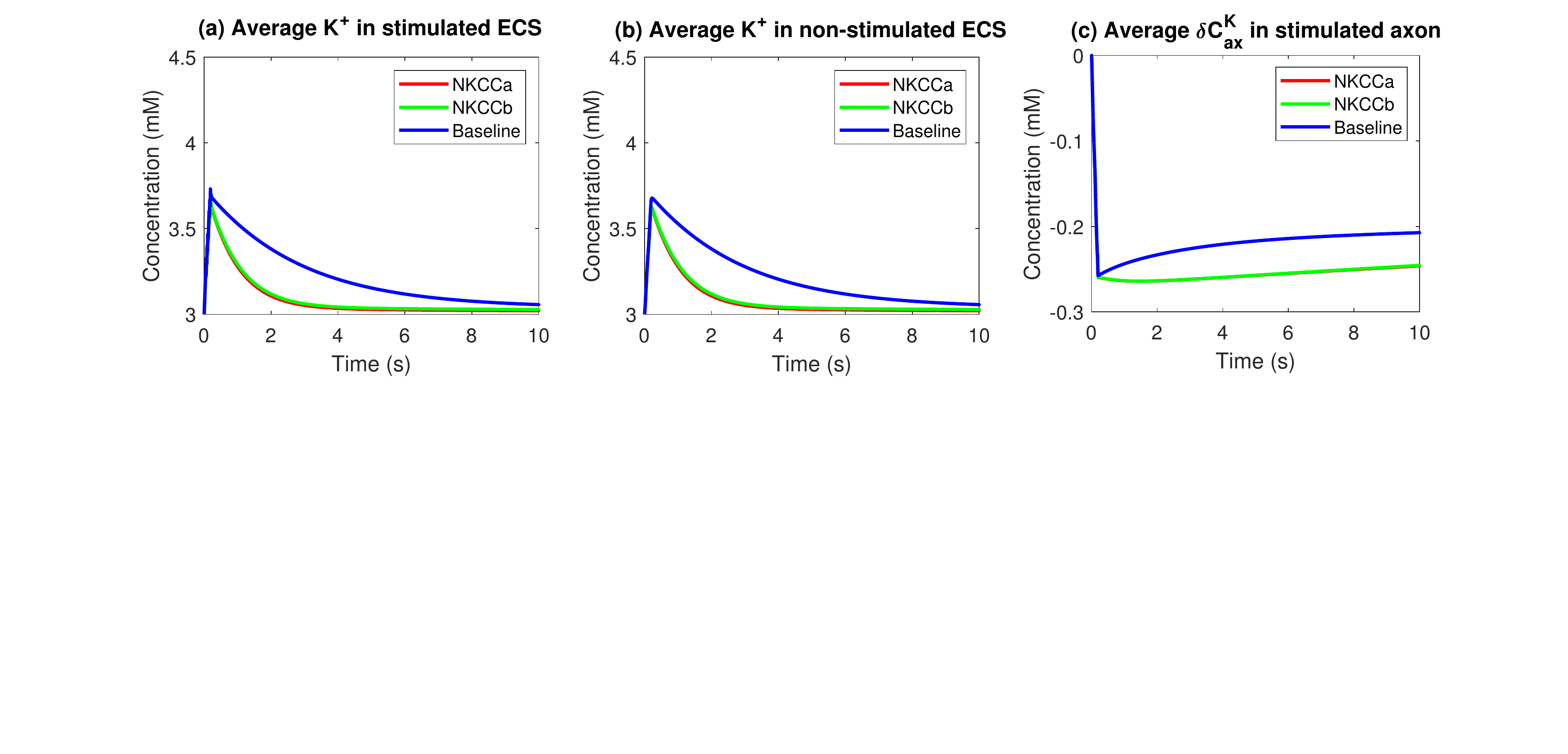}
	\caption{a-b: extracellular potassium concentration variation comparison between the model with NKCC and baseline model (without NKCC). c: average potassium variation in the axon stimulated region.}
	\label{Fig::4d13}
\end{figure}

\subsubsection{Non-selective pathway through the pia matter}
In this section, we consider the effect of a non-selective pathway across the pia boundary. The nonselective pathway allows both ion and water fluid transport through the cleft between the cells in the pia mater. We assume that the water fluid goes through the non-selective pathway only depends on the hydrostatic pressure difference. Therefore, the fluid condition on pia boundary $(\Gamma_7)$ in  (\ref{P_ex_boundary_cd}) become 
\begin{equation*}
\boldsymbol{u}_{ex}^{OP} \cdot \hat{\boldsymbol{r}}=\boldsymbol{u}_{ex}^{SAS} \cdot \hat{\boldsymbol{r}}=u_{pia}^{m}+u_{pia}^{n s}, \quad \text { on } \Gamma_{7},
\end{equation*}
where
\begin{equation*}
\begin{aligned}
u_{pia}^{m}&=L_{pia}^{m}\left(p_{ex}^{OP}-p_{ex}^{SAS}-\gamma_{pia}k_{B} T\left(O_{ex}^{OP}-O_{ex}^{SAS}\right)\right), \\ u_{pia}^{ns}&=L_{pia}^{ns}\left(p_{ex}^{OP}-p_{ex}^{SAS}\right).
\end{aligned}
\end{equation*}

The non-selective pathway between the cell clefts provides an additional pathway for diffusion ($\nabla C_l^i$), electric drift ($C_l^i\nabla\phi_l$) as well as convection for ions ($C_l^i\boldsymbol{u}_l$). We modify the boundary condition (\ref{C_ex_bc}) for ion on pia boundary $(\Gamma_7)$ as
\begin{equation*}
\begin{aligned}
&\boldsymbol{j}_{ex}^{i, OP} \cdot \hat{\boldsymbol{r}}=\boldsymbol{j}_{ex}^{i,SAS} \cdot \hat{\boldsymbol{r}}\\
&=\frac{G_{pia}^{i}+G^{ns}}{z^{i} e}\left(\phi_{ex}^{OP}-\phi_{ex}^{SAS}-E_{pia}^{i}\right)+c_{ex}^{i} u_{pia}^{ns},\ \text{ on } \Gamma_{7},
\end{aligned}
\end{equation*}
Here $G^{ns}$ is the additional conductance due to the non-selective pathway and $C_{ex}^{i}u_{pia}^{ns}$ is the convection flux through the non-selective pathway on pia boundary. 

In the simulation below, we compare the model with non-selective pathway with the model without the non-selective pathway. We choose the comparison parameter to be
\begin{equation*}
\frac{G^{ns}}{G_{pia}^{K}}=\frac{L_{pia}^{ns}}{L_{pia}^{m}}=10.
\end{equation*}

\begin{figure}[hpt]
	\centering
	\includegraphics[bb=25 5 530 330, clip=true,width=3.25in,height=6.2 cm]{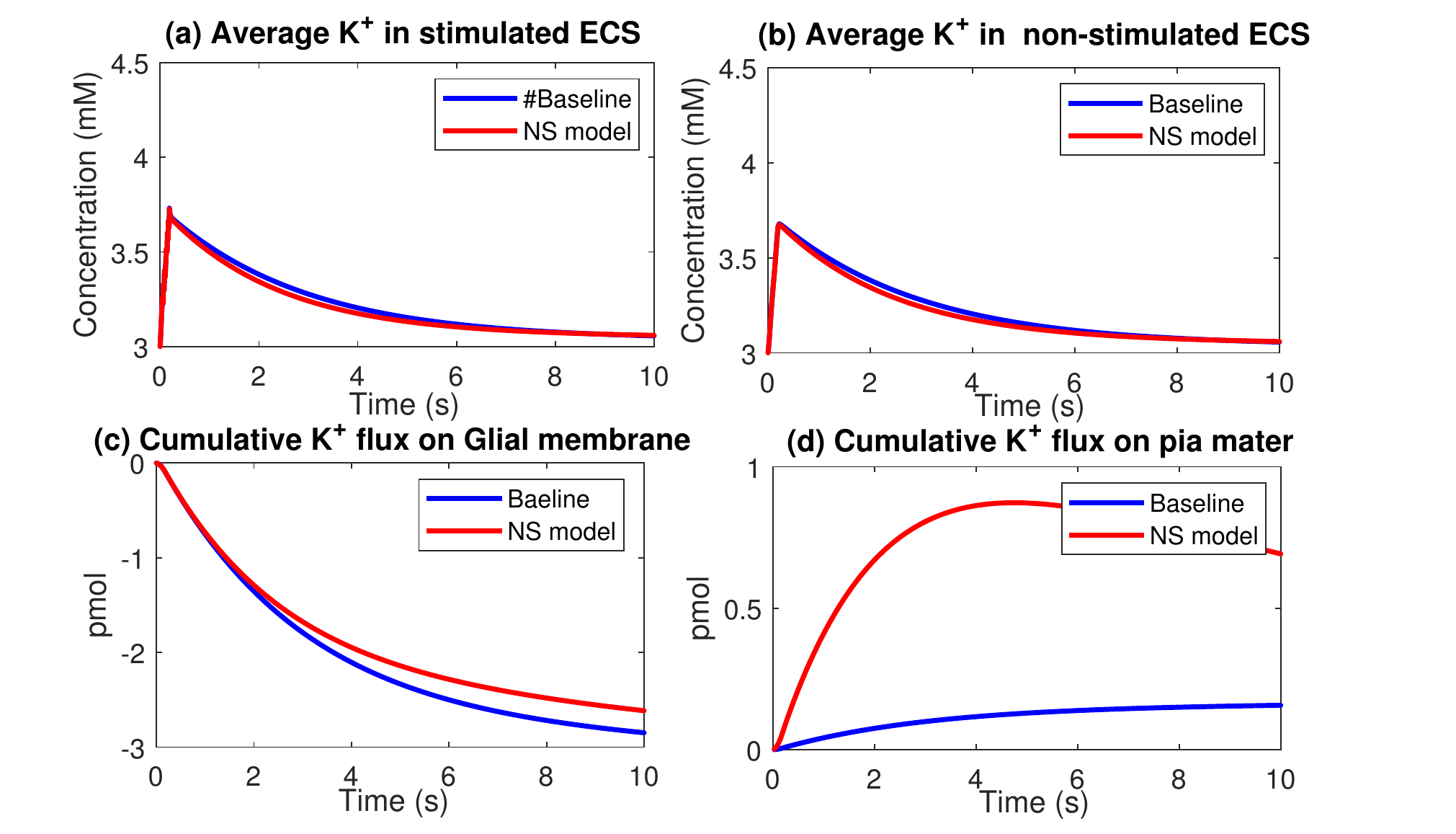}
	\caption{a-b: extracellular potassium concentration variation comparison between the model with non-selective pathway on pia boundary and baseline model (without non-selective pathway).c: total cumulative potassium flux through glial membrane. d: total cumulative potassium flux though pia boundary. }
	\label{Fig::4d14}
\end{figure}

In the Fig. \ref{Fig::4d14}, we show the potassium variation in extracellular space and the cumulative potassium flux through the pia mater and glial membrane. 

The amount of potassium leak out of the optic nerve through the pia boundary is dramatically increased when the pathway is present in Fig. \ref{Fig::4d14}d. However, the dominant pathway of potassium clearance is still through the glial membrane as previously in Fig. \ref{Fig::4d14}c. This is because the total glial membrane area is much larger than the surface area on pia boundary, 

\begin{equation*}
\frac{2 \pi R_a L}{V_{op} \mathcal{M}_{gl}}=O\left(10^{-3}\right),
\end{equation*}
where the $V_on$ is the total optic nerve volume and $M_gl$ is the glial membrane per unit volume. Fig. \ref{Fig::4d14}a\&b shows that with the non-selective pathway on the pia boundary does not accelerate the potassium clearance rate very much.   

\section{Historical Comments and Discussion}  \label{sec:discussion}
There may be a concern about the large number of parameters in this and similar models. Models that contain large numbers of parameters can be difficult to compare and thus difficult to use in the scientific process of conjecture and refutation needed to understand these complex systems.

The need to deal with the complexity of the system as it is presented to us by evolution and used by animals should be clearly understood. Perhaps it is best understood by comparison with engineering systems. Engineering systems use enormous complexity---consider the $10^{13}$ components in our computers and their connections---to perform specific functions. If the complexity is not described in a model, the model cannot describe the functions.
Biological systems perform definite functions and use structures to do so. They are often devices in the exact engineering sense of the word. Complexity is needed to define these systems because without that complexity, the systems do not function.

Of course, the complexity cannot be uniquely defined in either engineering or biological systems. One can always appeal to an atomic description in the desperation of ignorance, but it is not at all clear that much detail helps, and the problems of dealing with irrelevant thermal motion, and difficulties in computation, have made such an approach so difficult that it is rarely used in engineering. Here we do the best we can by choosing structures and parameters that are needed considering their role in other similar biological systems.

The work here is done in the spirit of the structural analysis of physiological problems started by Falk and Fatt \cite{falk1964linear}, continued by their student Eisenberg and his colleagues \cite{schneider1970linear,valdiosera1974impedance}  and applied to the lens of the eye with Rae and Mathias. This approach uses measured anatomical parameters (best by \textcolor{black}{application of} statistical sampling methods \textcolor{black}{to biological systems, which was} pioneered by Eisenberg   and then extended generally by Brenda Eisenberg \cite{eisenberg1968selective}), and impedance spectroscopy  \cite{eisenberg2015electrical}, to determine parameters exploiting the invariance of the capacitance (per unit area) of biological membranes. Mathias and his group showed how to extend these methods to include water flow \cite{mathias1979electrical,mathias1981lens,rae1982physiological} in a bidomain tissue, the lens of the eye. We extend that structural approach here to a tridomain model of an optic nerve.

In the electrical case, the parameters of a structural model are quite well specified by this approach. When dealing with water flow, it is important to include measurements of flow and pressure. In our situation the electrical information is available from the enormous knowledge of the properties of nerve fibers and action potential conduction developed since Hodgkin showed \cite{hodgkin1937evidence} that nerve conduction is electrical and not chemical \cite{hill1932chemical}. Measurements of pressure and flow are not available and they are surely needed if the model is to be further refined or extended to other analogous systems of the central nervous system, of great clinical importance. Measurements of a crucial property modified by water flow are possible thanks to the work of the Harvard group (Orkand et al). That work allows us to define our system as well as we have, but surely not well enough.

As this work is developed to become a model of the important phenomena of the recently discovered glymphatic system \cite{jessen2015glymphatic, mestre2020brain}, measurements of flow and pressure, as well as stereological measurements of structure and biophysical measurements of channel and pump distribution will assume crucial importance, in our view. Extensions of our theory and appropriate simplifications will also be helpful.

\section{Conclusion} \label{sec:Conclusion}
In this work, we propose a tridomain model to study potassium clearance in the optic nerve of Necturus in a series of experiments from Richard Orkand and the Harvard group  \cite{orkand1966effect,kuffler1966physiological}. Our model, analysis, and simulations provide a detailed picture of the role of glial cells in buffering the concentration of ions, mostly in the narrow extracellular space. While the nerve axons are being stimulated, both the extracellular space and glial cells play important roles. They both clear extra potassium from the narrow extracellular space while the axon is firing action potentials. After the action potentials stop, the potassium remaining in the extracellular space is cleared by the glial compartment.

Our model shows that the longitudinal electrical syncytium of the glial cells is critical for clearing potassium (from the extracellular space) when the neuron fires. The inward glial transmembrane potassium flux in the stimulated region is almost the same as the outward potassium flux out to the extracellular space in the non-stimulated region, in response to the change in potassium concentration between the extracellular space in the stimulated and unstimulated regions. This is because the electric potential spreads through the connected cells in the glial compartment. The glial electric potential in the unstimulated region becomes more positive in response to the depolarization of the glial electric potential in the stimulus region. The ‘syncytial properties’ of the glial compartment are a feature that separates our field model with its partial differential equations from the compartment models in the literature that use ordinary differential equations. Compartment models lack the combination of space and time dependence needed to describe the temporal and spatial spread of potential. The combined temporal and spatial spread of potential is a crucial property of neurons. Here we show that the combined temporal and spatial spread of potential is a crucial property of the glia, and the narrow extracellular space between glia and neurons. In a sense, we extend the electrical cable theory used to describe the spatial spread of the electrical potential of neurons to a theory of spatial distribution of flow in  neuron, glia, and the narrow extracellular space between them.

We discuss the effect of enhanced potassium conductance in the glial membrane and nerve membranes of the pia mater. On the one hand, incorporating NKCC channels into the glial membrane increases potassium clearance. Potassium clearance time is much shorter than that predicted by the baseline model (without NKCC channels in the glial membrane). On the other hand, an additional non-selective pathway in another location—in the pia mater—does not have significant effect on potassium clearance. This is not surprising since the total membrane area of the glial membrane in the optic nerve is much greater than the effective surface membrane of nerves in the pia mater.

Finally,  our analysis of the model for the optic nerve is just a first small step towards the understanding of the mechanisms of   the glial compartment buffering effect during the potassium clearance
and microcirculation patterns of water and ions. The axons   considered here is without myelin sheath. 
Myelin can be included   by combining the model of myelinated nerve proposed in  \cite{song2018electroneutral} as a first next step. 
Our distributed model and its partial differential equations can be generalized to describe ionic and water transport in tissues with more complicated and heterogeneous structures and with glymphatic pathways connected to the circulatory system. We expect that the spatially nonuniform distribution of ion and water channels and transporters will be used in many structures in the central nervous system to control flow.  Obviously, at higher resolution much more \textcolor{black}{detailed} experimental observations and structure measurements ( like pump distributions and membrane permeabilities) will be needed.  

\noindent\textbf{Author Contributions}
Y.Z., S.X., and H.H. did the model derivations and carried out the numerical simulations.
R.S.E. and H.H. designed the study, coordinated the study, and commented on the
manuscript. All authors gave final approval for publication.

\noindent\textbf{Acknowledgments}
This research is supported in part by National Natural Science Foundation of China
12071190 and the Fundamental Research Funds for the Central Universities (S.X), the Fields Institute for Research in mathematical Science (S.X., R.S.E.,
and H.H.) and the Natural Sciences and Engineering Research Council of Canada
(H.H.).





\bibliographystyle{plain}
\bibliography{Mybib}

\begin{thebibliography}{10}

\bibitem{abbott2018role}
N~Joan Abbott, Michelle~E Pizzo, Jane~E Preston, Damir Janigro, and Robert~G
  Thorne.
\newblock The role of brain barriers in fluid movement in the cns: is there a
  ‘glymphatic’system?
\newblock {\em Acta neuropathologica}, 135(3):387--407, 2018.

\bibitem{ayata2015spreading}
Cenk Ayata and Martin Lauritzen.
\newblock Spreading depression, spreading depolarizations, and the cerebral
  vasculature.
\newblock {\em Physiological reviews}, 95(3):953--993, 2015.

\bibitem{band2009intracellular}
Leah~R Band, Cameron~L Hall, Giles Richardson, Oliver~E Jensen, Jennifer~H
  Siggers, and Alexander~JE Foss.
\newblock Intracellular flow in optic nerve axons: a mechanism for cell death
  in glaucoma.
\newblock {\em Investigative ophthalmology \& visual science},
  50(8):3750--3758, 2009.

\bibitem{bellinger2008submyelin}
SC~Bellinger, G~Miyazawa, and PN~Steinmetz.
\newblock Submyelin potassium accumulation may functionally block subsets of
  local axons during deep brain stimulation: a modeling study.
\newblock {\em Journal of neural engineering}, 5(3):263, 2008.

\bibitem{bellot2017astrocytic}
Alba Bellot-Saez, Orsolya Kekesi, John~W Morley, and Yossi Buskila.
\newblock Astrocytic modulation of neuronal excitability through k+ spatial
  buffering.
\newblock {\em Neuroscience \& Biobehavioral Reviews}, 77:87--97, 2017.

\bibitem{bracho1975further}
H~Bracho, PM~Orkand, and RK~Orkand.
\newblock A further study of the fine structure and membrane properties of
  neuroglia in the optic nerve of necturus.
\newblock {\em Journal of neurobiology}, 6(4):395--410, 1975.

\bibitem{chang2013mathematical}
Joshua~C Chang, Kevin~C Brennan, Dongdong He, Huaxiong Huang, Robert~M Miura,
  Phillip~L Wilson, and Jonathan~J Wylie.
\newblock A mathematical model of the metabolic and perfusion effects on
  cortical spreading depression.
\newblock {\em PLoS One}, 8(8):e70469, 2013.

\bibitem{chen2000spatial}
Kevin~C Chen and Charles Nicholson.
\newblock Spatial buffering of potassium ions in brain extracellular space.
\newblock {\em Biophysical journal}, 78(6):2776--2797, 2000.

\bibitem{eisenberg2010energy}
Bob Eisenberg, Yunkyong Hyon, and Chun Liu.
\newblock Energy variational analysis of ions in water and channels: Field
  theory for primitive models of complex ionic fluids.
\newblock {\em The Journal of Chemical Physics}, 133(10):104104, 2010.

\bibitem{eisenberg1968selective}
Brenda Eisenberg and Robert~S Eisenberg.
\newblock Selective disruption of the sarcotubular system in frog sartorius
  muscle: a quantitative study with exogenous peroxidase as a marker.
\newblock {\em The Journal of cell biology}, 39(2):451--467, 1968.

\bibitem{eisenberg2015electrical}
Robert Eisenberg.
\newblock Electrical structure of biological cells and tissues: impedance
  spectroscopy, stereology, and singular perturbation theory.
\newblock {\em arXiv preprint arXiv:1511.01339}, 2015.

\bibitem{falk1964linear}
G~Falk and Paul Fatt.
\newblock Linear electrical properties of striated muscle fibres observed with
  intracellular electrodes.
\newblock {\em Proceedings of the Royal Society of London. Series B. Biological
  Sciences}, 160(978):69--123, 1964.

\bibitem{feher2017quantitative}
Joseph~J Feher.
\newblock {\em Quantitative human physiology: an introduction}.
\newblock Academic press, 2017.

\bibitem{feola2016finite}
Andrew~J Feola, Jerry~G Myers, Julia Raykin, Lealem Mulugeta, Emily~S Nelson,
  Brian~C Samuels, and C~Ross Ethier.
\newblock Finite element modeling of factors influencing optic nerve head
  deformation due to intracranial pressure.
\newblock {\em Investigative ophthalmology \& visual science},
  57(4):1901--1911, 2016.

\bibitem{filippidis2012permeability}
Aristotelis~S Filippidis, Sotirios~G Zarogiannis, Maria Ioannou, Konstantinos
  Gourgoulianis, Paschalis-Adam Molyvdas, and Chrissi Hatzoglou.
\newblock Permeability of the arachnoid and pia mater. the role of ion channels
  in the leptomeningeal physiology.
\newblock {\em Child's Nervous System}, 28(4):533--540, 2012.

\bibitem{fitzhugh1960thresholds}
Richard Fitzhugh.
\newblock Thresholds and plateaus in the hodgkin-huxley nerve equations.
\newblock {\em The Journal of general physiology}, 43(5):867--896, 1960.

\bibitem{frankenhaeuser1956after}
B~Frankenhaeuser and AL~Hodgkin.
\newblock The after-effects of impulses in the giant nerve fibres of loligo.
\newblock {\em The Journal of physiology}, 131(2):341--376, 1956.

\bibitem{gakuba2018general}
Clement Gakuba, Thomas Gaberel, Suzanne Goursaud, Jennifer Bourges, Camille
  Di~Palma, Aur{\'e}lien Quenault, Sara~Martinez de~Lizarrondo, Denis Vivien,
  and Maxime Gauberti.
\newblock General anesthesia inhibits the activity of the “glymphatic
  system”.
\newblock {\em Theranostics}, 8(3):710, 2018.

\bibitem{gao2000isoform}
Junyuan Gao, X~Sun, V~Yatsula, RS~Wymore, and RT~Mathias.
\newblock Isoform-specific function and distribution of na/k pumps in the frog
  lens epithelium.
\newblock {\em The Journal of membrane biology}, 178(2):89--101, 2000.

\bibitem{gardiner2010computational}
Bruce~S Gardiner, David~W Smith, Michael Coote, and Jonathan~G Crowston.
\newblock Computational modeling of fluid flow and intra-ocular pressure
  following glaucoma surgery.
\newblock {\em PLoS One}, 5(10):e13178, 2010.

\bibitem{hayreh1984sheath}
Sohan~Singh Hayreh.
\newblock The sheath of the optic nerve.
\newblock {\em Ophthalmologica}, 189(1-2):54--63, 1984.

\bibitem{hayreh2009ischemic}
Sohan~Singh Hayreh.
\newblock Ischemic optic neuropathy.
\newblock {\em Progress in retinal and eye research}, 28(1):34--62, 2009.

\bibitem{hill1932chemical}
Archibald~Vivian Hill.
\newblock {\em Chemical wave transmission in nerve}.
\newblock CUP Archive, 1932.

\bibitem{hodgkin1937evidence}
AL~Hodgkin.
\newblock Evidence for electrical transmission in nerve: Part i.
\newblock {\em The Journal of physiology}, 90(2):183--210, 1937.

\bibitem{hodgkin1949ionic}
AL~im HODGKIN.
\newblock Ionic currents underlying activity in the giant axon of the squid.
\newblock {\em Arch. Sci. Physiol.}, 3:129--150, 1949.

\bibitem{hodgkin1952quantitative}
Alan~L Hodgkin and Andrew~F Huxley.
\newblock A quantitative description of membrane current and its application to
  conduction and excitation in nerve.
\newblock {\em The Journal of physiology}, 117(4):500--544, 1952.

\bibitem{hodgkin1952measurement}
Alan~L Hodgkin, Andrew~F Huxley, and Bernard Katz.
\newblock Measurement of current-voltage relations in the membrane of the giant
  axon of loligo.
\newblock {\em The Journal of physiology}, 116(4):424, 1952.

\bibitem{hodgkin1952currents}
Allan~L Hodgkin and Andrew~F Huxley.
\newblock Currents carried by sodium and potassium ions through the membrane of
  the giant axon of loligo.
\newblock {\em The Journal of physiology}, 116(4):449--472, 1952.

\bibitem{hou2016intracranial}
Ruowu Hou, Zheng Zhang, Diya Yang, Huaizhou Wang, Weiwei Chen, Zhen Li,
  Jinghong Sang, Sumeng Liu, Yiwen Cao, Xiaobin Xie, et~al.
\newblock Intracranial pressure (icp) and optic nerve subarachnoid space
  pressure (onsp) correlation in the optic nerve chamber: the beijing
  intracranial and intraocular pressure (icop) study.
\newblock {\em brain research}, 1635:201--208, 2016.

\bibitem{hua2018cerebrospinal}
Yi~Hua, Andrew~P Voorhees, and Ian~A Sigal.
\newblock Cerebrospinal fluid pressure: revisiting factors influencing optic
  nerve head biomechanics.
\newblock {\em Investigative ophthalmology \& visual science}, 59(1):154--165,
  2018.

\bibitem{jessen2015glymphatic}
Nadia~Aalling Jessen, Anne Sofie~Finmann Munk, Iben Lundgaard, and Maiken
  Nedergaard.
\newblock The glymphatic system: a beginner’s guide.
\newblock {\em Neurochemical research}, 40(12):2583--2599, 2015.

\bibitem{jiang2017impairment}
Quan Jiang, Li~Zhang, Guangliang Ding, Esmaeil Davoodi-Bojd, Qingjiang Li, Lian
  Li, Neema Sadry, Maiken Nedergaard, Michael Chopp, and Zhenggang Zhang.
\newblock Impairment of the glymphatic system after diabetes.
\newblock {\em Journal of Cerebral Blood Flow \& Metabolism}, 37(4):1326--1337,
  2017.

\bibitem{jonas2003anatomic}
Jost~B Jonas, Eduard Berenshtein, and Leonard Holbach.
\newblock Anatomic relationship between lamina cribrosa, intraocular space, and
  cerebrospinal fluid space.
\newblock {\em Investigative ophthalmology \& visual science},
  44(12):5189--5195, 2003.

\bibitem{killer2006cerebrospinal}
HE~Killer, GP~Jaggi, J~Flammer, Neil~R Miller, AR~Huber, and A~Mironov.
\newblock Cerebrospinal fluid dynamics between the intracranial and the
  subarachnoid space of the optic nerve. is it always bidirectional?
\newblock {\em Brain}, 130(2):514--520, 2006.

\bibitem{kofuji2004potassium}
Paulo Kofuji and Eric~A Newman.
\newblock Potassium buffering in the central nervous system.
\newblock {\em Neuroscience}, 129(4):1043--1054, 2004.

\bibitem{kuffler1966physiology}
Stephen~W Kuffler and John~G Nicholls.
\newblock The physiology of neuroglial cells.
\newblock In {\em Ergebnisse der physiologie biologischen chemie und
  experimentellen pharmakologie}, pages 1--90. Springer, 1966.

\bibitem{kuffler1966physiological}
SW~Kuffler, JG~Nicholls, and RK~Orkand.
\newblock Physiological properties of glial cells in the central nervous system
  of amphibia.
\newblock {\em Journal of Neurophysiology}, 29(4):768--787, 1966.

\bibitem{lauf2000k}
Peter~K Lauf and Norma~C Adragna.
\newblock K-cl cotransport: properties and molecular mechanism.
\newblock {\em Cellular Physiology and Biochemistry}, 10(5-6):341--354, 2000.

\bibitem{lu2006viscoelastic}
Yun-Bi Lu, Kristian Franze, Gerald Seifert, Christian Steinh{\"a}user, Frank
  Kirchhoff, Hartwig Wolburg, Jochen Guck, Paul Janmey, Er-Qing Wei, Josef
  K{\"a}s, et~al.
\newblock Viscoelastic properties of individual glial cells and neurons in the
  cns.
\newblock {\em Proceedings of the National Academy of Sciences},
  103(47):17759--17764, 2006.

\bibitem{malcolm2006computational}
Duane Tearaitoa~Kingwell Malcolm.
\newblock {\em A computational model of the ocular lens}.
\newblock PhD thesis, ResearchSpace@ Auckland, 2006.

\bibitem{mathias1985steady}
RICHARD~T Mathias.
\newblock Steady-state voltages, ion fluxes, and volume regulation in syncytial
  tissues.
\newblock {\em Biophysical journal}, 48(3):435, 1985.

\bibitem{mathias1979electrical}
RT~Mathias, JL~Rae, and RS~Eisenberg.
\newblock Electrical properties of structural components of the crystalline
  lens.
\newblock {\em Biophysical Journal}, 25(1):181--201, 1979.

\bibitem{mathias1981lens}
RT~Mathias, JL~Rae, and RS~Eisenberg.
\newblock The lens as a nonuniform spherical syncytium.
\newblock {\em Biophysical journal}, 34(1):61--83, 1981.

\bibitem{mclaughlin1985electro}
STUART McLAUGHLIN and RICHARD~T Mathias.
\newblock Electro-osmosis and the reabsorption of fluid in renal proximal
  tubules.
\newblock {\em The Journal of general physiology}, 85(5):699--728, 1985.

\bibitem{mestre2020brain}
Humberto Mestre, Yuki Mori, and Maiken Nedergaard.
\newblock The brain’s glymphatic system: current controversies.
\newblock {\em Trends in Neurosciences}, 2020.

\bibitem{miura2007cortical}
Robert~M Miura, Huaxiong Huang, and Jonathan~J Wylie.
\newblock Cortical spreading depression: An enigma.
\newblock {\em The European Physical Journal Special Topics}, 147(1):287--302,
  2007.

\bibitem{morgan2016cerebrospinal}
William~H Morgan, Chandrakumar Balaratnasingam, Christopher~RP Lind, Steve
  Colley, Min~H Kang, Philip~H House, and Dao-Yi Yu.
\newblock Cerebrospinal fluid pressure and the eye.
\newblock {\em British Journal of Ophthalmology}, 100(1):71--77, 2016.

\bibitem{mori2015multidomain}
Yoichiro Mori.
\newblock A multidomain model for ionic electrodiffusion and osmosis with an
  application to cortical spreading depression.
\newblock {\em Physica D: Nonlinear Phenomena}, 308:94--108, 2015.

\bibitem{nedergaard2020glymphatic}
Maiken Nedergaard and Steven~A Goldman.
\newblock Glymphatic failure as a final common pathway to dementia.
\newblock {\em Science}, 370(6512):50--56, 2020.

\bibitem{nicholson2001diffusion}
Charles Nicholson.
\newblock Diffusion and related transport mechanisms in brain tissue.
\newblock {\em Reports on progress in Physics}, 64(7):815, 2001.

\bibitem{nicholson2017brain}
Charles Nicholson and Sabina Hrab{\v{e}}tov{\'a}.
\newblock Brain extracellular space: the final frontier of neuroscience.
\newblock {\em Biophysical journal}, 113(10):2133--2142, 2017.

\bibitem{o2016effects}
Rosemary O'Connell and Yoichiro Mori.
\newblock Effects of glia in a triphasic continuum model of cortical spreading
  depression.
\newblock {\em Bulletin of mathematical biology}, 78(10):1943--1967, 2016.

\bibitem{orkand1966effect}
RK~Orkand, JG~Nicholls, and SW~Kuffler.
\newblock Effect of nerve impulses on the membrane potential of glial cells in
  the central nervous system of amphibia.
\newblock {\em Journal of neurophysiology}, 29(4):788--806, 1966.

\bibitem{ostby2009astrocytic}
Ivar {\O}stby, Leiv {\O}yehaug, Gaute~T Einevoll, Erlend~A Nagelhus, Erik
  Plahte, Thomas Zeuthen, Catherine~M Lloyd, Ole~P Ottersen, and Stig~W Omholt.
\newblock Astrocytic mechanisms explaining neural-activity-induced shrinkage of
  extraneuronal space.
\newblock {\em PLoS computational biology}, 5(1):e1000272, 2009.

\bibitem{pache2006morphological}
Mona Pache and Peter Meyer.
\newblock Morphological changes of the retrobulbar optic nerve and its
  meningeal sheaths in glaucoma.
\newblock {\em Ophthalmologica}, 220(6):393--396, 2006.

\bibitem{perez1995extracellular}
MA~P{\'e}rez-Pinz{\'o}n, LIAN Tao, and CHARLES Nicholson.
\newblock Extracellular potassium, volume fraction, and tortuosity in rat
  hippocampal ca1, ca3, and cortical slices during ischemia.
\newblock {\em Journal of Neurophysiology}, 74(2):565--573, 1995.

\bibitem{pilgrim1982volume}
CH~Pilgrim, I~Reisert, and D~Grab.
\newblock Volume densities and specific surfaces of neuronal and glial tissue
  elements in the rat supraoptic nucleus.
\newblock {\em Journal of Comparative Neurology}, 211(4):427--431, 1982.

\bibitem{postnov2007functional}
Dmitry~E Postnov, Ludmila~S Ryazanova, and Olga~V Sosnovtseva.
\newblock Functional modeling of neural--glial interaction.
\newblock {\em BioSystems}, 89(1-3):84--91, 2007.

\bibitem{rae1982physiological}
JL~Rae, RT~Mathias, and RS~Eisenberg.
\newblock Physiological role of the membranes and extracellular space within
  the ocular lens.
\newblock {\em Experimental eye research}, 35(5):471--489, 1982.

\bibitem{reichenbach1993retinal}
A~Reichenbach, J-U Stolzenburg, W~Eberhardt, TI~Chao, D~Dettmer, and L~Hertz.
\newblock What do retinal m{\"u}ller (glial) cells do for their neuronal
  ‘small siblings’?
\newblock {\em Journal of chemical neuroanatomy}, 6(4):201--213, 1993.

\bibitem{schneider1970linear}
Martin~F Schneider.
\newblock Linear electrical properties of the transverse tubules and surface
  membrane of skeletal muscle fibers.
\newblock {\em The Journal of general physiology}, 56(5):640--671, 1970.

\bibitem{sibille2015neuroglial}
J{\'e}r{\'e}mie Sibille, Khanh~Dao Duc, David Holcman, and Nathalie Rouach.
\newblock The neuroglial potassium cycle during neurotransmission: role of
  kir4. 1 channels.
\newblock {\em PLoS computational biology}, 11(3):e1004137, 2015.

\bibitem{sigal2005factors}
Ian~A Sigal, John~G Flanagan, and C~Ross Ethier.
\newblock Factors influencing optic nerve head biomechanics.
\newblock {\em Investigative ophthalmology \& visual science},
  46(11):4189--4199, 2005.

\bibitem{song2018electroneutral}
Zilong Song, Xiulei Cao, and Huaxiong Huang.
\newblock Electroneutral models for dynamic poisson-nernst-planck systems.
\newblock {\em Physical Review E}, 97(1):012411, 2018.

\bibitem{tuckwell1978mathematical}
Henry~C Tuckwell and Robert~M Miura.
\newblock A mathematical model for spreading cortical depression.
\newblock {\em Biophysical Journal}, 23(2):257--276, 1978.

\bibitem{tuttle2019computational}
Austin Tuttle, Jorge~Riera Diaz, and Yoichiro Mori.
\newblock A computational study on the role of glutamate and nmda receptors on
  cortical spreading depression using a multidomain electrodiffusion model.
\newblock {\em PLoS computational biology}, 15(12):e1007455, 2019.

\bibitem{ullah2015role}
Ghanim Ullah, Yina Wei, Markus~A Dahlem, Martin Wechselberger, and Steven~J
  Schiff.
\newblock The role of cell volume in the dynamics of seizure, spreading
  depression, and anoxic depolarization.
\newblock {\em PLoS Comput Biol}, 11(8):e1004414, 2015.

\bibitem{vaghefi2012development}
Ehsan Vaghefi, Duane~TK Malcolm, Marc~D Jacobs, and Paul~J Donaldson.
\newblock Development of a 3d finite element model of lens microcirculation.
\newblock {\em Biomedical engineering online}, 11(1):69, 2012.

\bibitem{valdiosera1974impedance}
R~Valdiosera, C~Clausen, and RS~Eisenberg.
\newblock Impedance of frog skeletal muscle fibers in various solutions.
\newblock {\em The Journal of general physiology}, 63(4):460--491, 1974.

\bibitem{villegas1960characterization}
Raimundo Villegas and Gloria~M Villegas.
\newblock Characterization of the membranes in the giant nerve fiber of the
  squid.
\newblock {\em The Journal of general physiology}, 43(5):73, 1960.

\bibitem{wan2014self}
Li~Wan, Shixin Xu, Maijia Liao, Chun Liu, and Ping Sheng.
\newblock Self-consistent approach to global charge neutrality in
  electrokinetics: A surface potential trap model.
\newblock {\em Physical Review X}, 4(1):011042, 2014.

\bibitem{wang2019intraocular}
Ningli Wang.
\newblock {\em Intraocular and Intracranial Pressure Gradient in Glaucoma},
  volume~1.
\newblock Springer, 2019.

\bibitem{wei2014unification}
Yina Wei, Ghanim Ullah, and Steven~J Schiff.
\newblock Unification of neuronal spikes, seizures, and spreading depression.
\newblock {\em Journal of Neuroscience}, 34(35):11733--11743, 2014.

\bibitem{xu2018osmosis}
Shixin Xu, Bob Eisenberg, Zilong Song, and Huaxiong Huang.
\newblock Osmosis through a semi-permeable membrane: a consistent approach to
  interactions.
\newblock {\em arXiv preprint arXiv:1806.00646}, 2018.

\bibitem{yao2011continuum}
Wei Yao, Huaxiong Huang, and Robert~M Miura.
\newblock A continuum neuronal model for the instigation and propagation of
  cortical spreading depression.
\newblock {\em Bulletin of mathematical biology}, 73(11):2773--2790, 2011.

\bibitem{zhu2019bidomain}
Yi~Zhu, Shixin Xu, Robert~S Eisenberg, and Huaxiong Huang.
\newblock A bidomain model for lens microcirculation.
\newblock {\em Biophysical journal}, 116(6):1171--1184, 2019.

\end{thebibliography}



\newpage
\appendix
\textcolor{black}{
	\section{Supporting Information}
	\subsection{Randomly distributed stimulation}
	\label{Randomly stimulated cases}
	Table \ref{Table_1} compares the decay time between the spatially uniform radial (inner and outer) and spatially random stimulated case (Case 1 – Case 4). The stimulated volume ratio is the ratio of the volume of stimulated axons divided by the total axon volume. We calculate how much time the potassium concentration takes to decay (50\%, 70\% and 90\%) from its maximum in each case.}

\begin{table}[h!p]
	\begin{tiny}
		\centering
		\caption{Stimulated Volume Ratio and Potassium Decay Time}
		\label{Table_1} 
		\begin{threeparttable}
			\begin{tabular}{c r r r r}
				\hline & \text { Stimulated Volume  Ratio } & \text { 50\% decay } & \text { 70\% decay } & \text { 90\% decay } \\
				\hline 
				\text { Inner case } & 0.25 & 0.2 $\mathrm{~s}$ & 1.0 $\mathrm{~s}$ & 4.5 $\mathrm{~s}$ \\
				\text { Outer case } & 0.75 & 1.7 $\mathrm{~s}$ & 3.4 $\mathrm{~s}$ & 7.5 $\mathrm{~s}$\\
				\text { Random case 1 } & 0.55 & 2.0 $\mathrm{~s}$ & 3.6 $\mathrm{~s}$ & 7.8 $\mathrm{~s}$\\
				\text { Random case 2 } & 0.30 & 1.5 $\mathrm{~s}$ & 3.1 $\mathrm{~s}$ & 6.9 $\mathrm{~s}$ \\
				\text { Random case 3 }  & 0.73 & 2.0 $\mathrm{~s}$ & 3.7 $\mathrm{~s}$ & 7.9 $\mathrm{~s}$ \\
				\text { Random case 4 }  & 0.70 & 2.0 $\mathrm{~s}$ & 3.7 $\mathrm{~s}$ & 7.6 $\mathrm{~s}$ \\
				\hline
			\end{tabular}
		\end{threeparttable}
	\end{tiny}
\end{table}

\begin{figure}[!hbt]
	\centering \includegraphics[width=3.25in,height=4.5cm ]{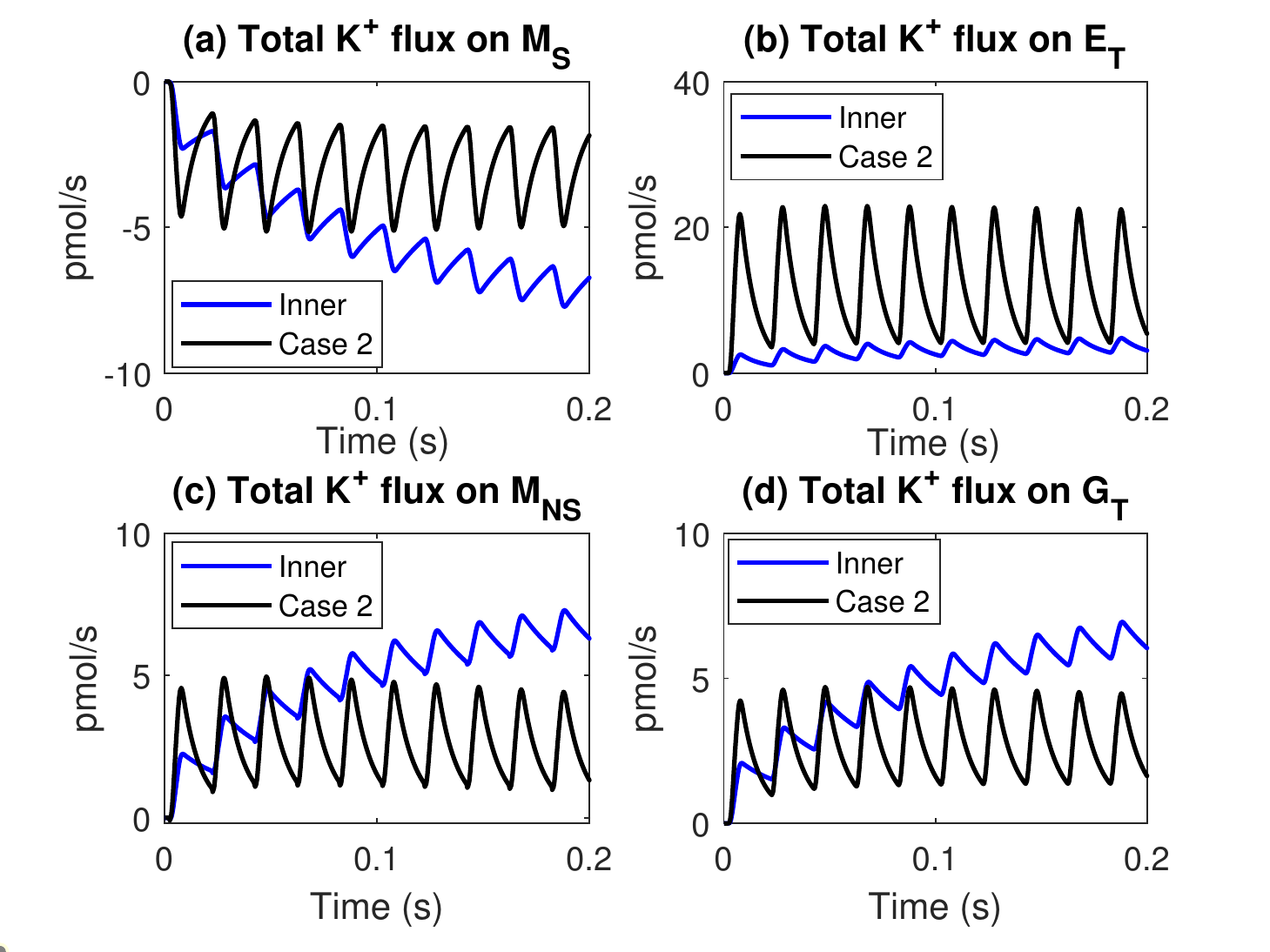}
	\includegraphics[width=3.25in,height=4.5cm]{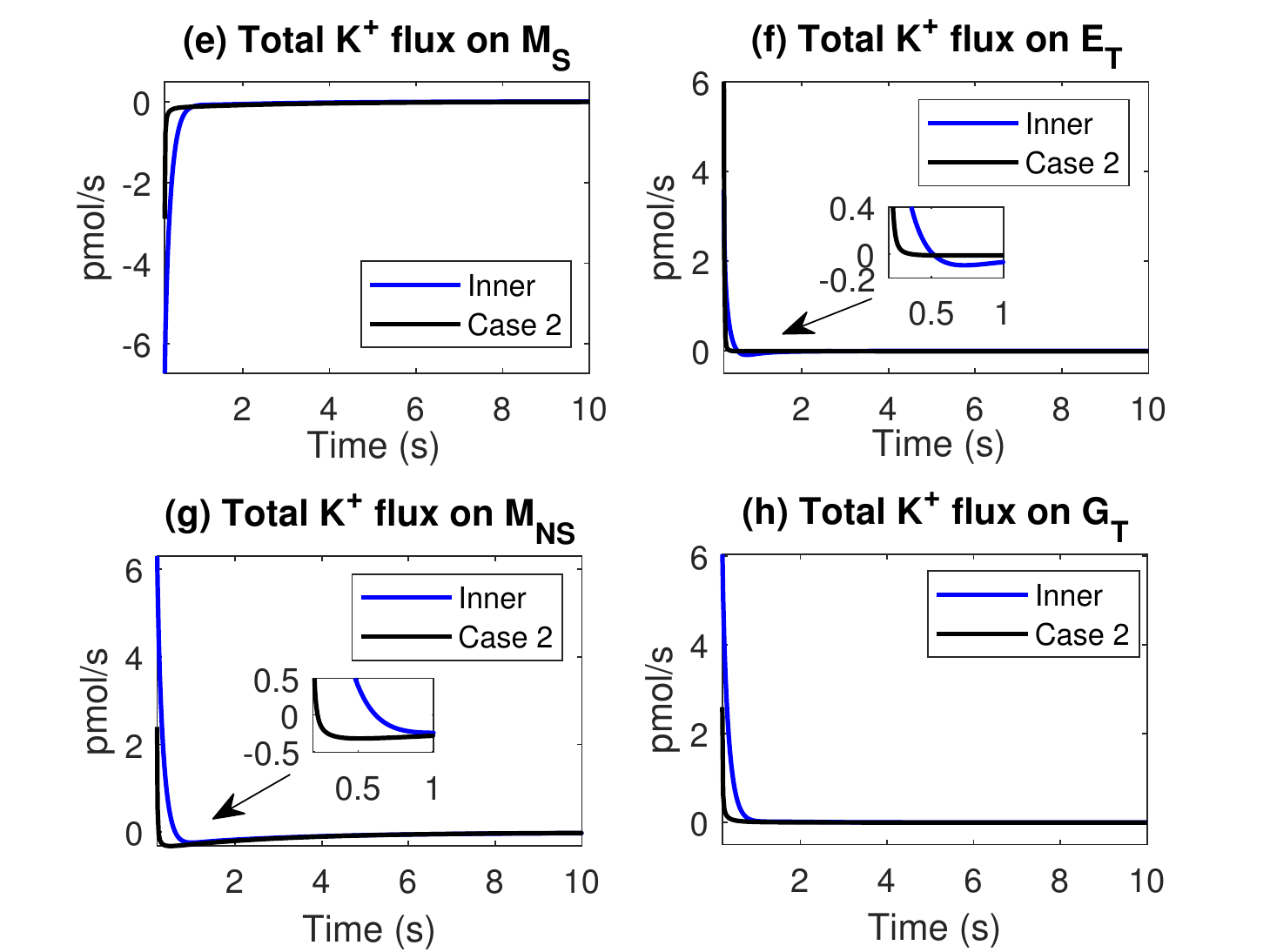}
	\caption{Comparison between spatially randomly stimulated case 2 with the spatially uniform  (inner) case.}
	\label{Fig_B_1}
\end{figure}

\begin{figure}[!hbt]
	\centering
	\includegraphics[width=3.25in,height=4.5cm ]{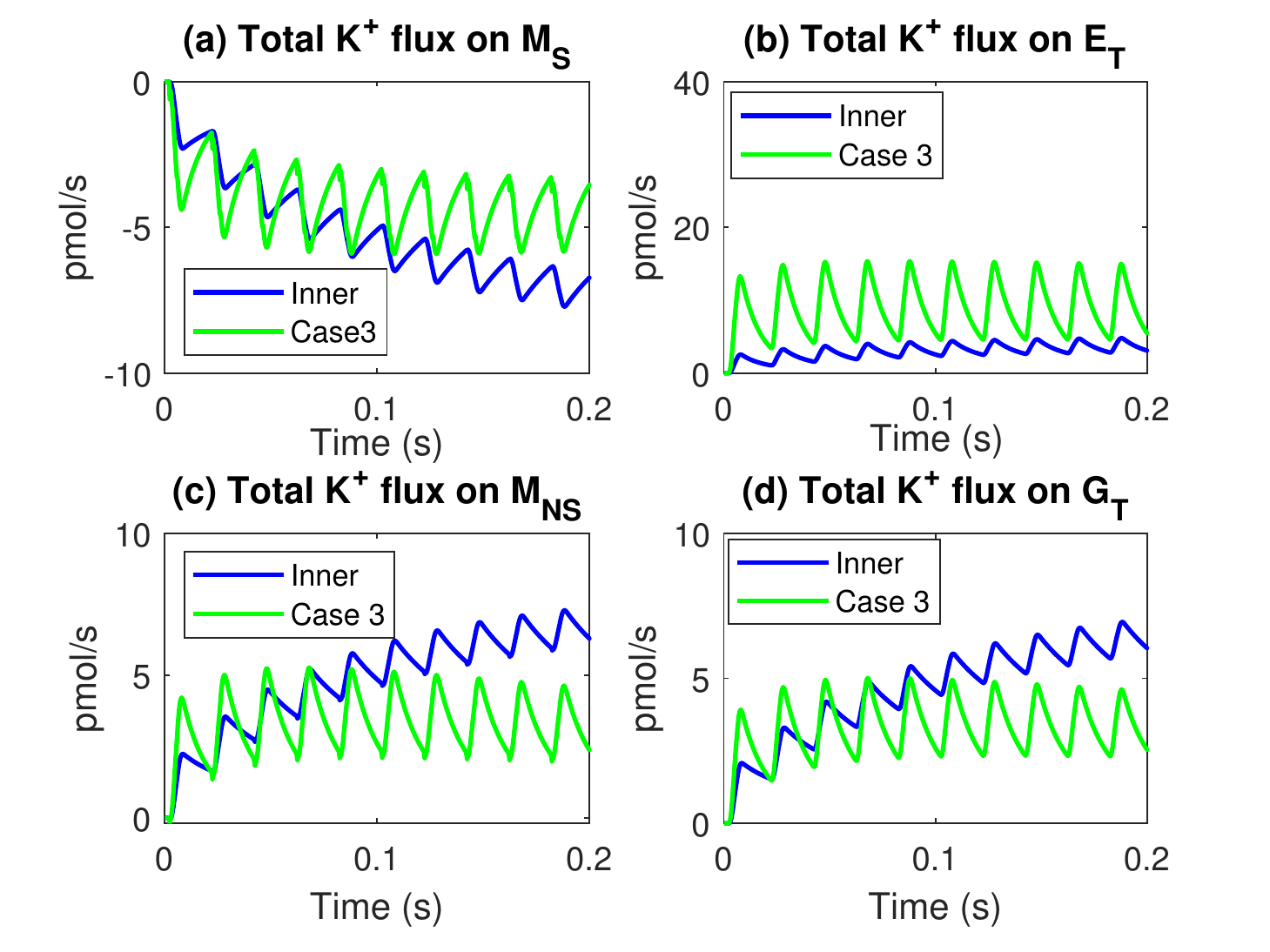} \includegraphics[width=3.25in,height=4.5cm]{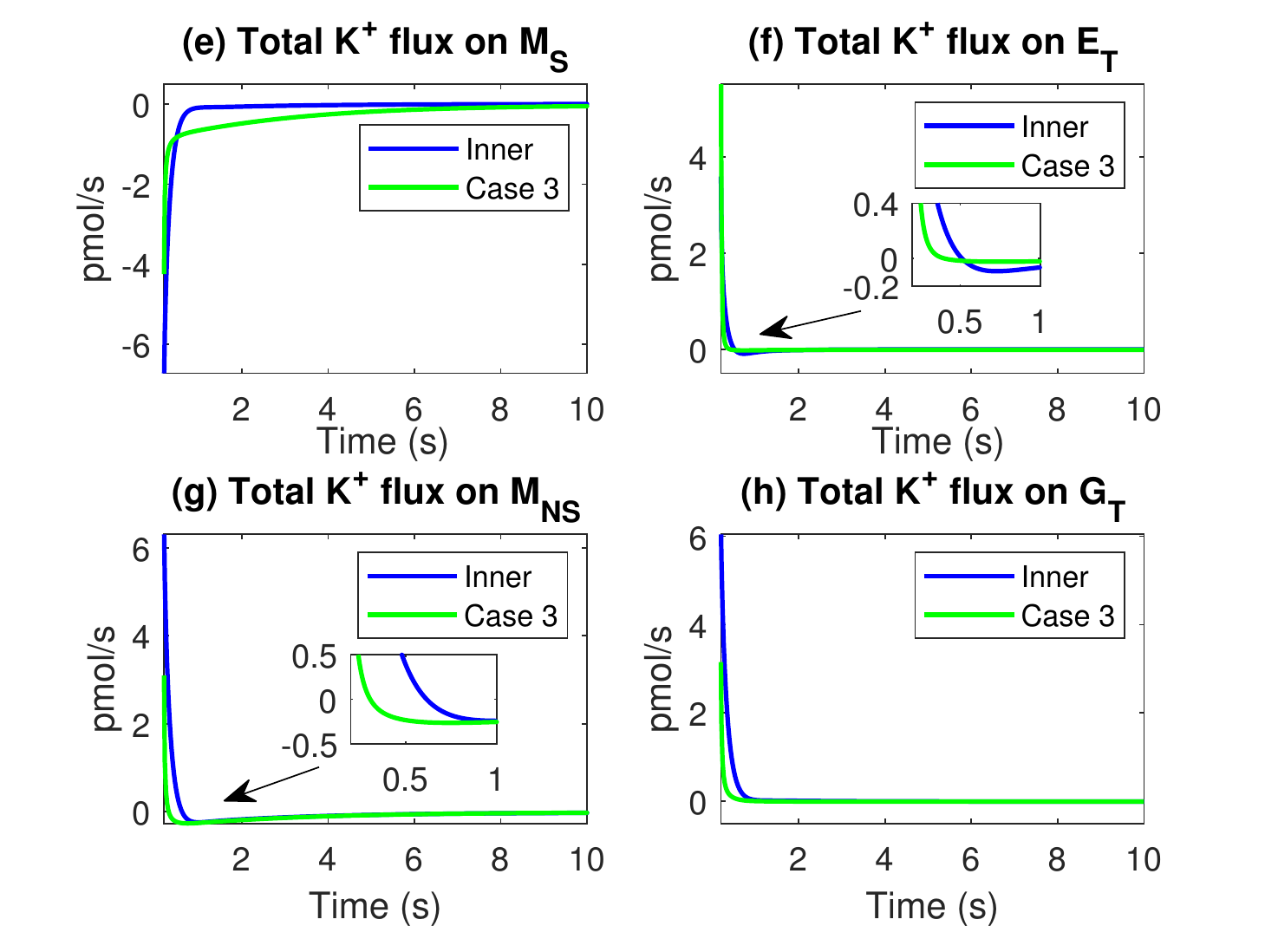}
	\caption{Comparison between spatially randomly stimulated case 3 with the  spatially uniform  (inner) case.}
	\label{Fig_B_2}
\end{figure}

\begin{figure}[!hbt]
	\centering
	\includegraphics[width=3.25in,height=4.5cm ]{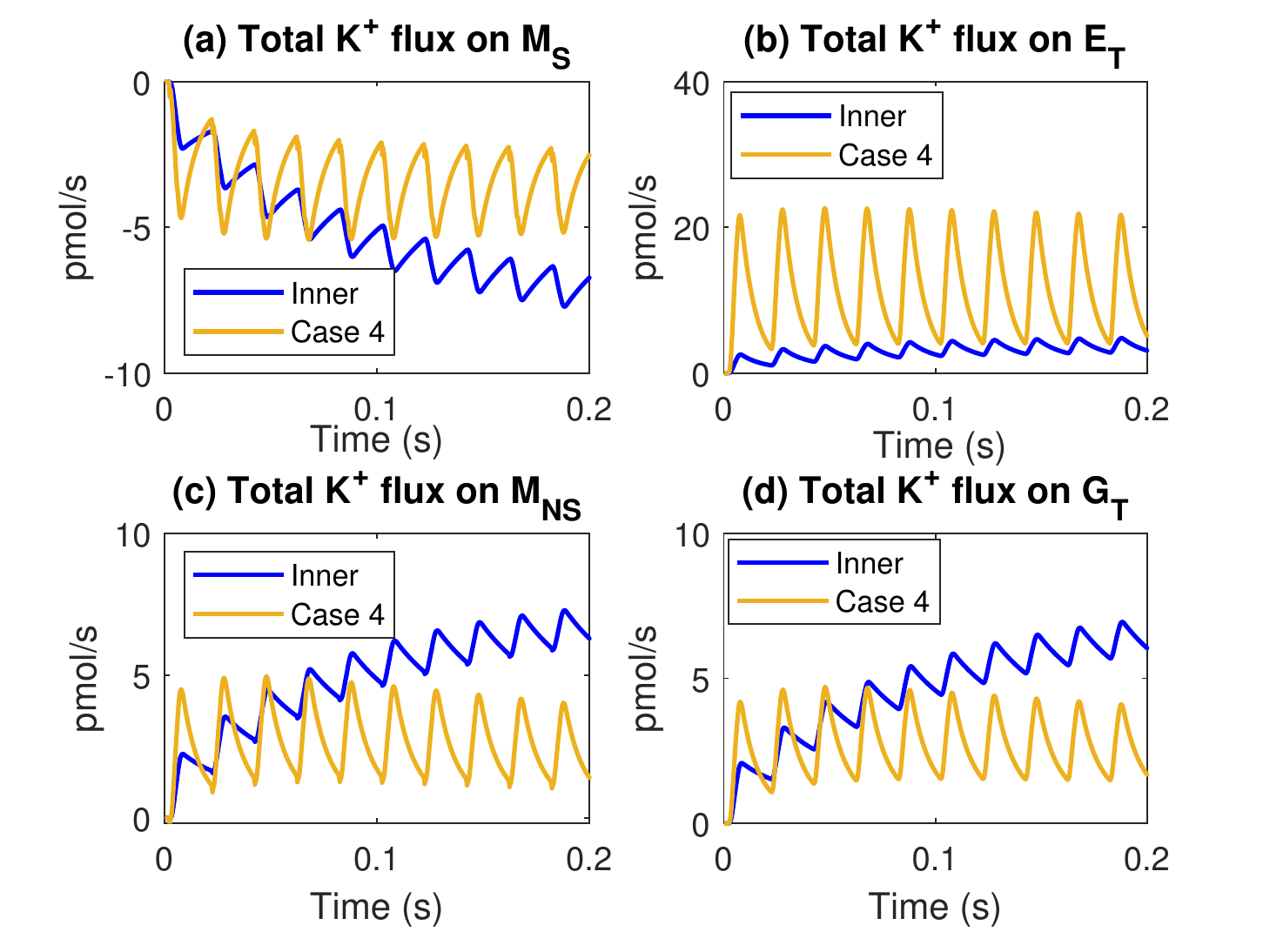} \includegraphics[width=3.25in,height=4.5cm]{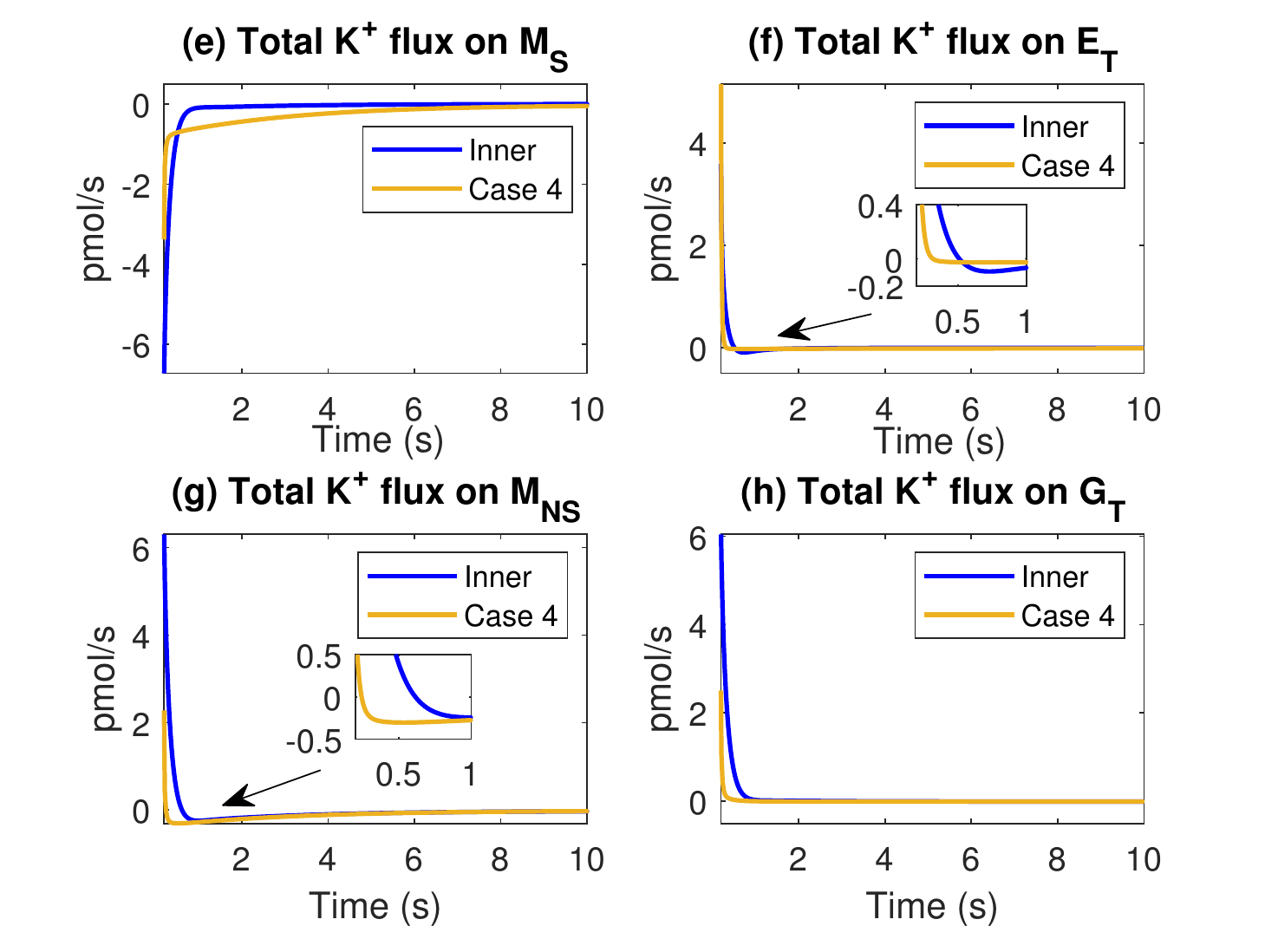}
	\caption{Comparison between spatially randomly stimulated case 4 with the spatially uniform  (inner) case.}
	\label{Fig_B_3}
\end{figure}

\textcolor{black}{Fig. \ref{Fig_B_1}-\ref{Fig_B_3} shows the potassium flux through $M_S,E_T,M_{NS}$ and $G_T$ in the spatially random stimulated case (Case 2 – Case 4) during a train of axon firing and after the axon stops firing. When the axon is firing, the pattern of periodic oscillation is hardly changed but the magnitude of the oscillation is lager than in the inner stimulated case.}

\textcolor{black}{\subsection{NKCC Parameters}\label{NKCC}
	In the table \ref{Table_2} below, we provide the parameters of the glial membrane when the NKCC channel is present. In resting state, the concentration of potassium and sodium and the electric potential in both glial compartment and extracellular space is the same in both cases, NKCC and baseline. }

\begin{table}[h]
	\centering
	\caption{NKCC Parameters}
	\label{Table_2}
	\begin{tiny}
		\begin{tabular}{c r r r r r }
			\hline
			& $I_{\max }^{N K C C}$ & $g_{g l}^{K}$ & $g_{g l}^{N a}$ & $I_{g l, 1}$ & $A_{g l}$ \\
			\hline 
			\text { Baseline } &  $0 \quad \mathrm{~A} / \mathrm{m}^{2} $ & $2.1 \mathrm{~S} / \mathrm{m}^{2}$ & $ 2.2 \times 10^{-3} \mathrm{~S} / \mathrm{m}^{2}$ & $4.78 \times 10^{-4} \mathrm{~A} / \mathrm{m}^{2}$ & $105 \mathrm{mM}$ \\
			\text { NKCCa } & $2 \times 10^{-3} \mathrm{~A} / \mathrm{m}^{2}$ & $2.88 \mathrm{~S} / \mathrm{m}^{2}$ & $1.65 \times 10^{-3} \mathrm{~S} / \mathrm{m}^{2}$ & $4.78 \times 10^{-4} \mathrm{~A} / \mathrm{m}^{2}$ & $34.7 \mathrm{mM} $ \\
			\text { NKCCb } & $2 \times 10^{-3} \mathrm{~A} / \mathrm{m}^{2}$ & $2.1 \mathrm{~S} / \mathrm{m}^{2}$ & $8.36 \times 10^{-4} \mathrm{~S} / \mathrm{m}^{2}$ & $2.49 \times 10^{-4} \mathrm{~A} / \mathrm{m}^{2}$ & $34.7 \mathrm{mM}$ \\
			\hline
		\end{tabular}
	\end{tiny}
\end{table}

\textcolor{black}{Table \ref{Table_3} shows the potassium concentration decay time in the extracellular stimulated region in the model with/without NKCC. The model with the NKCC channel in the glial membrane produces a faster decay of the potassium concentration in the extracellular space. }

\begin{table}[h]
	\centering
	\caption{Potassium Concentration Decay Time with/without NKCC}
	\label{Table_3} 
	\begin{tabular}{ c r r r }
		\hline & \text { 50\% decay } & \text { 70\% decay } & \text { 90\% decay } \\
		\hline 
		\text { Baseline } & $1.95 \mathrm{~s}$ & $3.58 \mathrm{~s}$ & $7.83 \mathrm{~s}$ \\
		\text { NKCCa } & $0.60 \mathrm{~s}$ & $1.11 \mathrm{~s}$ & $2.41 \mathrm{~s}$ \\
		\text { NKCCb } & $0.63 \mathrm{~s}$ & $1.17 \mathrm{~s}$ & $2.66 \mathrm{~s}$ \\
		\hline
	\end{tabular}
\end{table}

\subsection{Parameters }
\begin{table}[hb]
	\begin{tiny}
		\caption{Parameters in optic nerve model}
		\begin{tabular}{cccc}
			Parameters  &  Value   & Parameters  &  Value  
			\\
			\hline
			$R_{a}$ & $4.8 \times 10^{-5} \mathrm{~m}$ \ (Ref.\cite{kuffler1966physiological,bracho1975further}) & $\mu$ & $7 \times 10^{-4} \mathrm{~Pa} \cdot \mathrm{s}$ \ (Ref.\cite{mathias1985steady})\\
			\hline 
			$R_{b}$	 & $6\times 10^{-5}  \mathrm{m} $ \ (Ref.\cite{wang2019intraocular}) &	$c_{csf,IOP}^{Na}$ &	$111 \ \mathrm{mM}$\ (Ref.\cite{kuffler1966physiological}) \\
			\hline  
			$L$ & $1.5 \times 10^{-2} \mathrm{~m}$ (Ref.\cite{kuffler1966physiological}) & $c_{\text {csf,IOP }}^{\text {K }}$ & $3 \ \mathrm{mM}$\ (Ref.\cite{kuffler1966physiological})\\
			\hline 
			$e$ & $1.69 \times 10^{-19} \mathrm{~A} \cdot \mathrm{s}$ & $c_{gl}^{\text {Na,re }}$ & $7.57\ \mathrm{mM}$ (*) \\
			\hline
			$k_{B}$ & $1.38 \times 10^{-23} \mathrm{~J} / \mathrm{K}$ &  $c_{gl}^{K, re}$ & $100.84\  \mathrm{mM}$ (*,Ref.\cite{kuffler1966physiological}) \\
			\hline
			$T $& $296.15 \mathrm{~K}$ (Ref.\cite{kuffler1966physiological}) & $c_{ax}^{\text {Na,re}}$ & $10.17\  \mathrm{mM}$ (*) \\
			\hline
			$\eta_{ax}^{re}$ & $5 \times 10^{-1}$  (Ref.\cite{kuffler1966physiological}) & $c_{ax}^{K, re}$ & $100.04 \  \mathrm{mM}$ (*)\\
			\hline 
			$\eta_{gl}^{re}$ & $4 \times 10^{-1}$ \ (Ref.\cite{kuffler1966physiological}) & $A^{re}_{ax, gl}$ & $105 \ \mathrm{mM}$ (*)\\
			\hline
			$\eta_{ex}^{re}$ & $1 \times 10^{-1}$ (Ref.\cite{kuffler1966physiological}) &  $\tau_{ex}^{OP}$ & $0.16$ \ (Ref.\cite{mathias1985steady,malcolm2006computational})  \\
			\hline 
			$\mathcal{M}_{ax}$ & $5.9 \times 10^{6} \mathrm{~m}^{-1}$ \ (Ref.\cite{pilgrim1982volume}) & $\tau_{ex}^{SAS}$ & $1$ (*)  \\
			\hline 
			$\mathcal{M}_{gl}$ & $1.25 \times 10^{7} \mathrm{~m}^{-1}$\  (Ref.\cite{pilgrim1982volume}) & $\tau_{gl}$ & $0.5$ (*)  \\
			\hline
			$z^{Na, K}$ & $1$ & $p_{CSF}$ & $1.3 \times 10^{3} \mathrm{~Pa}$ \ (Ref.\cite{band2009intracellular}) \\
			\hline 
			$z^{Cl}$ & $-1$ & $p_{IOP}$ & $4 \times 10^{3} \mathrm{~Pa}$ \ (Ref.\cite{band2009intracellular}) \\
			\hline
			$z^{ax, gl}$ & $-1$\ (*) & $p_{OBP}$ & $0 \mathrm{~Pa}$\  \ (Ref.\cite{band2009intracellular}) \\
			\hline

			$ \gamma_{\text {ax,gl}}$ & $1$ \ (Ref.\cite{mathias1985steady,malcolm2006computational})&$ D_{ex,ax}^{Na}$ & $1.39 \times 10^{-9} \mathrm{~m}^{2} / \mathrm{s}$ (Ref.\cite{mathias1985steady})\\
			\hline
			$\gamma_{pia}$ & $1$ (Ref.\cite{mathias1985steady,malcolm2006computational}) & $D_{ex, ax}^{K}$ & $2.04 \times 10^{-9} \mathrm{~m}^{2} / \mathrm{s} $ \ (Ref.\cite{mathias1985steady}) \\
			\hline 
			$K_{\text {Na1,Na2}}$ & $2.3393 \mathrm{mM}$ (Ref.\cite{zhu2019bidomain})& $D_{ex,ax}^{Cl}$ & $2.12 \times 10^{-9} \mathrm{~m}^{2} / \mathrm{s}$ (Ref.\cite{mathias1985steady}) \\
			\hline
			$K_{K1}$ & $1.6154 \mathrm{mM}$ (Ref.\cite{zhu2019bidomain}) & $D_{gl}^{N a}$ & $1.39 \times 10^{-11} \mathrm{~m}^{2} / \mathrm{s}$ (Ref.\cite{mathias1985steady}) \\
			\hline 
			$K_{K 2}$ & $0.1657 \mathrm{mM}$ (Ref.\cite{zhu2019bidomain})& $D_{gl}^{K}$ & $2.04 \times 10^{-11} \mathrm{~m}^{2} / \mathrm{s}$ (Ref.\cite{mathias1985steady}) \\
			\hline
			$I_{gl,1}$ & $4.78 \times 10^{-4} \mathrm{~A} / \mathrm{m}^{2}$ (**,Ref.\cite{zhu2019bidomain})  & $D_{gl}^{Cl}$ & $2.12 \times 10^{-11} \mathrm{~m}^{2} / \mathrm{s}$ (Ref.\cite{mathias1985steady}) \\
			\hline 
			$I_{gl, 2}$ & $6.5 \times 10^{-5} \mathrm{~A} / \mathrm{m}^{2}$ (**,Ref.\cite{zhu2019bidomain}) & $k_{ex}^{OP}$  & $1.3729 \times 10^{-8} \mathrm{~m}^{2} / \cdot \mathrm{s}$ (Ref.\cite{malcolm2006computational}) \\
			\hline
			$I_{ax, 1}$ & $9.56 \times 10^{-4} \mathrm{~A} / \mathrm{m}^{2}$ (**,Ref.\cite{zhu2019bidomain})& $k_{ex}^{SAS}$ & $0 \mathrm{~m}^{2} / \mathrm{V} \cdot \mathrm{s}$\ (*) \\
			\hline 
			$I_{ax, 2}$ & $1.3 \times 10^{-4} \mathrm{~A} / \mathrm{m}^{2}$ (**,Ref.\cite{zhu2019bidomain})& $K_{ax}$ & $1.67 \times 10^{6} \mathrm{~Pa}$\ (Ref.\cite{hua2018cerebrospinal,lu2006viscoelastic})  \\
			\hline
			$g_{gl}^{N a}$ & $2.2 \times 10^{-3} \mathrm{~S} / \mathrm{m}^{2}$ (Ref.\cite{mathias1985steady})& $K_{gl}$ & $8.33 \times 10^{5} \mathrm{~Pa}$\ (Ref.\cite{hua2018cerebrospinal,lu2006viscoelastic}) \\
			\hline 
			$g_{gl}^{K}$ & $2.1 \mathrm{~S} / \mathrm{m}^{2}$ (Ref.\cite{mathias1985steady}) & $L_{dr}^{m}$ & $8.89 \times 10^{-13} \mathrm{~m} / \mathrm{Pa} \cdot \mathrm{s}$ (Ref.\cite{malcolm2006computational,zhu2019bidomain}) \\
			\hline 
			$g_{gl}^{Cl}$ & $2.2 \times 10^{-3} \mathrm{~S} / \mathrm{m}^{2}$ (Ref.\cite{mathias1985steady}) & $L_{pia}^{m}$ & $8.89 \times 10^{-13} \mathrm{~m} / \mathrm{Pa} \cdot \mathrm{s}$ (Ref.\cite{malcolm2006computational,zhu2019bidomain})\\
			\hline
			$g_{leak}^{Na}$ & $4.8 \times 10^{-3} \mathrm{~S} / \mathrm{m}^{2}$ (**,Ref.\cite{song2018electroneutral})& $L_{gl}^{m}$ & $1.34 \times 10^{-13} \mathrm{~m} / \mathrm{Pa} \cdot \mathrm{s}$ (Ref.\cite{malcolm2006computational,zhu2019bidomain}) \\
			\hline 
			$g_{leak}^{K}$ & $2.2 \times 10^{-2} \mathrm{~S} / \mathrm{m}^{2}$ (**,Ref.\cite{song2018electroneutral})& $L_{ax}^{m}$ & $7.954 \times 10^{-14} \mathrm{~m} / \mathrm{Pa} \cdot \mathrm{s}$ (Ref.\cite{villegas1960characterization}) \\
			\hline 
			$\bar{g}^{Na}$ & $1.357 \times 10^{1} \mathrm{~S} / \mathrm{m}^{2}$ (**,Ref.\cite{song2018electroneutral})& $\kappa_{g l}$ & $9.366 \times 10^{-19} \mathrm{~m}^{2}$ (Ref.\cite{malcolm2006computational,zhu2019bidomain}) \\
			\hline
			$\bar{g}^{K}$ & $2.945 \mathrm{~S} / \mathrm{m}^{2}$ (**,Ref.\cite{song2018electroneutral})& $\kappa_{ax}$ & $1.33 \times 10^{-16} \mathrm{~m}^{2}$ (Ref.\cite{malcolm2006computational,zhu2019bidomain})  \\
			\hline 
			$g_{ax}^{Cl}$ & $1.5 \times 10^{-1} \mathrm{~S} / \mathrm{m}^{2}$ (*) & $\kappa_{ex}^{OP}$ & $3.99 \times 10^{-16} \mathrm{~m}^{2}$ (**,Ref.\cite{malcolm2006computational,zhu2019bidomain})\\
			\hline 
			$G_{pia}^{Na,K,Cl}$ & $3 \mathrm{~S} / \mathrm{m}^{2}$ (*)& $\kappa_{ex}^{SAS}$ & $1.33 \times 10^{-14} \mathrm{~m}^{2}$ (**,Ref.\cite{malcolm2006computational,zhu2019bidomain}) 
		\end{tabular}
		\begin{tablenotes}
			\item[a] Note: the `*' estimated or induced from the concentration balance.
			\item[b] Note: the `**' deduct proportional from reference.
		\end{tablenotes}
	\end{tiny}
\end{table}

\end{document}